\documentclass[a4paper,10pt,openany,twocolumn,twoside]{book}

\pdfoutput=1 %for arxiv to recognise pdflatex

\usepackage{amsmath}
\usepackage{latexsym,amssymb}
\usepackage[nodayofweek]{datetime}
\usepackage{algorithmic} 
\usepackage[svgnames]{xcolor}
\usepackage{balance}


\usepackage{titlesec}
\titleformat{\chapter}[display]
  {\LARGE\bfseries}
  {}
  {0pt}
  {\thechapter.\ }
\titleformat{name=\chapter,numberless}[display]
  {\LARGE\bfseries}
  {}
  {0pt}
  {}

\titlespacing\chapter{0pt}{0pt plus 0pt minus 0pt}{12pt plus 2pt minus 2pt}
\titlespacing\section{0pt}{12pt plus 4pt minus 2pt}{6pt plus 2pt minus 2pt}
\titlespacing\subsection{0pt}{12pt plus 4pt minus 2pt}{4pt plus 2pt minus 2pt}
\titlespacing\subsubsection{0pt}{12pt plus 4pt minus 2pt}{2pt plus 2pt minus 2pt}

\usepackage[nottoc,numbib]{tocbibind}  % bib in toc

\usepackage{fancyhdr}
\usepackage[export]{adjustbox}% to allow vertical alignment
\usepackage{tikz}
\usetikzlibrary{arrows}

\usepackage{alltt}

\usepackage{url}
\usepackage[small,nooneline]{caption}

\usepackage{algorithm} % Algorithm formatting
\usepackage{algorithmic} % Floating thingy with algorithm within
\usepackage{paralist} % compact lists
\usepackage{color}

\usepackage[framemethod=TikZ]{mdframed}
\mdfdefinestyle{defn}{nobreak=true,linecolor=Blue,outerlinewidth=0.2,innerrightmargin=12pt,innerleftmargin=12pt,leftmargin=0pt,rightmargin=0pt,backgroundcolor=Blue!1,skipabove=12pt,skipbelow=12pt,roundcorner=5pt,frametitle={Definition},splitbottomskip=6pt,splittopskip=12pt}

\mdfdefinestyle{exmpl}{nobreak=true,linecolor=Orange,outerlinewidth=0.2,innerrightmargin=12pt,innerleftmargin=12pt,leftmargin=0pt,rightmargin=0pt,backgroundcolor=Yellow!1,skipabove=12pt,skipbelow=12pt,roundcorner=5pt,frametitle={Example},splitbottomskip=6pt,splittopskip=12pt}

\makeatletter
\newcommand{\twocolumntoc}{%
  \chapter*{\contentsname
    \@mkboth{%
      \MakeUppercase\contentsname}{\MakeUppercase\contentsname}}%
  \footnotesize%
  \@starttoc{toc}%
	\normalsize%
}
\makeatother

\begin{document}
\frontmatter
%\maketitle
\thispagestyle{empty}

\onecolumn
\begin{mdframed}[style=defn,frametitle={}]

\begin{centering}
\vspace{5mm}
{\LARGE \bf Visualising high-dimensional state spaces\\[2mm] with ``Tuple Plots''}

\vspace{5mm}
{\bf \large Susan Stepney}

\vspace{3mm}
University of York, UK

\vspace{3mm}
12 May 2019

\vspace{5mm}
\end{centering}

%\noindent\makebox[\textwidth]{\rule{\textwidth}{0.4pt}}
\end{mdframed}

\vspace{20mm}

\begin{quotation}
\parindent=0pt
  \rightskip=2cm
  \leftskip=2cm

\noindent
{\Large \bf Abstract}
\vspace{1mm}

Complex systems are described with high-dimensional data that is hard to visualise.
Inselberg's parallel coordinates are one representation technique for visualising high-dimensional data. Here we generalise Inselberg's approach, and use it for visualising trajectories through high dimensional state spaces.
%, including the case where the dimensionality and structure of the state space are time-dependent.
We introduce two geometric projections of parallel coordinate representations -- `plan tuple plots' and `side tuple plots' -- and demonstrate a link between state space and ordinary space representations.
We provide examples from many domains to illustrate use of the approach, including Cellular Automata, Random Boolean Networks, coupled logistic maps, reservoir computing, search algorithms, Turing Machines, and flocking.
\end{quotation}

\twocolumn

%\tableofcontents
\balance
\twocolumntoc

\mainmatter
%==========================================================
\chapter{Introduction}
%\balance

Complex systems can have complex and high dimensional state spaces.
As we try to understand such systems,
it is valuable to be able to visualise their state spaces.
But high-dimensional spaces are hard to visualise using conventional orthogonal coordinate systems.

Inselberg \cite{Inselberg85,Inselberg09} introduces parallel coordinates as a technique for visualising
high dimensional spaces.
Parallel coordinates have been exploited in two main application areas: geometry in higher dimensions ($\Re^N$) \cite{Inselberg85,Inselberg09}; visualising and exploring large data sets (thousands, or millions, of data points) in high dimensional ($N$ of a few tens) spaces \cite{MoustafaW06}.

Here, we generalise the technique to visualise both populations of multiple data points, and the trajectory of a single high dimensional point, in a potentially high dimensional ($N$ of 100s or more) state space.

The structure of this report is as follows.
\S\ref{sec:par} describes Inselberg's parallel coordinates, and some existing generalisations.
\S\ref{sec:plan} introduces plan tuple coordinates, and illustrates them in the context of cellular automata, random boolean networks, coupled logistic maps, reservoir computing, and population-based search.
\S\ref{sec:side} introduces side tuple coordinates, and illustrates them in the context of coupled logistic maps.
\S\ref{sec:hybrid} discusses hybrid tuple coordinates, and illustrates them in the context of Turing machines, and flocking visualisations.
\S\ref{sec:further} discusses further generalisations that could be developed.

%==========================================================
\chapter{Inselberg's Parallel coordinate plots}\label{sec:par}
\nobalance
%----------------------------------------------------------
\section{Introduction}
Conventionally, $N$D coordinate systems are drawn with $N$ orthogonal axes, with the obvious difficulties of visualisation when $N > 3$, or even $N=3$ on paper;
each datum is represented as a single point, with its coordinates given by the projections onto the various axes.

In 1985 Inselberg \cite{Inselberg85} introduced parallel coordinates.
In parallel coordinates, the $N$ axes are drawn in 2D space as a set of $N$ equally spaced parallel lines (usually arranged vertically like fence posts or a series of $y$ axes arranged along a single $x$ axis, or occasionally horizontally like ladder rungs or a series of $x$ axes up a single $y$ axis);  each datum is represented as a polyline joining the relevant coordinate value on each axis.

\begin{figure}[b]
\centering
% trim l b r t
$(a)$\includegraphics[clip, trim= 1cm 0.5cm 0.8cm 0.6cm, scale=0.5]{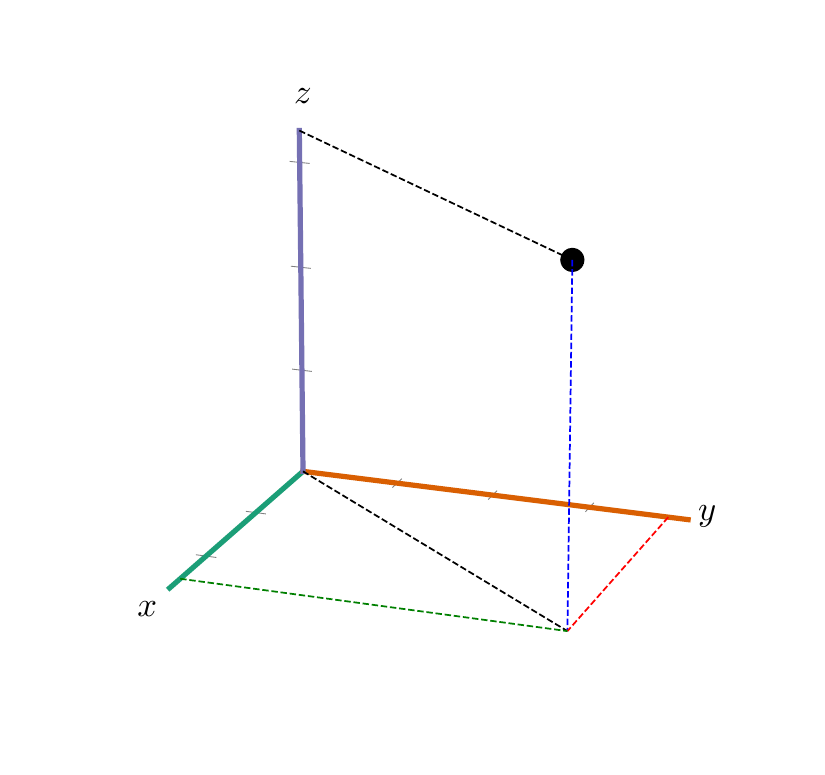}
$(b)$\includegraphics[clip, trim= 1cm 0.5cm 0.8cm 0.6cm, scale=0.5]{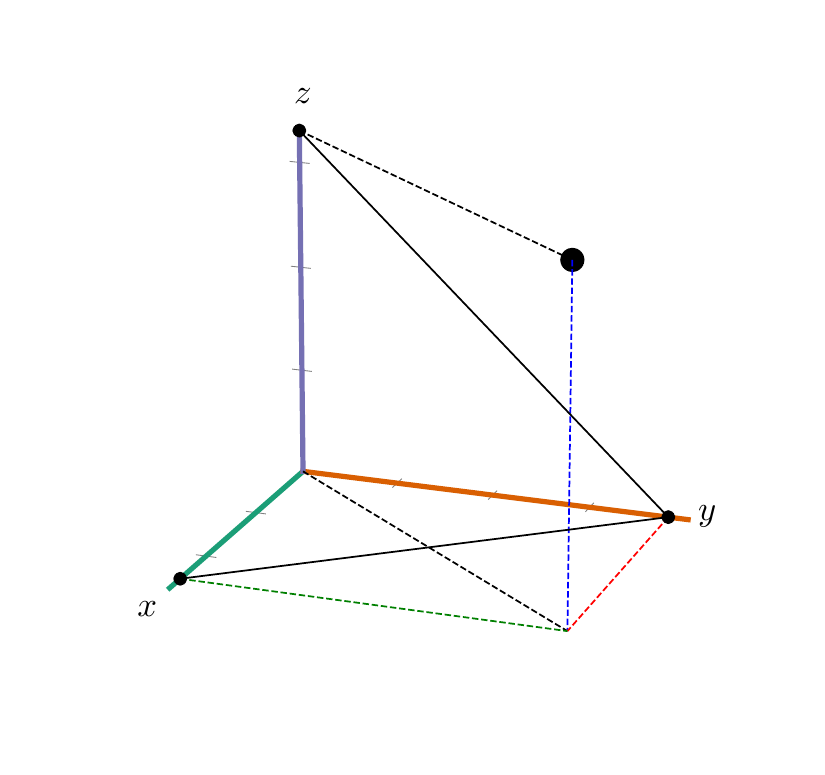}
$(c)$\includegraphics[clip, trim= 1cm 0.5cm 0.8cm 0.6cm, scale=0.5]{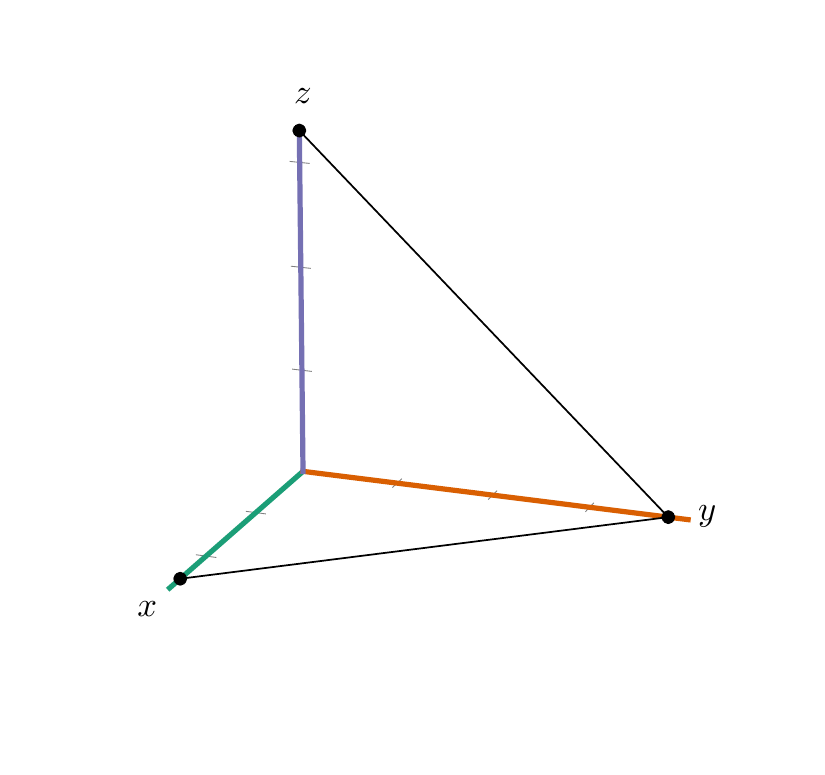}
$(d)$~\includegraphics[scale=0.7]{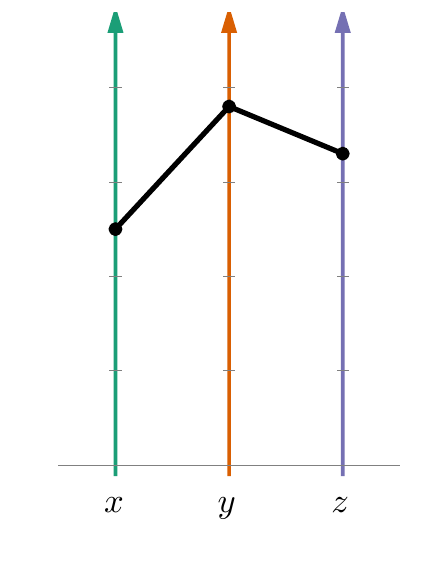}
\caption{\label{fig:parc}%
Plotting a 3D point $p_3$: $(a)$ in orthogonal coordinates, showing guide lines;
$(b)$ $p_3$ also projected onto axes, joined with a polyline;
$(c)$ removing the unprojected $p_3$ and guide lines;
$(d)$ ``flattening'' into parallel coordinates.
}
\end{figure}

\begin{figure}[t]
\centering
%~~~~~~~~~~\includegraphics[clip=true,trim= 1in 2in 0 0]{plan.pdf}
\includegraphics[width=\columnwidth]{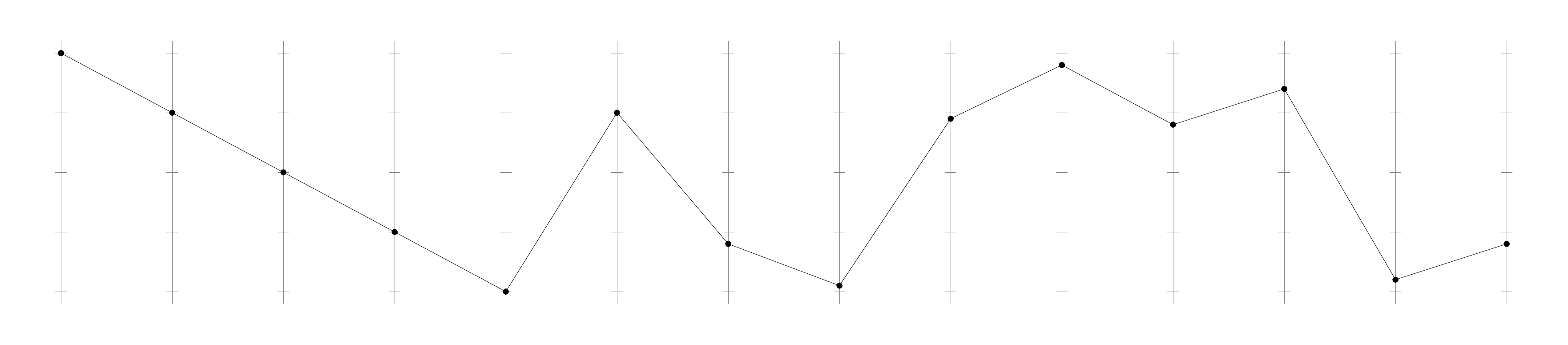}
\caption{Plotting a 14D point $p_{14}$ in parallel coordinates. 
}
\label{fig:parallel}
\end{figure}

Figure~\ref{fig:parc} demonstrates the approach.
There the 3D point $p_3$ is shown in both orthogonal and parallel coordinates. 
In orthogonal coordinates the point $p \in \Re^3$ is represented by the black disc, with dashed construction lines added for clarity. In parallel coordinates the point $p \in \Re \times \Re \times \Re$ is represented by the polyline passing through the three coordinate values.
Unlike the orthogonal form, the parallel coordinate form can be readily extended to higher dimensions, $\Re^N, N > 3$ (see figure~\ref{fig:parallel}).

%......................................................
\subsection{Notation}

Here and later we use the following notation to define the various plot types:
\begin{compactenum}
\item the real interval $[a,b] = \{ x\in\Re \,|\, a \leq x \leq b \}$
\item the integer interval $a..b = \{ k\in {\mathbb{Z}} \,|\, a \leq k \leq b \}$
\item a data type $V$: the set of values that the various components of the point to be plotted can take;
wlg we assume $V \subseteq \Re$, if not, assume an injective function $scale : V \rightarrow \Re$ is used to make the data values numerical
\item an $N$D point to be plotted: ${\bf p} = (p_1, p_2, \ldots, p_{N}) = (p_n)_{n = 1}^{N} \in V^N$
\item a set of $M$ such $N$-D points to be plotted: $\{\, {\bf p}_m \,\}_{m = 1}^{M} = \{\, p_{mn} \,\}_{m = 1;}^{M}{}_{n = 1}^{N}$
\item a $2$D position in the plotting plane: ${\bf x} = (x, y)  \in \Re^2$
\item a set of $M$ such $2$D positions in the plotting plane: $\{\, {\bf x}_m \,\}_{m = 1}^{M} = \{\, (x_m, y_m) \,\}_{m = 1}^{M}$
\item a set of symbols $S$ used to plot points in the plotting plane
\end{compactenum}

%......................................................
\subsection{Definition}

Given this notation, we can define the standard parallel coordinates plot as:

\begin{mdframed}[style=defn,frametitle={Definition: Parallel coordinates plot}]
Given
\begin{compactenum}
\item an $N$D point ${\bf p} = (p_n)_{n = 1}^{N} \in V^N$
\item a set of $M$ such $N$-D points: $\{\, {\bf p}_m \,\}_{m = 1}^{M}$
\end{compactenum}
\vspace{2mm}
A {\bf parallel coordinates plot} displays the point ${\bf p}$ in the rectangular-coordinate plane as the set of points 
\[ \{ {\bf x}_n \}_{n = 1}^{N} = \left\{\,  (n,p_n)  \,\right\}_{n = 1}^{N} \]
together with a polyline joining the ${\bf x}_n$ points.
See figure~\ref{fig:parallel}.

A {\bf parallel coordinates plot} displays the set of $M$ points $\{\, {\bf p}_m \,\}_{m = 1}^{M}$ as a collection of $M$ individual point ${\bf p}$ plots overlayed in the same rectangular-coordinate plane.
See figure~\ref{fig:parallel-hyper}.
\end{mdframed}

%.....................................................
\subsection{Visualising high dimensional geometry}

\begin{figure}[tp]
\centering
\includegraphics[trim = 15mm 8mm 11mm 3mm, clip,width=0.95\columnwidth]{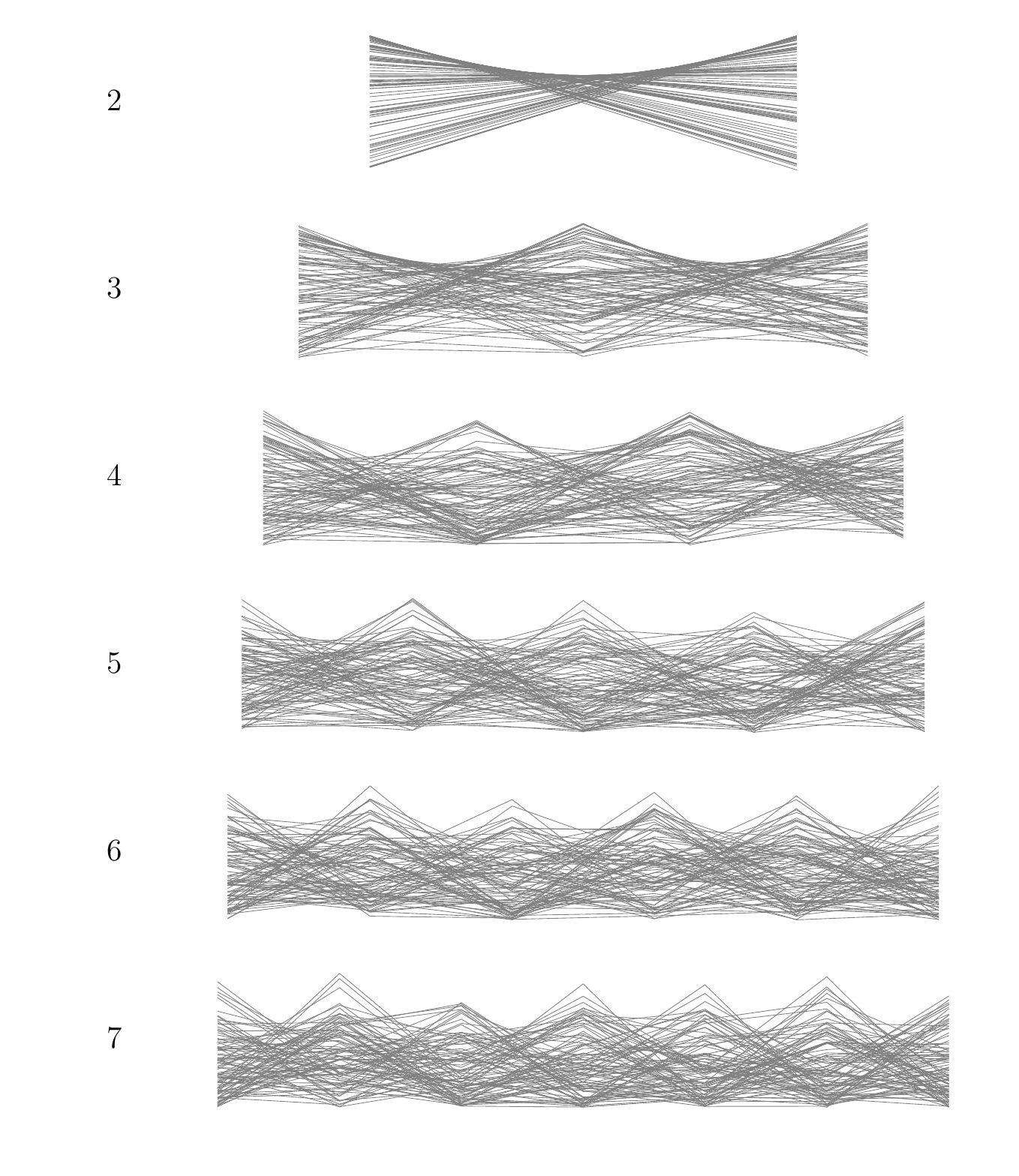}
\caption{100 random points on the surface of an $n$D hypersphere, for $n = 2$--$7$.
}
\label{fig:parallel-hyper}
\end{figure}

\begin{figure}[tp]
\centering
\includegraphics[trim = 15mm 6mm 11mm 3mm, clip,width=0.95\columnwidth]{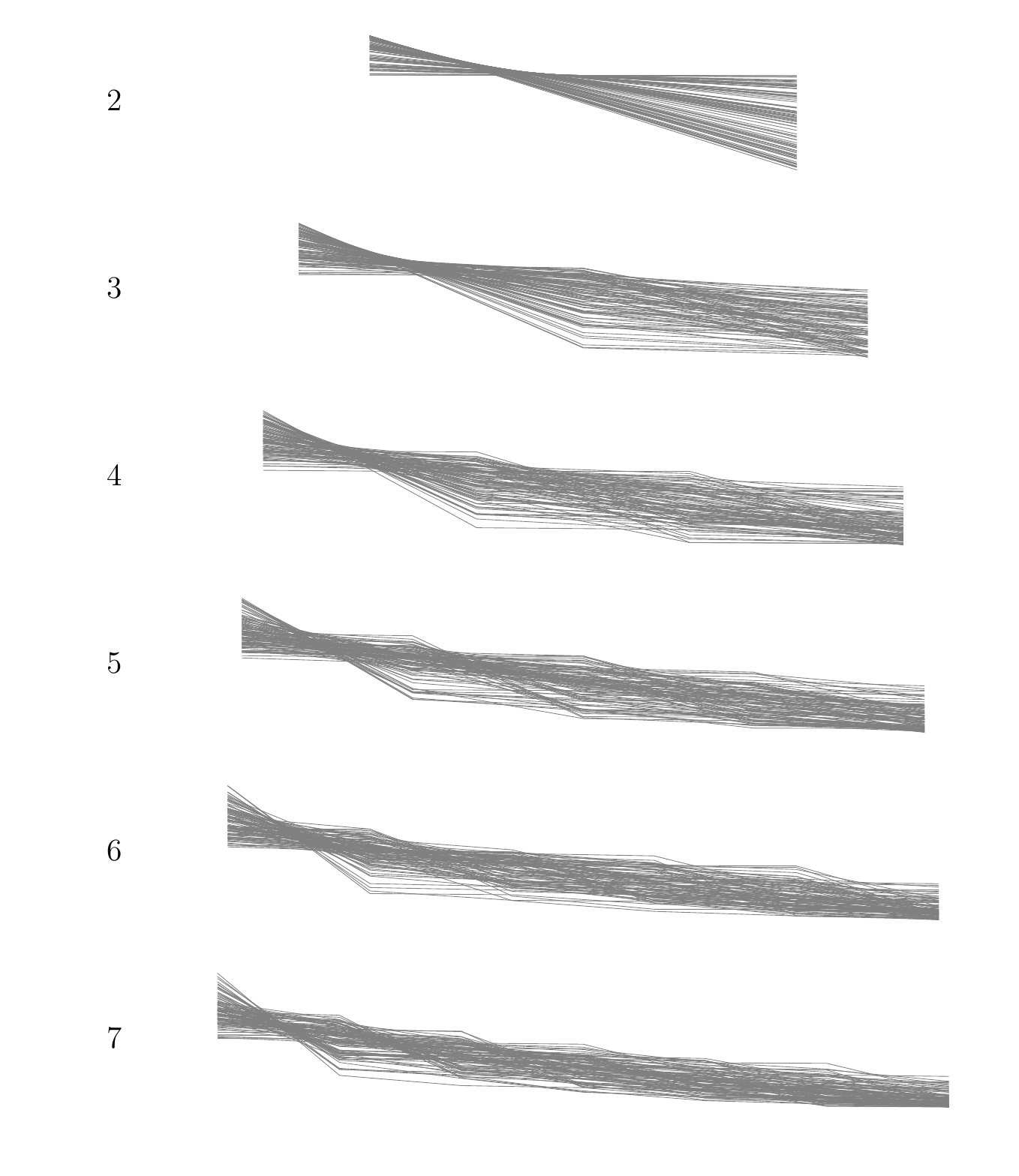}
\caption{100 random points on the surface of an $n$D hypersphere, for $n = 2$--$7$.
Here, the coordinate values of each point are sorted in descending order
(by symmetry, this is also a point on the surface of the hypersphere);
It is clear that for points where one dimension has a large value, the other dimensions have small values.
}
\label{fig:parallel-hyper-sorted}
\end{figure}

Inselberg's original emphasis was on geometrical applications in high dimensions (plotting lines, surfaces, hyperspheres, etc, see figures~\ref{fig:parallel-hyper},~\ref{fig:parallel-hyper-sorted}), so the parallel axes are necessarily equally spaced, in order to maintain the relevant geometrical properties.

%.....................................................
%\clearpage
\subsection{Visualising data}

\begin{figure}[tp]
\centering
(a)\includegraphics[trim = 4mm 30mm 107mm 1mm, clip,width=0.45\columnwidth]{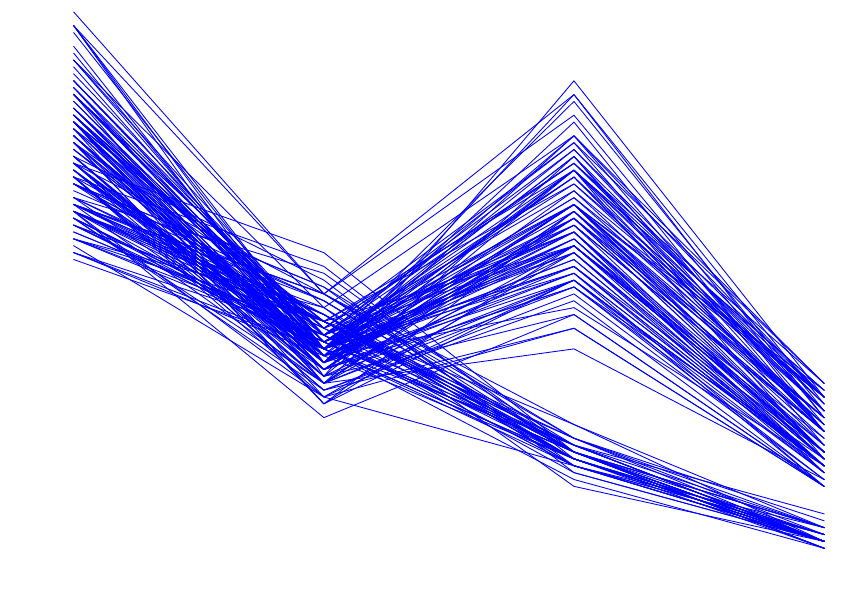}
(b)\includegraphics[trim = 105mm 10mm 2mm 1mm, clip,width=0.45\columnwidth]{parallel-iris.pdf}
\caption{\label{fig:parallel-iris}%
A parallel coordinates plot of the 4D iris data set, eliding the vertical axes:
$(a)$ plot of all 150 data points; $(b)$ colour coded plot by species.
}
\end{figure}

Parallel coordinates are also used for visualising large data sets.
Many data points correspond to many lines in the parallel system (see figure~\ref{fig:parallel-iris}).

\begin{figure}[tp]
%\hfil
%\includegraphics[scale=0.9]{figs/iris24.pdf}
\includegraphics[width=\columnwidth]{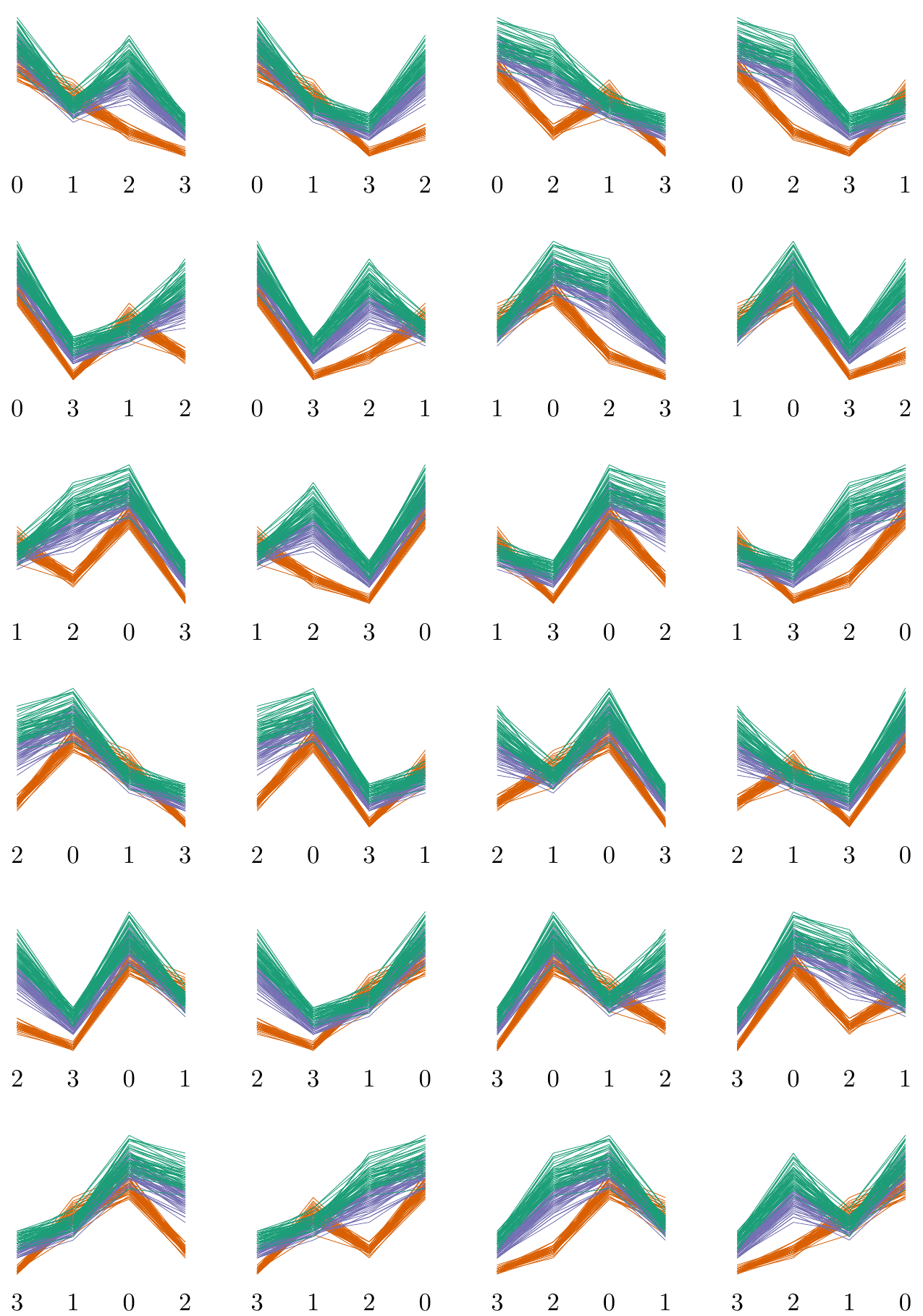}
%\hfil
\caption{A parallel coordinates plot, of the iris data, all 24 permutations of the $x_i$.
}
\label{fig:iris24}
\end{figure}

Such plots are used in data mining, where the coordinates can be heterogeneous.
This leads to techniques for scaling, reversing, and ordering the coordinates to expose
patterns and clusters in the data.  
(See figure~\ref{fig:iris24} for the effect of reordering the iris data axes.
Reordering is also used to expose structure in RBNs, \S\ref{sec:plan:rbn}.)
Smoothing the polylines to curves is also used to highlight clustering structure (rather than geometrical structures).  
This data mining use is often interactive, highlighting subsets of the plot, to help clarify and explore structure in the large number of overlapping data lines.

There are uses other than visualisation.
For example, Ye and Lin \cite{YeL06} use parallel coordinates to optimise a simulated annealing algorithm in $N$D. They express the $N$D search point in parallel coordinate space, and then approximate its polyline by an $M$D polynomial  (where $M < N$), thus reducing the dimensionality of the search space.

%\subsubsection{Representing permutations}\label{sec:perm}
%
%\todo{ Miller -- vector $->$ perm -- vis in parallel coords -- to do with order of co-ords -- monotonic increasing, axis order gives perm -- use as a repn -- see figure~\ref{fig:parallel-perm}}
%
%\todo{ ref SPV (smallest position value) paper}\newline
%\url{www.fatih.edu.tr/~msevkli/PSO%20for%20SMWTP.pdf}%
%
%
%\todo{ can see a small change to one number will result in a flip of two axes -- other permutation changes/mutations visualised in this repn.}
%
%

\begin{figure}[tp]
\includegraphics[trim = 8mm 9mm 0mm 2mm, clip,width=\columnwidth]{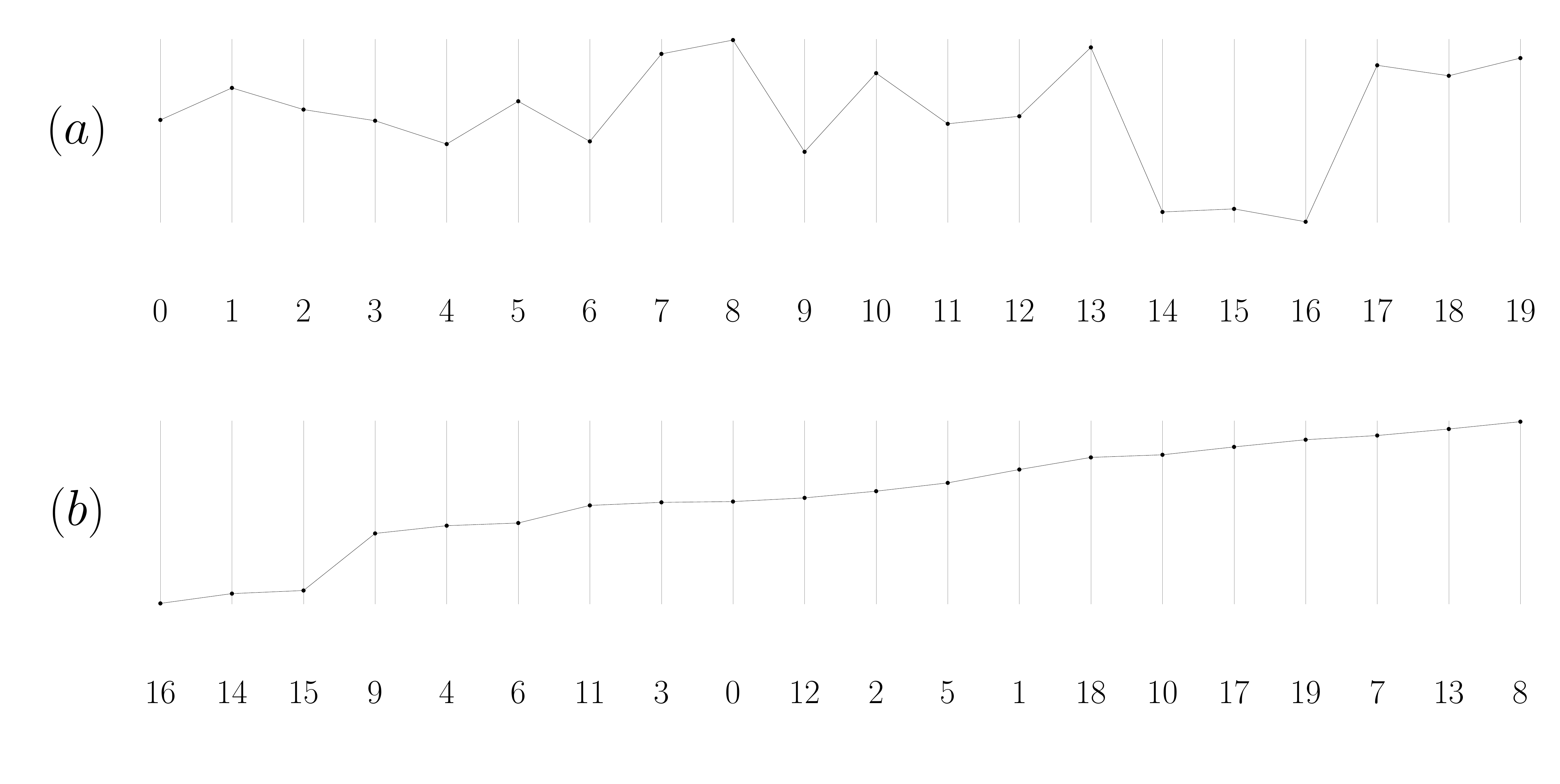}
\caption{A real vector representing a permutation; $N=20$.
(a) the unsorted real vector;
(b) the vector plotted with components in ascending order; the corresponding order of the axes gives the permutation represented by the real vector.}
\label{fig:parallel-perm}
\end{figure}

Another example of use is illustrating the algorithm for generating a permutation from a vector of real numbers~\cite{McCaskill-CompMatter-2018}: the desired permutation is that used to sort the vector into order (figure~\ref{fig:parallel-perm}).

%---------------------------------------------------------------------
\section{Polar parallel coordinate plots}
\label{sec:polar}
\begin{figure}[tp]
\centering
\includegraphics[trim = 6mm 5mm 2mm 1mm, clip]{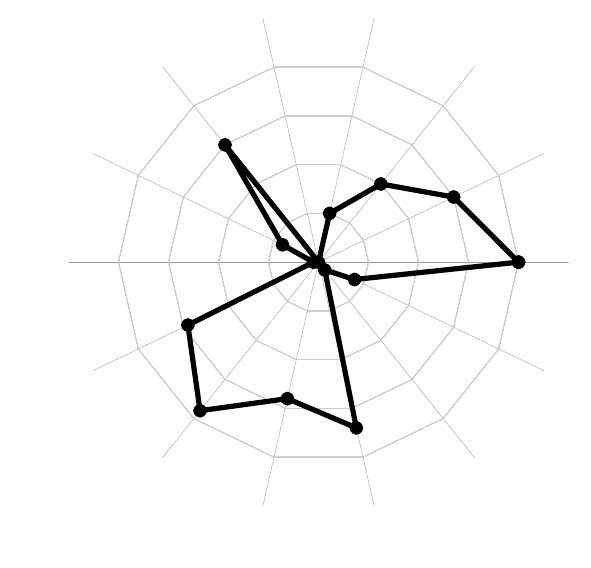}
\caption{Plotting a 14D point $p_{14}$ in radial coordinates.
}
\label{fig:radial14}
\end{figure}

Parallel coordinates lay out $N$ coordinate axes in parallel on a plane.
There are analogous ideas that use different layouts, such as using polar coordinates \cite[\S9.4.2.3]{Wilkinson05} to lay out the axes in a radial pattern.
Such examples are called \textit{star plots}, or \textit{spider plots}.
See figure~\ref{fig:radial14}.\footnote{%
Note that these are different from superficially similar \textit{radar plots} (different authors use different names for these plots; the names used here follow \cite[\S9.4.2.3, \S9.1.6.4]{Wilkinson05}).  A radar plot is a 2D plot,
in polar coordinates, where the angular dimension represents some continuous or discrete angular variable, such as time of day, month of year, or compass direction, and a single 2D datum is represented as a single point on the 2D plot.  In a spider plot,  in contrast, the $n$ radial lines represent the different axes of the different dimensions, and a single $n$D datum is represented by $n$ points on the 2D plot, one per axis, joined by a closed polyline. 
}

Wilkinson calls these multi-dimensional plots ``polar parallel coordinates'' plots \cite[\S 9.4.2.3]{Wilkinson05}, because in his graphics grammar, they are produced by passing a parallel coordinate plot through a polar transform.

\begin{mdframed}[style=defn,frametitle={Definition: Polar parallel coordinates plot}]
Given
\begin{compactenum}
\item an $N$D point ${\bf p} = (p_n)_{n=1}^{N} \in V^N$
\end{compactenum}
\vspace{2mm}
A {\bf polar parallel coordinates plot} displays ${\bf p}$ in the $(r,\theta)$ polar-coordinate plane as the set of points \[\{\, {\bf x}_n \,\}_{n = 1}^{N} = \{\, (p_n, 2\pi n/N)  \,\}_{n = 1}^{N}\] together with a closed polyline joining the ${\bf x}_n$ points.
(See figure~\ref{fig:radial14}.)
\end{mdframed}

%\balance
This approach has the advantage of not giving undue prominence to the first and last coordinates, and makes `shapes' that are readily comparable. 
It has the disadvantage of emphasising large values more than small ones, due to area effects.

%---------------------------------------------------------------------
\section{Generalisation}

Parallel coordinates are one possible way of visualising the large dimensional state spaces of dynamical systems, a formalism suitable for several unconventional computational models
\cite{SS-NCDynSys}.

\begin{figure*}[t]
\centering
\includegraphics[trim = 0mm 0mm 0mm 8mm, clip,width=0.85\textwidth]{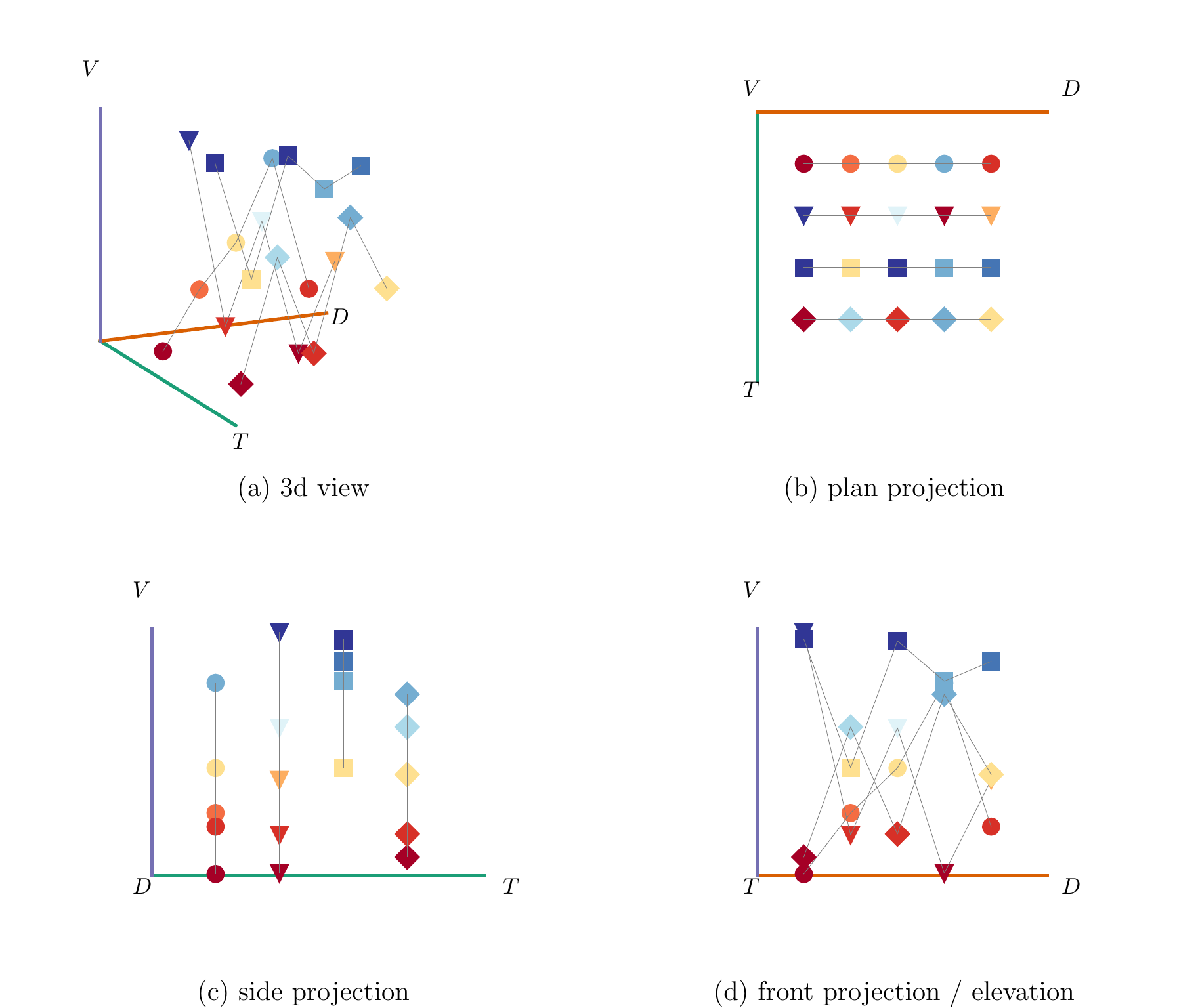}
\caption{A parallel coordinates plot of multiple $N$D points, each point in a separate time `plane' $T$.
Each point has its own marker shape, to help distinguish it here;
marker colour represents the value, the height on the $V$ axis.
(a) the full 3d plot; 
(b) the `plan projection', top view;
(c) the `side projection' view; 
(d) the `front projection' (or elevation) view, 
recovering the standard parallel coordinates plot of multiple points.
}
\label{fig:3d3projections}
\end{figure*}

The rest of this report modifies parallel coordinates
to visualise the state space trajectories of high dimensional dynamical systems.  
A trajectory is a time series of state space points.  
This series could be shown on a single parallel coordinates plot as a collection of lines, but this loses the time ordering.
However, many parallel coordinate plots can be reduced to a 1D view without loss of information, and then the time series can be visualised by plotting the time series as a sequence of 1D states.

%\balance
There are two conceptual steps in this visualisation process:
\begin{itemize}
\item a high dimensional space can be represented using coordinate axes that are ``flattened'', and laid out in parallel, rather than arranged orthogonally (Inselberg's parallel coordinates)
%\item These ``flattened'' coordinate axes do not need to be displayed in parallel (eg polar parallel coordinates, \S\ref{sec:polar})
\item the coordinate axes do not need to be drawn in the plane of the paper; they can be drawn in other projections (see figure \ref{fig:3d3projections})
%, as plan and side tuple plots, next
\end{itemize}

These plots exploit the trivial isomorphism between $X^N$, an $N$ dimensional point, and $X \times X \ldots X$ ($N$ times), of $N$ 1D points representing the tuple of the point's $N$ coordinates. 
Plots exploiting the latter structure are here referred to as ``tuple plots'', a term that covers all of parallel coordinates, polar parallel coordinates, and the plots developed here in the later chapters.
The name is modified to reference the geometry used to draw the individual axes: plan tuple plots (\S\ref{sec:plan}) and side tuple plots (\S\ref{sec:side}).

%===========================================================================
\chapter{Plan Tuple Plots}\label{sec:plan}
\nobalance
%---------------------------------------------------------------------
\section{Looking down}\label{sec:plan:look}
%.....................................................................
\subsection{Motivation}\label{sec:plan:look:motiv}

\begin{figure}[b]
\includegraphics[trim = 14mm 11mm 3mm 4mm, clip,width=\columnwidth]{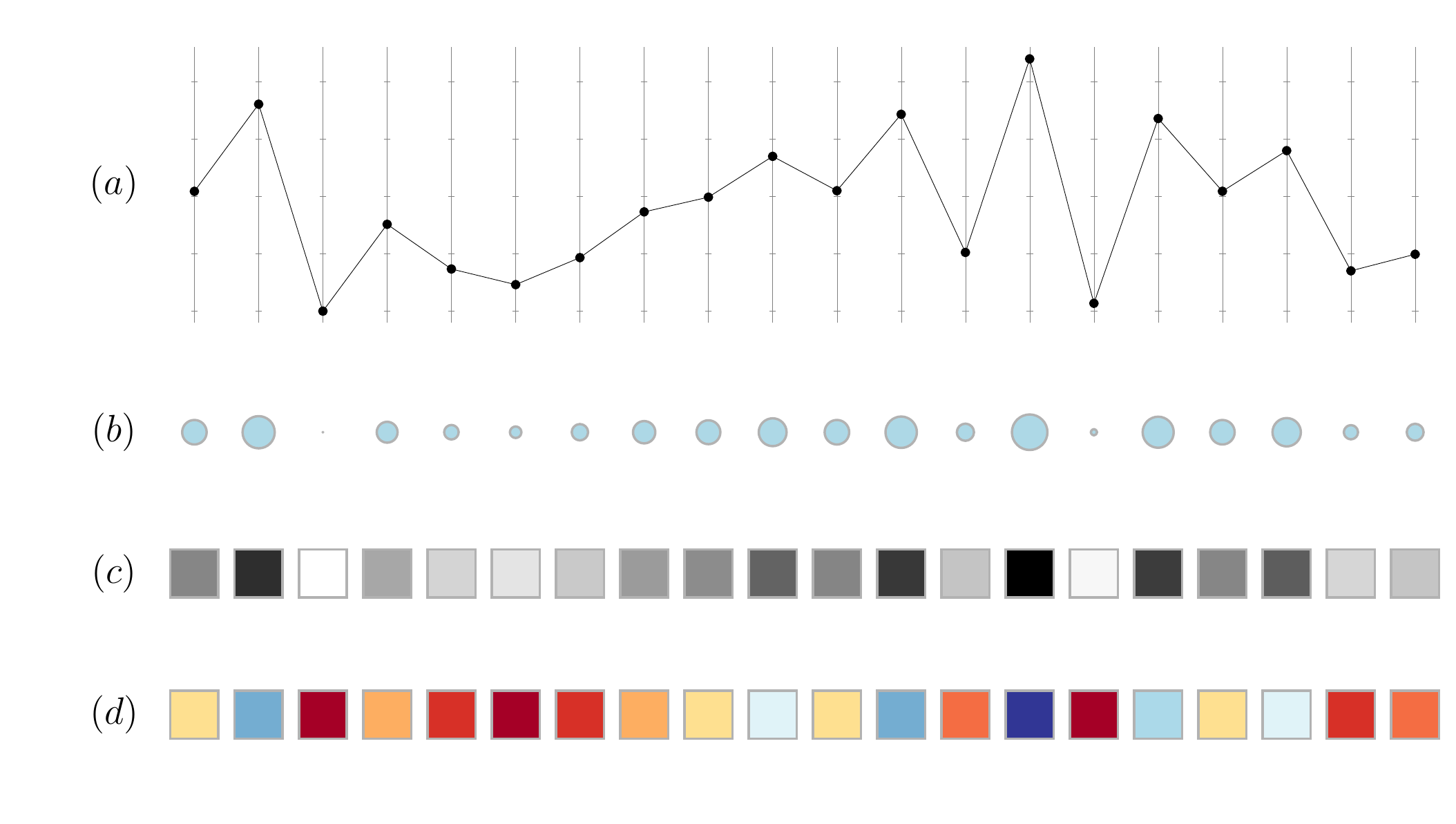}
\caption{20D real-valued random data (so $N=20, V=\Re$):
$(a)$ parallel coordinates plot;
$(b)$ the `plan' view, from projecting the values onto the horizontal axis: 
plan tuple plot using data points with sizes proportional to their value $v$; 
$(c)$ plan tuple plot, using grey levels instead of sizes;
$(d)$ plan tuple plot, using a diverging colour range, with low values being red, moving through yellow, to high values being blue, suitable for ranges with positive and negative values.
}
\label{fig:plan-motivate}
\end{figure}

Consider a system with $N$ identical dimensions each with values drawn from an ordered set $V$. 
We could draw this using $N$ parallel coordinates.  
Imagine these axes are lines sicking up from the ground, like fence posts; consider looking ``down'' on these coordinates from above (figures~\ref{fig:3d3projections}b, \ref{fig:plan-motivate}), so the line of each fence post is foreshortened into a point.  

The lower values are further away, so we might think that perspective would make the circles representing these points look smaller, with say size proportional to value $v \in V$ (figure~\ref{fig:plan-motivate}b).
 Alternatively, the further away values might appear to be ``fainter'', shaded differently, with 
 grey level proportional to value $v$ (figure~\ref{fig:plan-motivate}c).
 Or we could simply ``paint'' a suitable range of colours on the axes at different levels (figure~\ref{fig:plan-motivate}d).

%This basic approach works for ordered values $V$, where the `fence post' representing the axis is straight.
%For cyclic values, for example angles and vectors, the symbol could be chosen as an arrow or other marker of the angular direction (see figure~\ref{fig:plan-vector}).

We can reorder the axes to help expose properties of the data.
The permutation example from figure~\ref{fig:parallel-perm} can be redrawn in plan view, as figure~\ref{fig:plan-perm}.
See also \S\ref{sec:axisorder}.

%We can even place the axes at arbitrary points in the 2D plotting plane to expose properties of the data, for example, 
%Game of Life CA (\S\ref{sec:plan-ca})
%.....................................................................
\subsection{Definition of plan tuple plot}\label{sec:plan:defn}

We use this ``looking down'' idea to define plan tuple plots.
We plot one 2D point for each component of the tuple;
its position represents the axis index, and its symbol (size, shape, colour) represents its value.

\begin{figure}[tp]
\centerline{
\includegraphics[width=0.99\columnwidth]{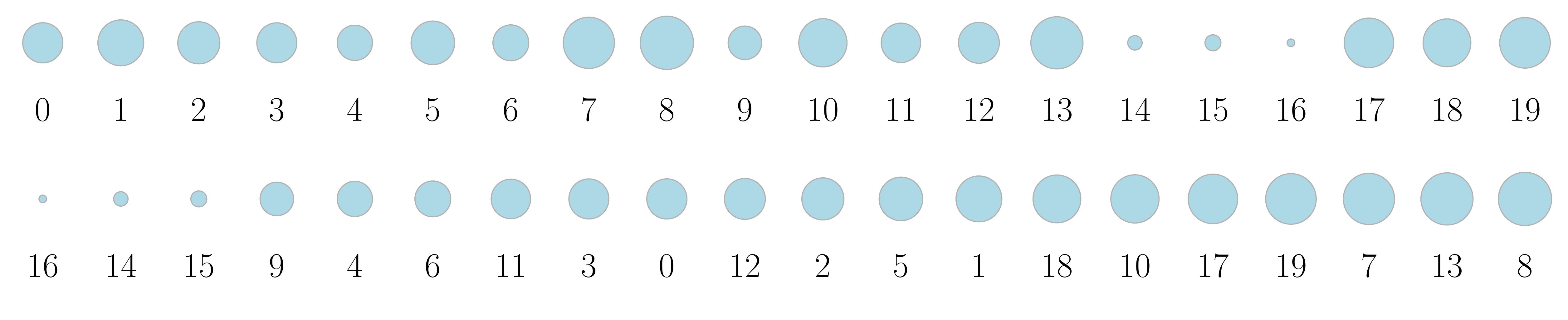}
}
\caption{The permutation example from figure~\ref{fig:parallel-perm}, drawn as plan tuple plot.
The upper view has the axes in their original order.
The lower view has the axes permuted to show the data in ascending order.
}
\label{fig:plan-perm}
\end{figure}

\begin{figure}[tp]
\centerline{
\includegraphics[width=0.99\columnwidth]{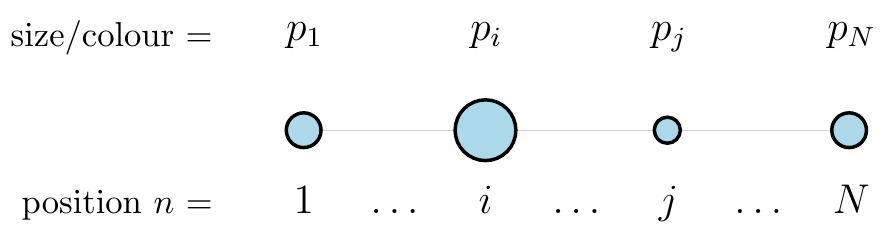}
}
\caption{Illustration of definition of plan tuple plot of the $N$D point
${\bf p} = (p_1, \ldots, p_i, \ldots, p_j, \ldots, p_{N}) \in V^N$.
Here the plotting function $sym$ maps each value $v \in V$ to a symbol $s \in S$ where the size of $s$ is proportional to $v$.
Each component $p_n$ is plotted on the horizontal axis line at position $(n,1)$ using symbol $sym(p_n)$.
The axis line itself may be elided.
}
\label{fig:plandef}
\end{figure}

\begin{mdframed}[style=defn,frametitle={Definition: plan tuple plot}]
Given
\begin{compactenum}
\item an $N$D state space $V^N$
\item an $N$D point ${\bf p} = (p_n)_{n=1}^{N} \in V^N$
\item a plotting function $sym:V\rightarrow S$ 
\item a position function $\varpi: 1..N \rightarrow \Re^2$ that maps the component indexes to positions in the plotting plane
\end{compactenum}
\vspace{2mm}
A {\bf plan tuple plot} displays the point ${\bf p}$ in the plotting plane as the set of points 
\[ \{\, {\bf x}_n \,\}_{n = 1}^{N} = \{\, \varpi(n) \,\}_{n = 1}^{N} \]
where each ${\bf x}_n$ point is plotted using the symbol $sym(p_{n})$.
See figure~\ref{fig:plandef}.
\end{mdframed}

%\vspace{3mm}

In the basic case, $\varpi(n) = (n,1)$: the points are plotted evenly along the $x$ axis, in axis order.

The axes may be permuted using $\varpi$, so that the points are plotted evenly along the $x$ axis, in permuted axis order.
Examples using a permuted order include: the sorted display of figure~\ref{fig:plan-perm};
Random Boolean Networks (\S\ref{sec:plan:rbn}).

In the general case, $\varpi$ can be a function that takes axis $n$ to some position in the $(x,y)$ plotting plane: the points are plotted at these positions.
Examples using a general position function include: Boolean hypercubes plotted on Hilbert curves \cite{Wiles2002};
Game of Life CA (\S\ref{sec:plan-ca}).

%.....................................................................
\subsection{Ensembles and trajectories}\label{sec:plan:traj}
With parallel coordinates, multiple points can be plotted as multiple overlapping polylines.
With plan tuple plots, such overlapping would obscure the data.

We might wish to display an \textit{ensemble} of points, representing some population in $N$D space,
or a \textit{time series} of points, representing some trajectory through $N$D space.
If we animated a plan tuple plot of a sequences of states, the $N$ symbols representing the $N$D point would stay fixed in position (on their axis position), and change in size or colour (as their value changes).
We can turn the animation into a static display if we use a 1D plan tuple plot for a single point, and use the second dimension of the plot to display the different points (see figure~\ref{fig:3d3projections}b).

%\balance

Assume we have some indexed set of $N$D points that we wish to plot: ${\bf p}_1, {\bf p}_2, \ldots, {\bf p}_M$.  We may wish to permute the points' indexes to expose structure, unless the indexes represent some intrinsic order, such as time.
We can then plot ${\bf p}_j$ along the $x$ axis that intercepts the $y$ axis at $j$.

\begin{figure}[tp]
\centerline{
\includegraphics[trim = 7mm 5mm 1mm 1mm, clip,scale=0.8]{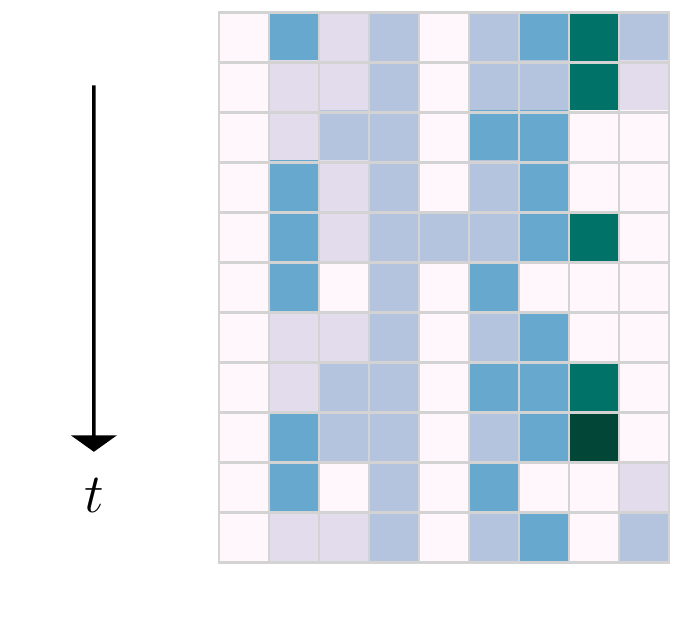}
}
\caption{A trajectory, or time series, as a plan tuple plot.
}
\label{fig:plan-traj}
\end{figure}

\begin{mdframed}[style=defn,frametitle={Definition: ensemble plan tuple plot}]
Given
\begin{compactenum}
\item an $N$D state space $V^N$
\item an indexed set of $M$ $N$D points: $\{\, {\bf p}_m \,\}_{m = 1}^{M} = \{\, p_{mn} \,\}_{m = 1;}^{M}{}_{n = 1}^{N}$ where each $N$-D point ${\bf p} = (p_n)_{n=1}^{N} \in V^N$
\item a plotting function, $sym:V\rightarrow S$  
\item a position function of the axes' indexes, $\varpi_n: 1..N \rightarrow \Re$
\item a position function of the points' indexes, $\varpi_m: 1..M \rightarrow \Re$ 
%\item  an injective function $posn: 1..M \times 1..N \rightarrow \Re^2$ that maps the component indexes to positions in the plotting plane
\end{compactenum}
\vspace{2mm}
An {\bf ensemble plan tuple plot} displays each of these ${\bf p}_m$ in the plotting plane as a plan tuple plot, with 
\[ \{ {\bf x}_{mn} \}_{m = 1;}^{M}{}_{n = 1}^{N} = \{ (\varpi_n(n),\varpi_m(m))  \}_{m = 1;}^{M}{}_{n = 1}^{N} \]
where each ${\bf x}_{mn}$ point is plotted using the symbol $sym(p_{mn})$.
\end{mdframed}

%This is isomorphic to plotting the $p_{ij}$ with $posn(ij) = (\varpi_m(i),\varpi_n(j))$.

%The same axis permutation $\varpi_n$ is used for all the points $p_m$.

In the basic case, $\varpi_n(n) = n, \varpi_m(m) = m$: each point is plotted evenly along the $x$ axis, in axis order, at a height $m$ up the $y$ axis.

If $M$ represents time, then $\varpi_m(m) = -m$ is conventionally used,
to maintain the temporal order, and to have time increase down the page
 (figure~\ref{fig:plan-traj}). 

This trajectory visualisation does not necessarily make cycles in the trajectory
immediately visible, but they can be inferred as repeated patterns of states.

%---------------------------------------------------------------------
\subsection{Vectors}\label{sec:plan-vector}

\begin{figure}[tp]
\centering
\includegraphics[width=0.95\columnwidth]{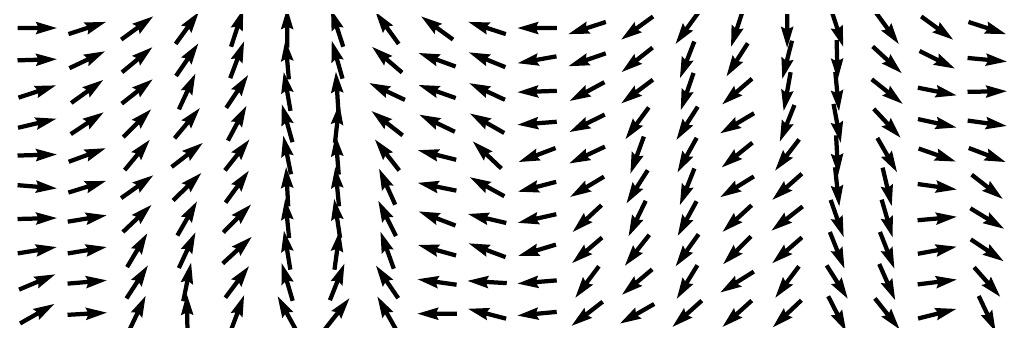}
\caption{An array of unit 2D vectors: the initial state is a uniform rotation;
each timestep each node has a small random rotation applied.
} 
\label{fig:plan-vector}
\end{figure}

If the values $V$ are vector, rather than scalar, we can adapt the symbol used with the plan plot.
Each 2D vector value (direction and size) can be plotted as a small arrow.
See for example, figure~\ref{fig:plan-vector}.

%\todo{Interpretation}
%---------------------------------------------------------------------
%\clearpage
\section{No information loss}
An important feature of the plan tuple plot is that it loses no information,
other than possible loss of resolution between the symbols.
Where the different values can be suitably distinguished (drawn from a small finite set, say), such as for the CA and RBN examples shown above, there is no loss of resolution.
Figure~\ref{fig:plan-clm} shows an example of a plan tuple plot used with continuous-valued data, $V = [0,1]$.  Here the grey scales encode the real values with finite resolution;
compare orthogonal and standard parallel coordinates, which use position to encode the real values also with finite resolution.

%---------------------------------------------------------------------
%\clearpage
\section{Examples}\label{sec:plan:egs}
%\nobalance
%........................................................
\subsection{Hypersphere surface}\label{sec:plan:hyper}
Given an $N$D hypersphere of radius $r$, a point ${\bf p} = (p_1,p_2,\ldots,p_N)$ on its surface obeys
\[ p_1^2 + p_2^2 + \ldots p_N^2 = r^2 \]

\begin{mdframed}[style=exmpl,frametitle={Example: hypersphere plan tuple plot}]
Given
\begin{compactenum}
\item an $N$D state space $V^N = [0,r]^N$
\item an indexed set of $M$ state space points $\{ {\bf p}_m \}$ sampled at random from the positive hyper-octant of the surface of a hypersphere
\item a plotting function $sym = cubehelix$ that maps a component value $p\in V$ to a colour scale with extremes $0 \mapsto white, r \mapsto black$
\item each of the $M$ point has its axes individually permuted so that the values are in descending order
\item an axis position function $\varpi_n = \mbox{Id}$
\item an index position function $\varpi_m$ on the indexes $m \in 1..M$ such that points ${\bf p}_m$ are sorted in ascending order of their maximum component values $max(\{ {p}_{mn} \}_{n=1}^N)$
\end{compactenum}
\vspace{2mm}
An {\bf ensemble plan tuple plot} displays each of these ${\bf p}_m$ in the plotting plane as a plan tuple plot, with 
\[ \{ {\bf x}_{mn} \}_{m = 1;}^{M}{}_{n = 1}^{N} = \{ (\varpi_m(m),n)  \}_{m = 1;}^{M}{}_{n = 1}^{N} \]
%where each ${\bf x}_{mn}$ point is plotted using the symbol $cubehelix(p_{mn})$.
\end{mdframed}
We additionally ensure that each point ${\bf p}$ has $p_i \ge p_j$ if $i>j$,
by sorting the point's components into descending order.
By symmetry, this is also a point on the sphere, and this sorting helps to expose the structure.

The use of $\varpi_m$ means that points closer to an orthogonal axis, which have one large component value, and the rest small, occur towards the top of the plot; points maximally distant from the orthogonal axes, which have all component values approximately equal, occur lower in the plot.

\begin{figure*}[tp]
\centering
\includegraphics[width=\textwidth]{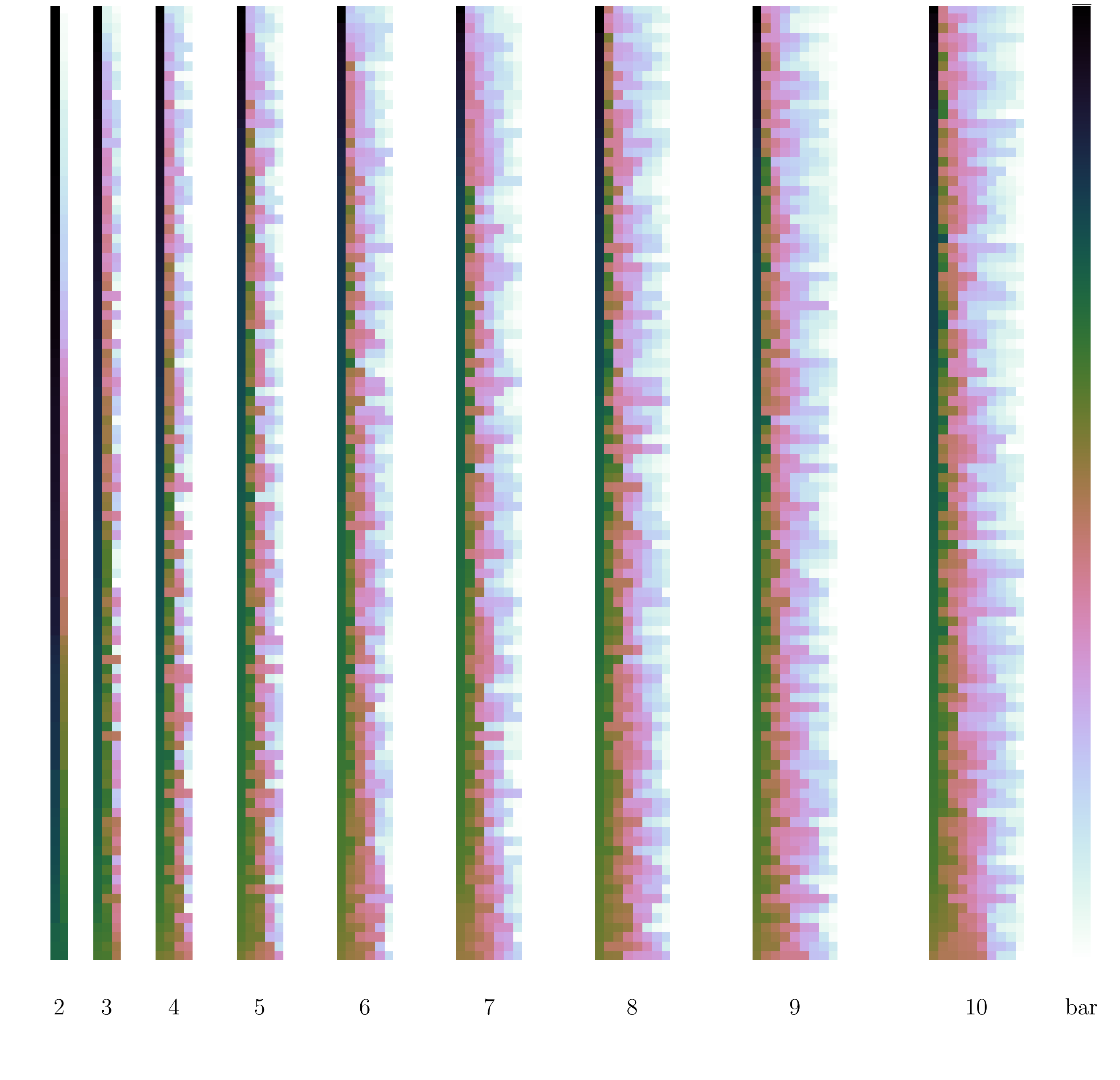}
\caption{Ensemble plan tuple plot of 100 points on the surface of the positive hyper-octant of an $N$-dimensional hypersphere, $N=2$--$10$.
The coordinate values of each point are sorted in descending order (by
symmetry, this is also a point on the surface of the hypersphere).
The ensemble of points are sorted in order of the maximum coordinate value.
From this is is clear that when there is one large value (see the points towards the top of the plot), visible as a single dark square, 
then all the other values are small, visible as very light coloured squares,
whereas when there is no very large value (see the points in the bottom part of the plot), there can be several medium sized values, with intermediate colour squares.
}
\label{fig:plan-hyper}
\end{figure*}

The resulting ensemble plan tuple plot for $M=100$, $N=2..10$ is shown in figure~\ref{fig:plan-hyper}. 

%\todo{version with the $n$ permuted individually, exploiting symmetry of hypersphere?}
%........................................................
\newpage
\subsection{Elementary cellular automata}\label{sec:plan-ca}

Consider an elementary cellular automaton (ECA) \cite{Wolfram1983-ECA} with $N$ nodes. 
Each node has 3 inputs, from its left and right neighbour, and from itself.
The specific ECA is defined by the choice of a boolean function of three inputs.
Each node has a binary-valued state. 
At each timestep, the state of each node is updated in parallel, 
by applying the boolean function to its inputs.

An ECA with $N$ nodes laid out in 1D line in (physical) space\footnote{%
The ``dimensionality'' of the layout in physical space (a 1D line of
nodes) is unrelated to the dimensionality of the state space (of $N$D, because there are $N$ nodes); if the same $N$ nodes were instead laid out in a 2D grid, or even a 27D grid, say, it would not affect the dimensionality of the state space. Instead, this ``1D'' refers to the {\sl topology} of the connections between
the nodes, and hence the potential information flow between the nodes.} 
has an (abstract) state space of ${\mathbb B}^N$.
The parallel coordinate view has one axis line per node,
each with two possible values, 0 or 1.

\begin{figure}[tp]
\centering
\includegraphics[width=0.98\columnwidth]{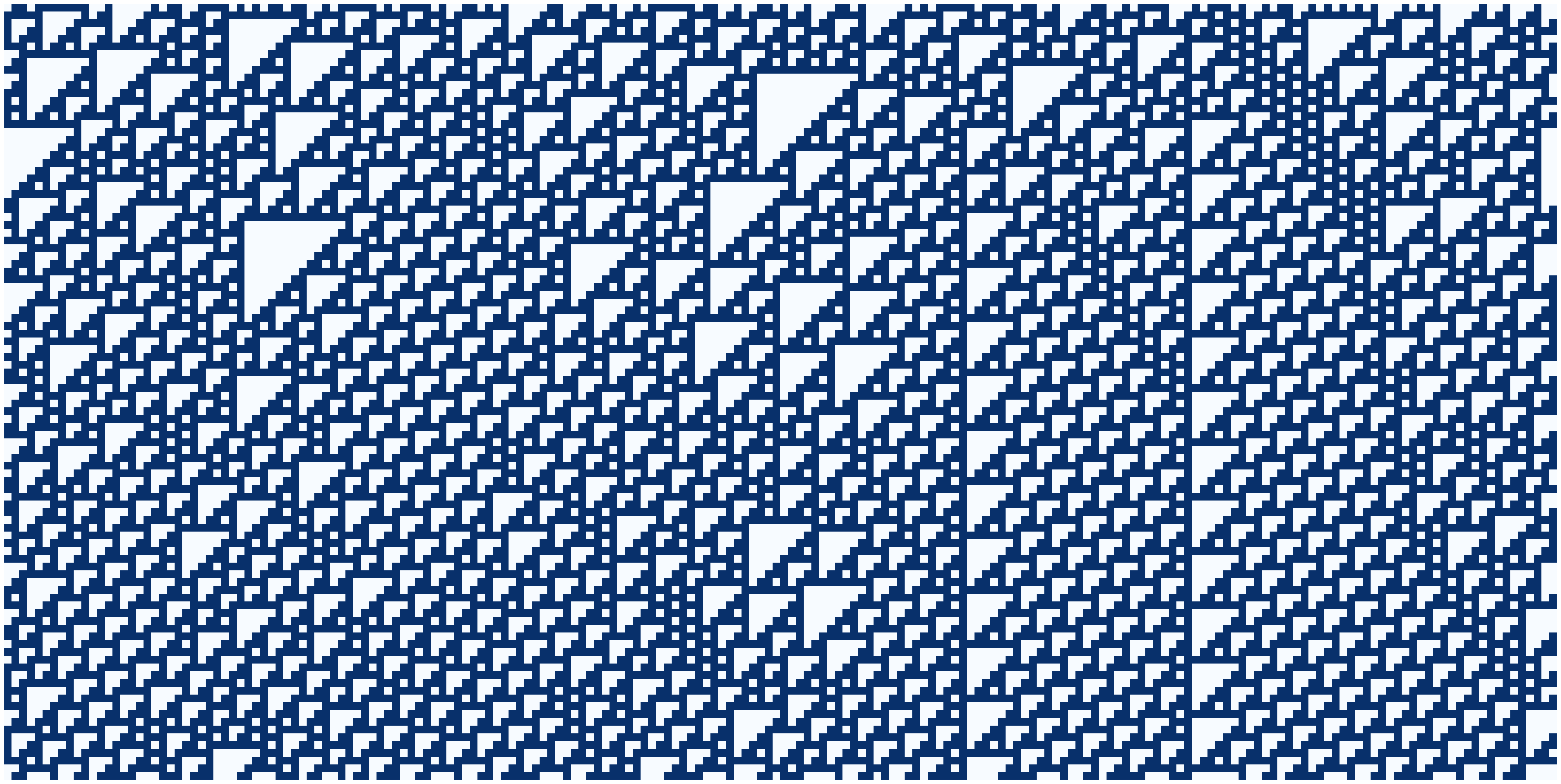}
\caption{Plan tuple plot of the time evolution of
Elementary CA rule 110, with $N=400$ cells, $T=200$ timesteps, 
and a random (50\% 0s, 50\% 1s) initial condition.
}
\label{fig:plan-ca110}
\end{figure}

\begin{mdframed}[style=exmpl,frametitle={Example: 1D ECA plan tuple plot}]
Consider
\begin{compactenum}
\item an $N$D state space $V^N = {\mathbb B}^N$
\item an indexed set of $T$ state space points $\{ {\bf p}_t \}$ forming a time series of the time evolution of the CA
\item a plotting function $sym = \{ 0 \mapsto white, 1 \mapsto black \}$
\item the identity axis position function on axes $n \in 1..N$
\item the reverse index position function on $t \in 1..T$, so that time runs down the plot.
\end{compactenum}
The resulting trajectory plan tuple plot for Elementary CA rule 110,
with N = 400 cells, T = 200 timesteps, and a random (50\% 0s, 50\% 1s)
initial condition ${\bf p}_1$ is shown in figure~\ref{fig:plan-ca110}.
\end{mdframed}

This is identical to the standard display of a 1D ECA.
Hence the standard representation is in fact a visualisation of the state space in a plan tuple plot,
and the usual time series representation, of consecutive lines of state,
is a picture of the trajectory through this state space. 

Similarly, one could consider the standard display of a 2D CA, 
such as Conway's Game of Life \cite{Gardner70}, as a plan tuple plot of the state space
formed by projecting down on the parallel coordinates arranged in a 2D grid, one coordinate line rising from the site of each cell.
A time series plot here is harder, as it would be 3D~\cite[fig.4.11]{Wuensche2016-DDLab}.  
Animation is customarily used to visualise trajectories.

\begin{figure}[tp]
\centering
\includegraphics[width=0.98\columnwidth]{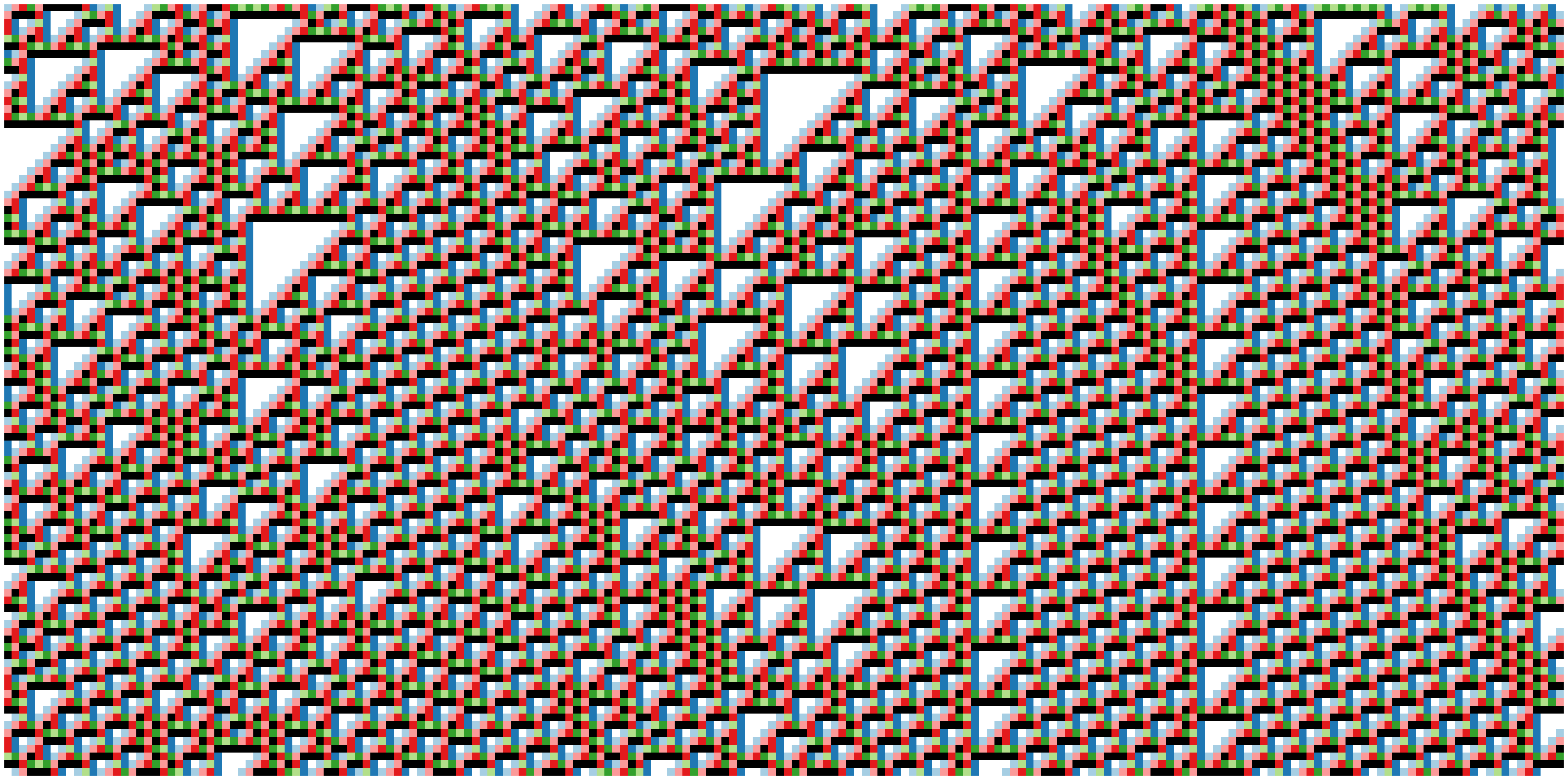}
\caption{Plan tuple plot of the time evolution of
Elementary CA rule 110 table accesses, for the ECA in figure~\ref{fig:plan-ca110}.
}
\label{fig:plan-ca110-table}
\end{figure}

We can use a plan tuple plot to visualise other aspects of the state,
such as filtering the plot to expose further structure~\cite{Wuensche2016-DDLab}.  
For example, figure~\ref{fig:plan-ca110-table} plots the lookup table entry used to produce the state value, in the following way.
\begin{mdframed}[style=exmpl,frametitle={Example: 1D ECA lookup plan tuple plot},nobreak=false]
Consider
\begin{compactenum}
\item an $N$D state space $V^N = (0..7)^N$
\item an indexed set of $T$ points (for calculation purposes) $\{ {\bf p}_t \}_{t=0}^T$ forming a time series of the time evolution of the CA, where ${\bf p}_t \in {\mathbb B}^N$;
and a further indexed set of $T-1$ state space points $\{ {\bf s}_t \}_{t=1}^T$, where $s_{nt} = 4p_{(n-1)(t-1)}+2p_{(n)(t-1)}+p_{(n+1)(t-1)}$ (index arithmetic performed modulo $N$)
\item a plotting function $sym$ that maps $0..7$ to a set of colours
\item the identity axis position function on axes $n \in 1..N$
\item the reverse index position function on $t \in 1..T$, so that time runs down the plot.
\end{compactenum}
This plot highlights structure in the generation process \cite[fig.3]{Wuensche1999}.
\end{mdframed}

\begin{figure}[t]
\centering
\includegraphics[trim = 9mm 2mm 2mm 0mm, clip,width=\columnwidth]{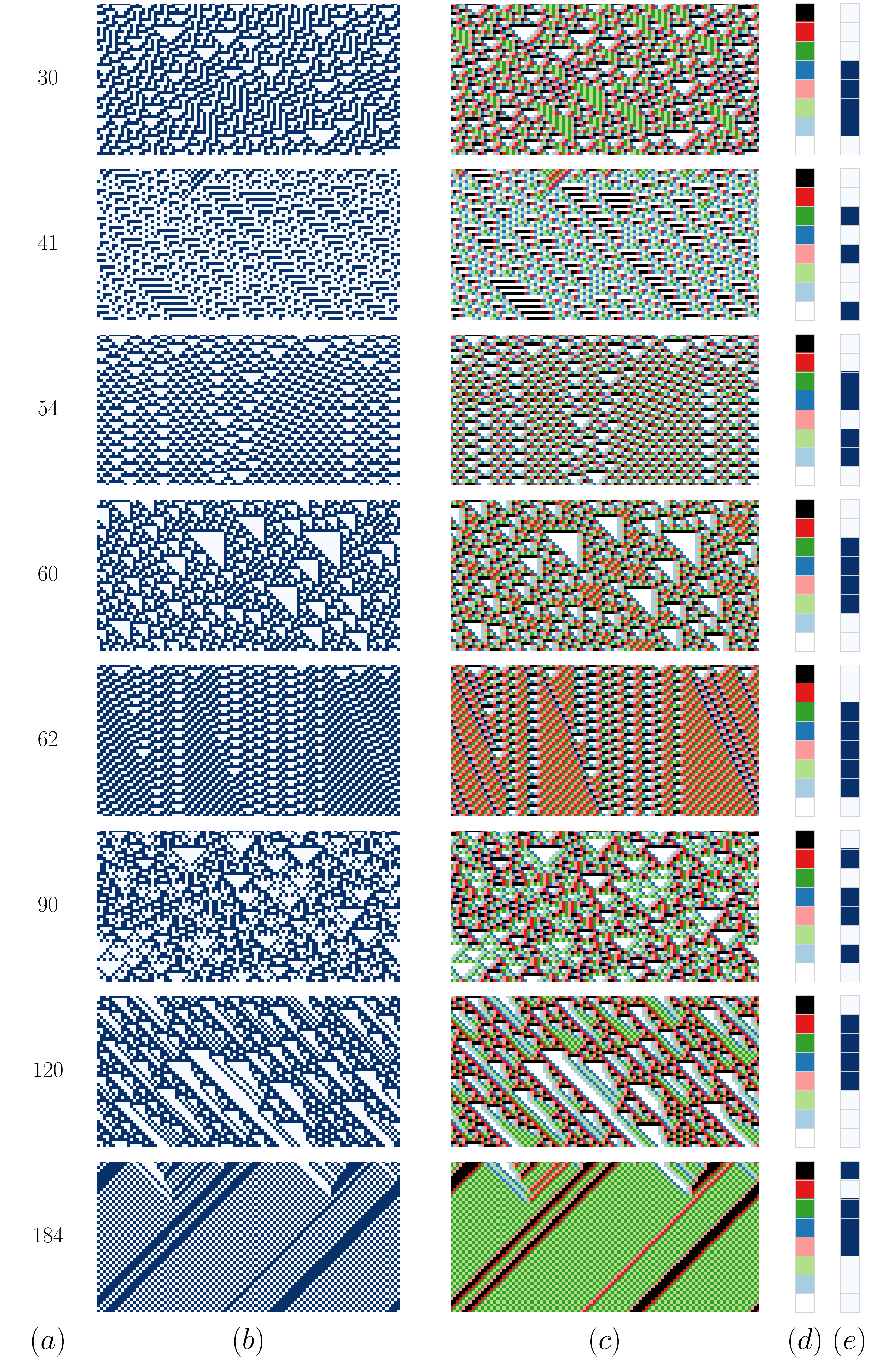}
\caption{Plan tuple plot of Elementary CA rules with $N=100$ cells, $T=40$ timesteps, 
and a random (50\% 0s, 50\% 1s) initial condition:
(a) ECA rule number;
(b) time evolution of the state;
(c) time evolution of the lookup table access;
(d) colour bar indicating the colour used to display each lookup table entry (000 at the bottom, 111 at the top); 
(e) bar indicating whether each lookup entry results in a 0 (white) or 1 (black) as the next state.
}
\label{fig:plan-ca-multi}
\end{figure}

Figure~\ref{fig:plan-ca-multi} shows both state, and lookup entry, plots for several elementary CAs.

We can use this style of visualisation to get an intuition for various behaviours of CAs.

\begin{figure*}[tp]\centering
%$\begin{array}{@{\hspace{0.1in}}c@{\hspace{0.1in}}c} 
\includegraphics[width=0.475\linewidth]{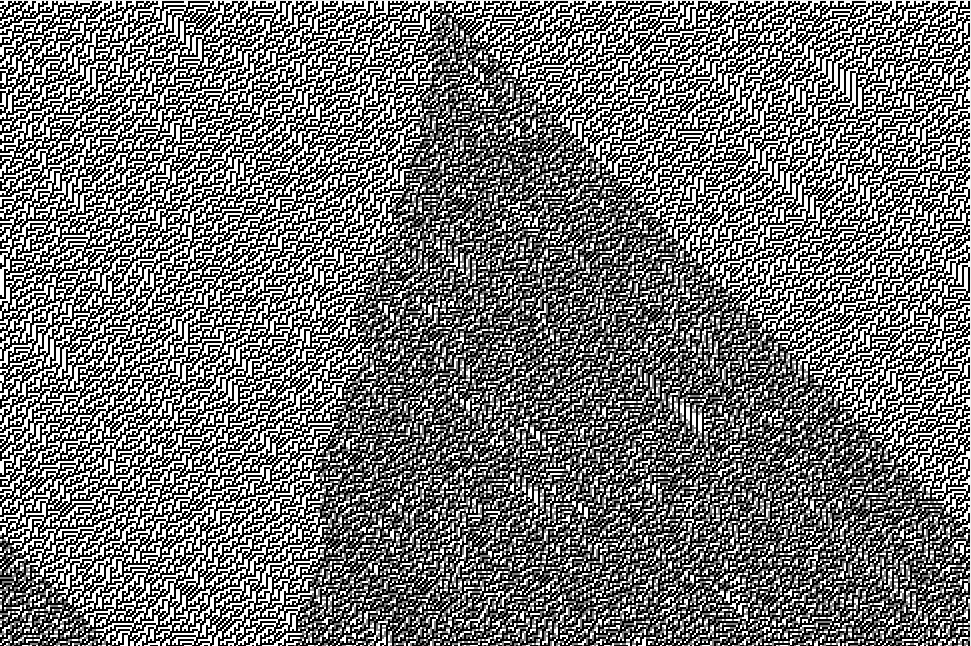} 
\includegraphics[width=0.475\linewidth]{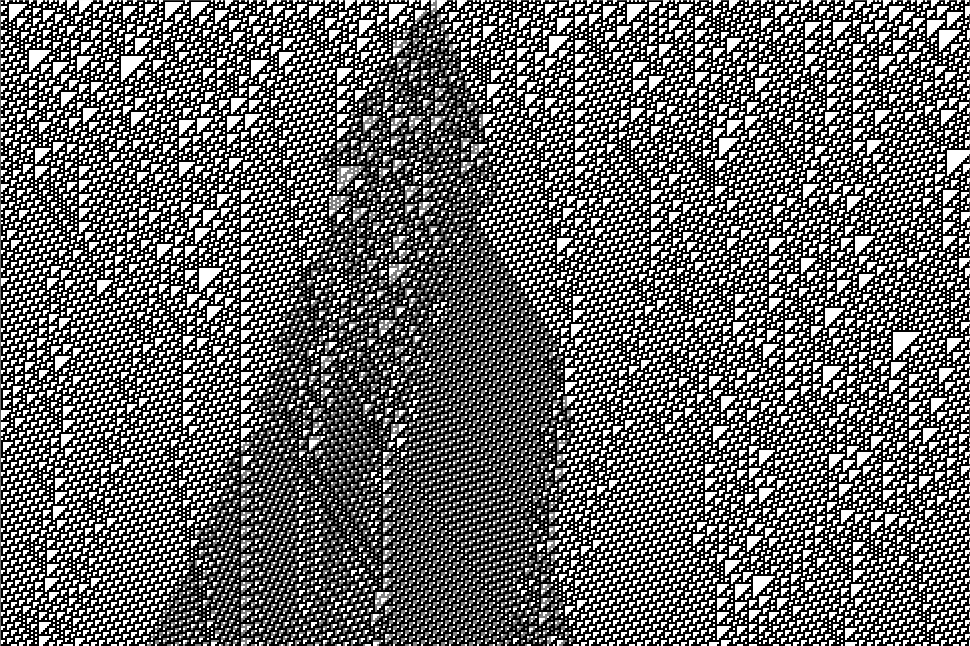}
%\end{array}$ 
\qquad (a) ECA rule 45 \hfil (b) ECA rule 110 \qquad
\caption{Sensitive dependence on initial conditions.
Each plot overlays the evolutions
of two initial states differing in only one bit:
the growing central dark region is different;
the outer regions are the same.
$N=971$, random initial state ${\bf s}_0$, over 646 timesteps. 
%(a) ECA rule 45; (b) ECA rule 110
% generated with EdP's applet + PaintShop Pro overlays
}
\label{fig:sensitive}
\end{figure*}

For example, some CA rules demonstrate {\sl sensitive dependence on initial conditions}:
the effect of a minimal (one cell state) change to the initial condition
propagates across the system,
eventually resulting in a completely different dynamics.
See figure~\ref{fig:sensitive}.

\begin{figure*}[tp]\centering
%$\begin{array}{@{\hspace{0.15in}}c@{\hspace{0.1in}}c} 
\includegraphics[width=0.475\linewidth]{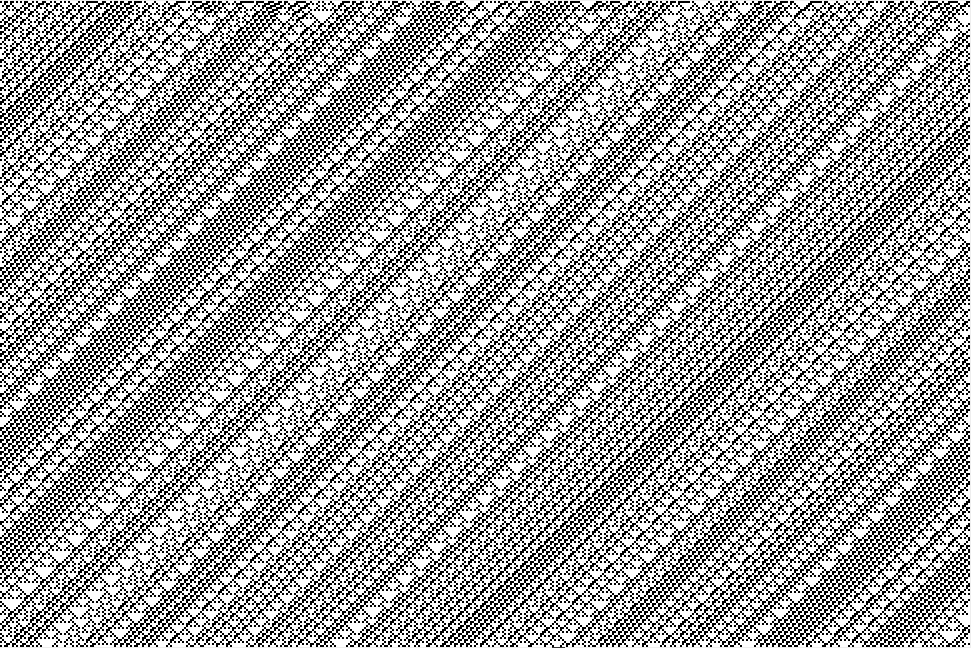} 
\includegraphics[width=0.475\linewidth]{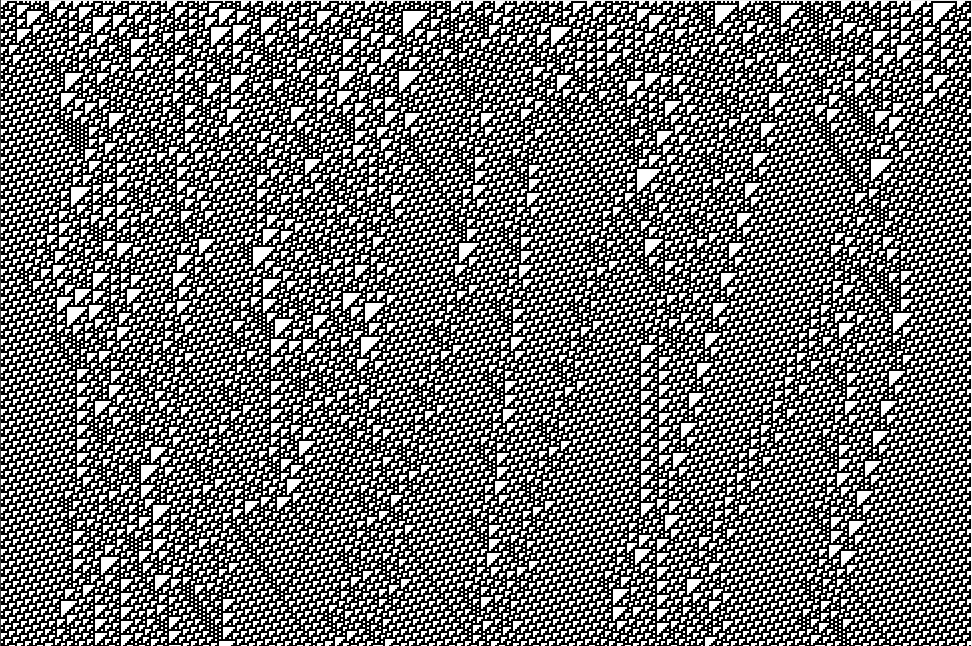} \\ \vspace{1mm} 
\includegraphics[width=0.475\linewidth]{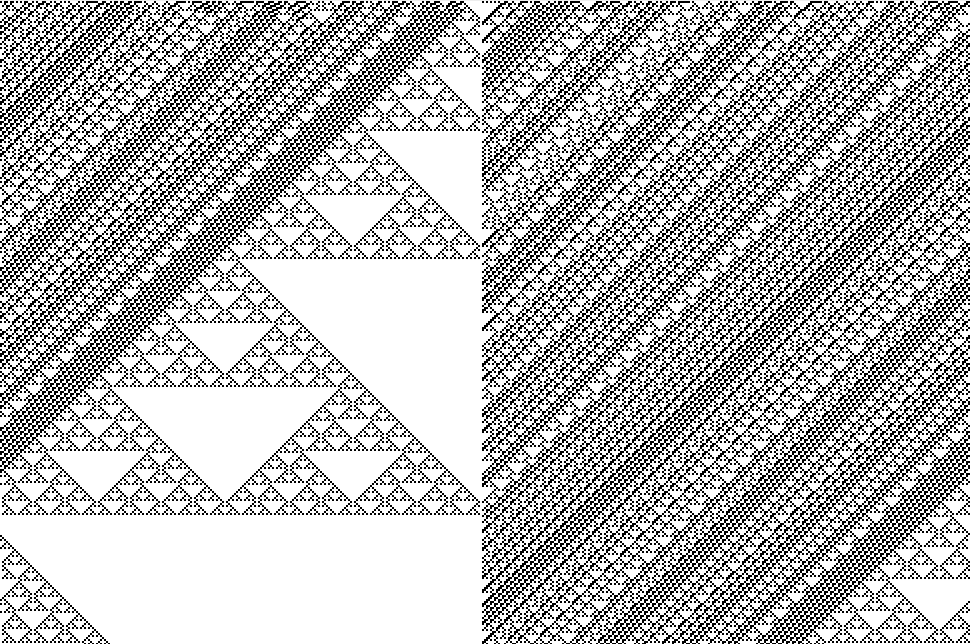} 
\includegraphics[width=0.475\linewidth]{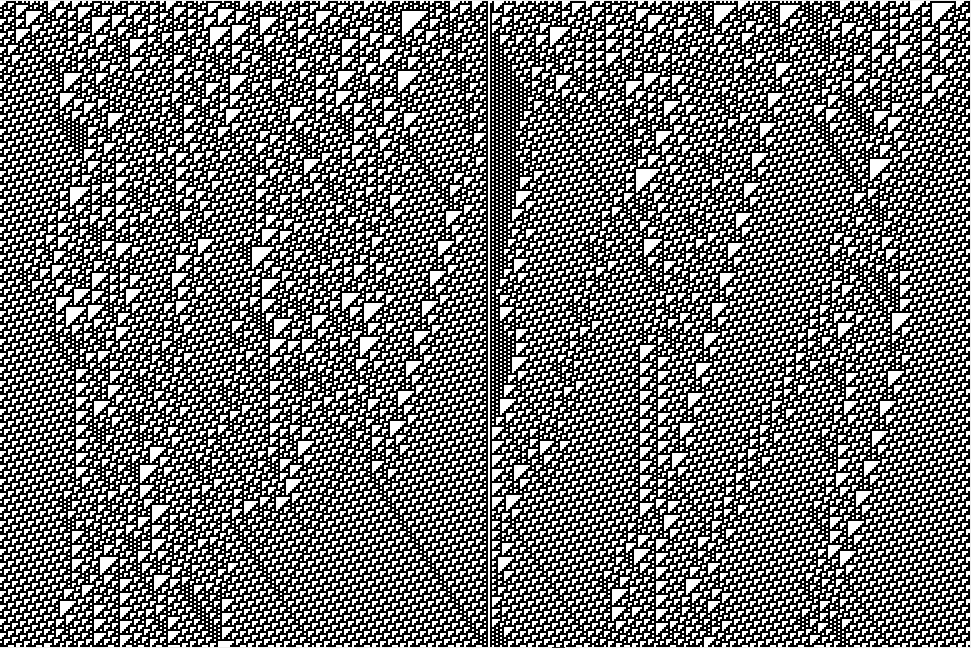}
%\end{array}$ 
\qquad (a) ECA rule 25 \hfil (b) ECA rule 110 \qquad
\caption{Dependence of CA dynamics on a ``clamped'' bit.
The upper plot shows ordinary periodic boundary conditions;
the lower plot shows the same initial conditions,
but with the central bit ``clamped'' to 0. 
%(a) ECA rule 26; (b) ECA rule 110
% generated with EdP's applet 
}
\label{fig:clamp}
\end{figure*}

For example, an input might ``clamp'' part of a CA into 
particular substate (by fixing the value of some bits for many timesteps). 
This not only perturbs the system at the point where the bits are clamped, 
but can also change the global dynamics. 
Clamping some bits in a CA can result in ``walls'' across which information 
cannot flow, isolating regions, and hence changing the dynamical structure 
of the system. See figure~\ref{fig:clamp}.

%........................................................
\clearpage
\nobalance
\subsection{Random boolean networks}\label{sec:plan:rbn}

A random boolean network (RBN) \cite{Drossel-2008,Kauffman93} comprises $N$ nodes. 
Each node has $K$ inputs assigned randomly from $K$
of the $N$ nodes (an input may be from the node itself); the wiring pattern is fixed throughout the lifetime of the network.
Each node has its own randomly chosen a boolean function of its $K$ inputs.
Each node has a binary valued state. At each timestep, the state of each node is updated in parallel, by applying the node's boolean function to its inputs.

\begin{figure}[tp]
\centering
\includegraphics[width=0.98\columnwidth]{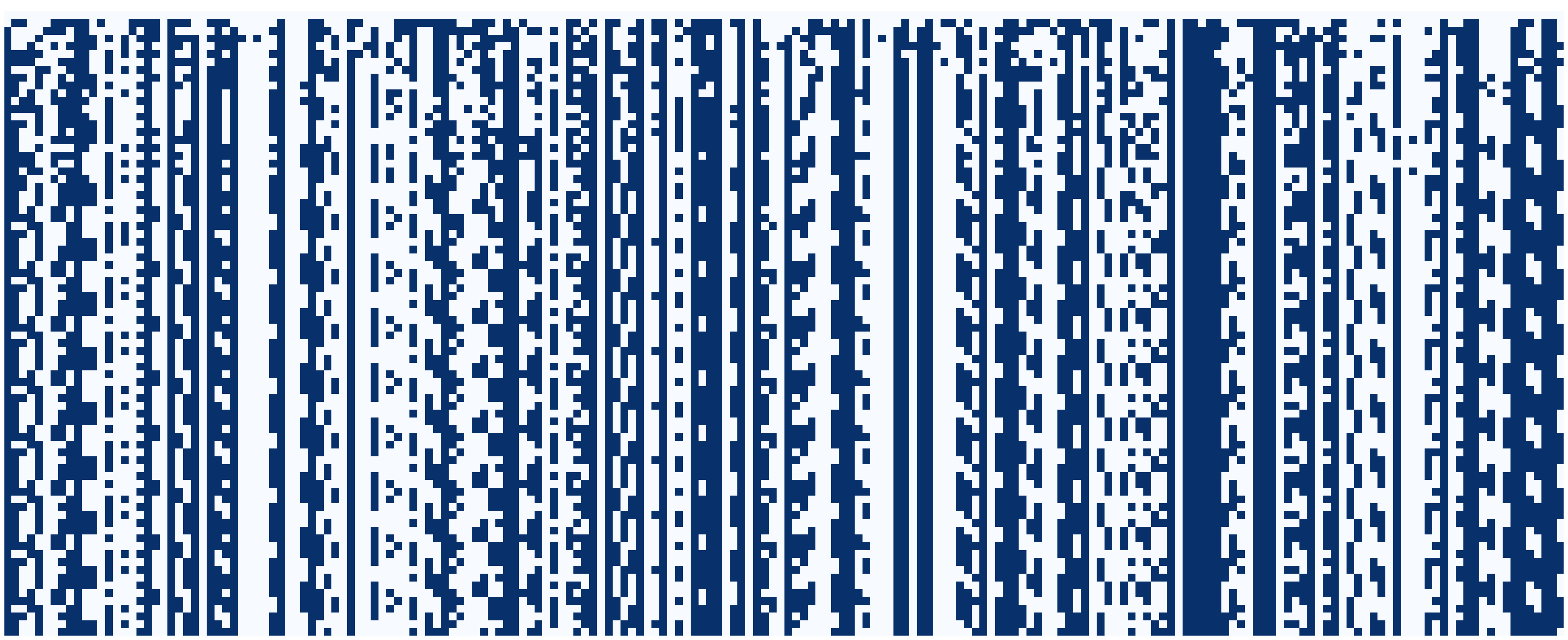}
\caption{Plan tuple plot of the time evolution of
a $K=2$ RBN with $N=200$ nodes, $T= 80$ timesteps:
nodes in arbitrary order.
}
\label{fig:plan-rbn}
\end{figure}

\begin{mdframed}[style=exmpl,frametitle={Example: unsorted RBN plan tuple plot}]
Consider
\begin{compactenum}
\item an $N$-D state space $V^N = {\mathbb B}^N$
\item an indexed set of $T$ state space points $\{ {\bf p}_t \}$ forming a time series of the time evolution of the RBN
\item a plotting function $sym = \{ 0 \mapsto white, 1 \mapsto black \}$
\item the identity position function on the axes' indexes $n \in 1..N$
\item the reverse position function on the points' indexes $t \in 1..T$, so that time runs down the plot page.
\end{compactenum}
The resulting trajectory plan tuple plot for a $K=2$ RBN
with N = 200 nodes, T = 80 timesteps, and an all-zeros
initial condition ${\bf p}_1$ is shown in figure~\ref{fig:plan-rbn}.
\end{mdframed}

%\newpage % new column
%This is identical to the standard display of an RBN.

For CAs, the order of the parallel coordinates respects the topology of the space.  With RBNs, there is no obvious ordering of the coordinates (nodes).
We can impose an ordering that highlights interesting structure, such as exposing the ``frozen core'' of unchanging values \cite{Stepney09,Stepney10}.

\begin{figure}[tp]
\centering
\includegraphics[width=0.98\columnwidth]{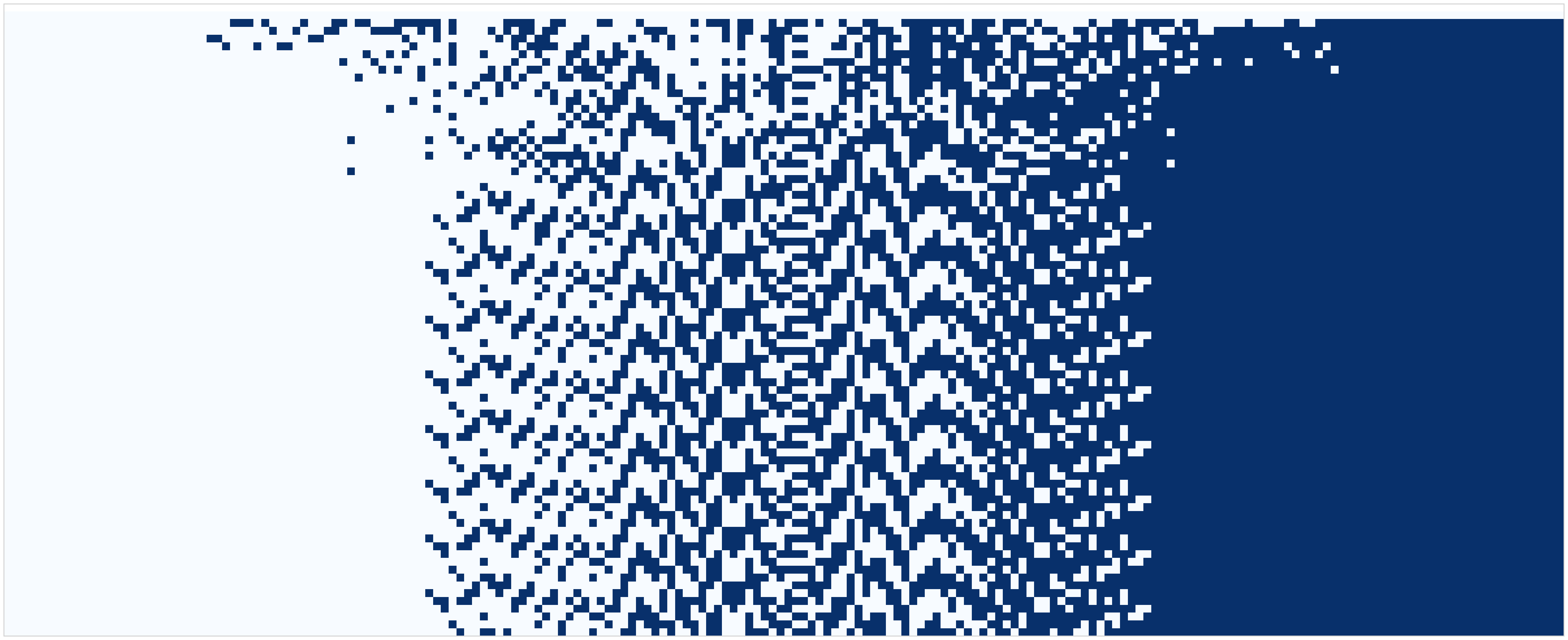}
\caption{Plan tuple plot of the time evolution of
a $K=2$ RBN with $N=200$ nodes, $T= 80$ timesteps:
nodes permuted to expose the `frozen core'.
}
\label{fig:plan-rbn-sorted}
\end{figure}

\begin{mdframed}[style=exmpl,frametitle={Example: sorted RBN plan tuple plot},nobreak=false]
Consider
\begin{compactenum}
\item an $N$D state space $V^N = {\mathbb B}^N$
\item an indexed set of $T$ state space points $\{ {\bf p}_t \}$ forming a time series of the time evolution of the RBN
\item a plotting function $sym = \{ 0 \mapsto white, 1 \mapsto black \}$
\item a position function that permutes the axes' indexes $n \in 1..N$ so that nodes that have a value of zero for more timesteps have a lower index 
\item the reverse position function on the points' indexes $t \in 1..T$, so that time runs down the plot.
\end{compactenum}
The resulting trajectory plan tuple plot 
is shown in figure~\ref{fig:plan-rbn-sorted}.
\end{mdframed}

\begin{figure*}[tp]
\centerline{%
\scalebox{0.5}{\includegraphics{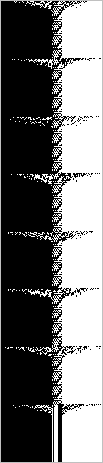}}~
\scalebox{0.5}{\includegraphics{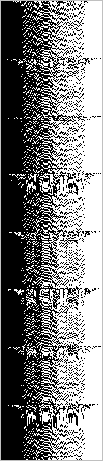}}~
\scalebox{0.5}{\includegraphics{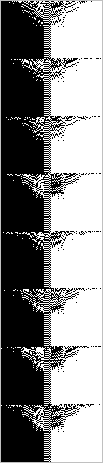}}~
\scalebox{0.5}{\includegraphics{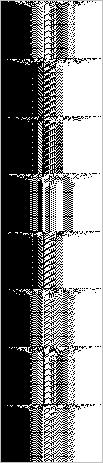}}~
\scalebox{0.5}{\includegraphics{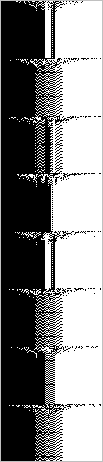}}~
\scalebox{0.5}{\includegraphics{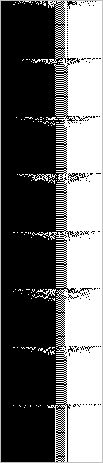}}
}\vspace{2mm}
\centerline{%
\scalebox{0.5}{\includegraphics{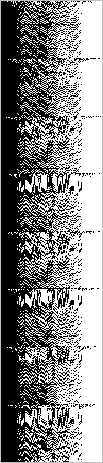}}~
\scalebox{0.5}{\includegraphics{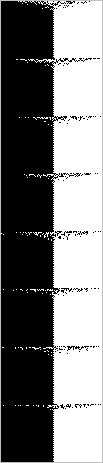}}~
\scalebox{0.5}{\includegraphics{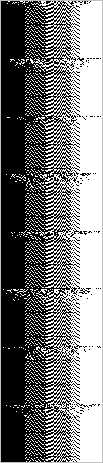}}~
\scalebox{0.5}{\includegraphics{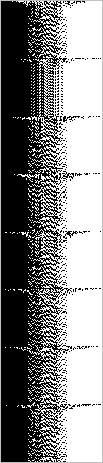}}~
\scalebox{0.5}{\includegraphics{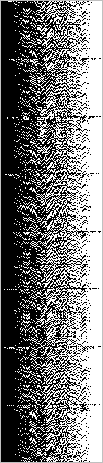}}~
\scalebox{0.5}{\includegraphics{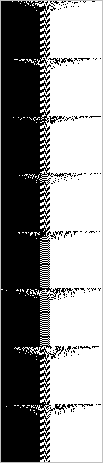}}
%\scalebox{0.5}{\includegraphics{figs/rbn-k2-n100-s12.png}}~
%\scalebox{0.5}{\includegraphics{figs/rbn-k2-n100-s13.png}}
}
\caption{Visualisation of the time evolution of 12 typical $K=2$ RBNs, 
with $N=100$.
Every 60 timesteps the nodes are reinitialised to a new random configuration,
to explore other attractors.
They exhibit ordered behaviour: 
short transients, and low period attractors.
%In each case, the nodes are sorted over the first four runs:
%note that this has not been sufficient to isolate the frozen core in the final example (bottom right).
}
\label{fig:100}
\end{figure*}

\begin{figure*}[tp]
\centerline{%
\scalebox{0.465}{\includegraphics{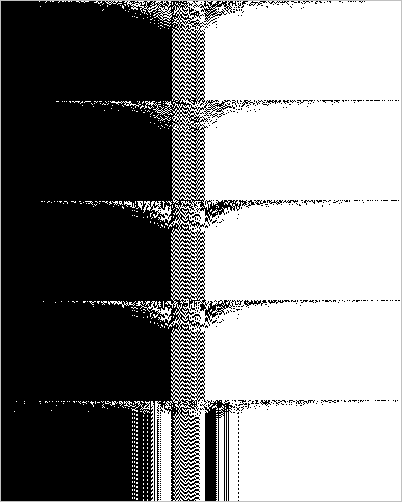}}~
\scalebox{0.465}{\includegraphics{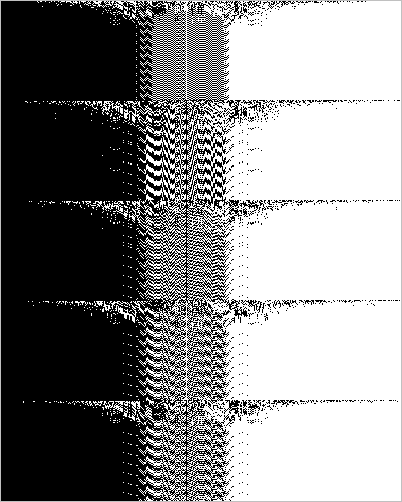}}
}\vspace{1mm}
\centerline{%
\scalebox{0.465}{\includegraphics{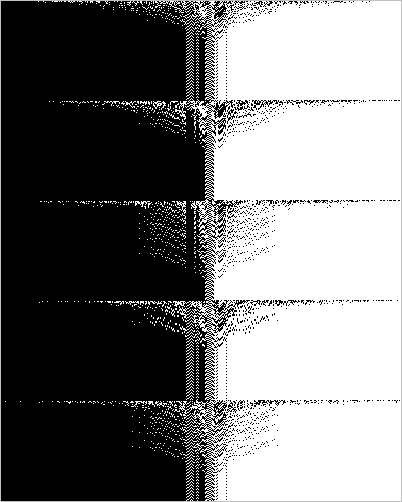}}~
\scalebox{0.465}{\includegraphics{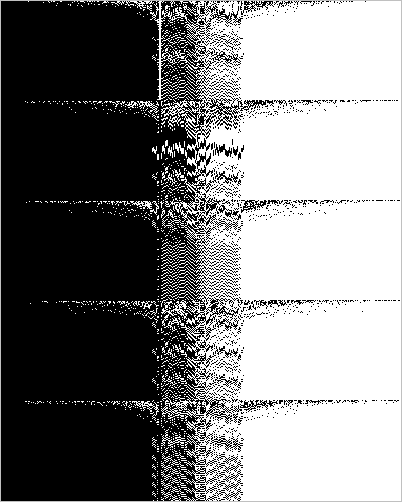}}
}
\caption{Visualisation of the time evolution of four typical $K=2$ RBNs, 
with $N=400$. 
Every 100 timesteps the nodes are reinitialised to a new random configuration,
to explore other attractors.
They exhibit ordered behaviour: 
short transients, and low period attractors.
\label{fig:400}
}
\end{figure*}

\begin{figure*}[p]
\centerline{%
\scalebox{0.58}{{\includegraphics{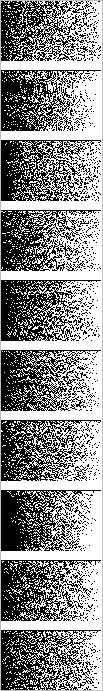}}}
\scalebox{0.58}{{\includegraphics{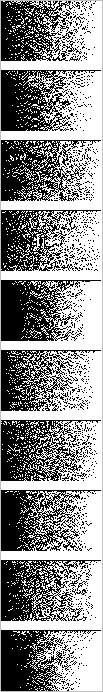}}}
\scalebox{0.58}{{\includegraphics{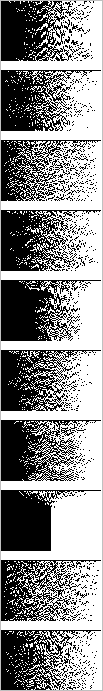}}}
\scalebox{0.58}{{\includegraphics{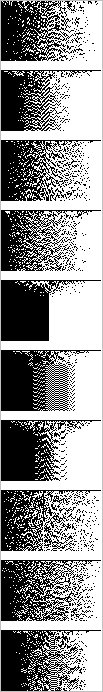}}}
\scalebox{0.58}{{\includegraphics{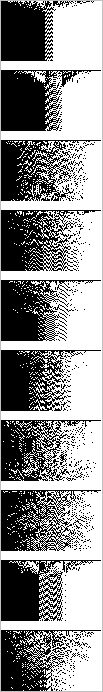}}}
\scalebox{0.58}{{\includegraphics{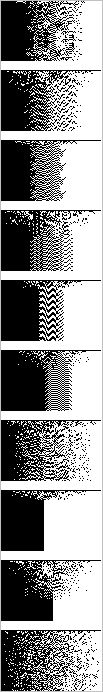}}}
\scalebox{0.58}{{\includegraphics{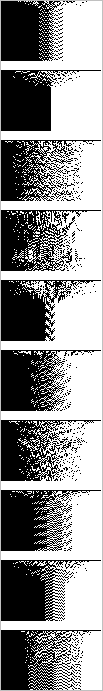}}}
}
\centerline{
~(a) 47 \hfill
(b) 64\hfill
(c) 84\hfill
(d) 90\hfill
(e) 92\hfill
(f) 95\hfill
(g) 100~~
}
\caption{Visualisation of the time evolution of typical $K=3$ RBNs,
with $N=100$, $t=80$, initial condition all nodes ``on'';
columns have an increasing amount of canalisation,
with the indicated number of canalised nodes.
}
\label{fig:k3-100}
\end{figure*}

We can use this visualisation to get an intuition for the documented behaviours of RBNs: their variability, their frozen cores, their attractor structure, their dependence on $K$ and canalised functions.  See figures~\ref{fig:100}--\ref{fig:k3-100};
see \cite{Stepney10} for further examples, including the visualisation of the effect of various network perturbations.

%\begin{figure}[tp]
%\centering
%\includegraphics[height=0.5\textheight]{figs/plan-rbn-multi.pdf}
%\caption{Plan tuple plots of three $K=2$ RBNs with $N=50$ nodes, $T= 25$ timesteps.
%The coloured strip under each plot indicates the fixed random boolean function in the respective node: functions are numbered $0$--$15$ in the usual encoding of the truth table; darker colours indicate higher numbers.
%}
%\label{fig:plan-rbn-multi}
%\end{figure}
%
%We can additionally augment the RBN plots with another, static, aspect of their state: the specific random boolean function located at each node (figure~\ref{fig:plan-rbn-multi}).

%........................................................
\newpage
\subsection{Coupled logistic maps}\label{sec:plan:clm}

The logistic map $x_{n+1} = \lambda x_n(1-x_n)$
is a well-known 1D discrete time dynamical system.
$N$ maps can be coupled together to form an $N$D system.
For example, Kaneko \cite{ref:kaneko_1985} discusses coupled map lattices,
and Sinha and Ditto \cite{ref:sinha_physlett_1993,ref:sinha_physlett_1998,ref:sudeshna_phyrev_1999}
consider one-way threshold coupled lattices (TCL).

Here we use the Sinha and Ditto example.
Consider a set of $N$ cells, each with a state $x \in [0,1]$.
At each timestep, each cell's state is updated by applying 
the logistic map at its fully chaotic value: $f(x) = 4x(1-x)$.
Next, any `excess' value is transferred: the cells are considered in order $1..N$;
if cell $n$ has a value greater than the threshold parameter $x_*$,
its value is reduced to $x_*$, and the excess $\delta = x_n - x_*$
is transferred to cell $n+1$, increasing its value to $x_{n+1}+\delta$.
Any excess from the last cell $x_N$ is removed from the system.
See algorithm~\ref{alg:threshold_lattice_relaxation}.

\begin{figure}[t]
\centering
\includegraphics[width=0.98\columnwidth]{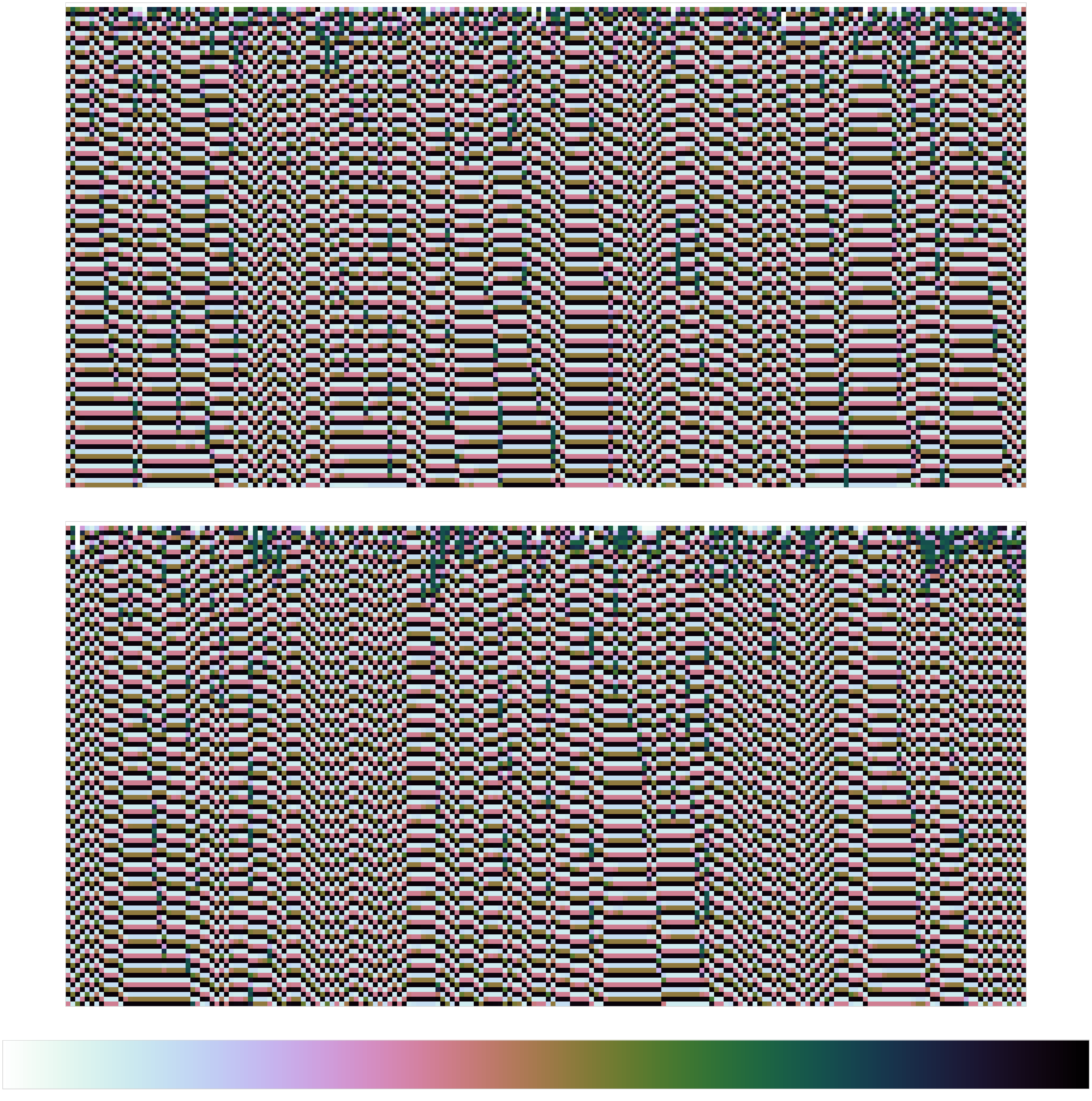}
\caption{Plan tuple plots of the threshold coupled lattice,
threshold $x_* = 0.971$, number of cells $N = 200$, $T=100$ timesteps:
the $x$ axis shows the individual cell; each cell's value $x_n \in [0,1]$ 
is indicated by its colour, as shown in the lower colour bar.
Two different random initial conditions are shown.
} 
\vspace{5cm}
\label{fig:plan-clm}
\end{figure}

\begin{mdframed}[style=exmpl,frametitle={Example: Threshold coupled lattice plan tuple plot},nobreak=false]
Consider
\begin{compactenum}
\item an $N$D state space $V^N = [0,1]^N$
\item an indexed set of $T$ state space points $\{ {\bf p}_t \}$ forming a time series of the time evolution of the TCL
\item a plotting function $grey$ that maps a component value $p\in [0,1]$ to a greyscale with extremes $0 \mapsto white, 1 \mapsto black$
\item the identity position function on the axes' indexes $n \in 1..N$
\item the reverse position function on the points' indexes $t \in 1..T$, so that time runs down the plot.
\end{compactenum}
The resulting trajectory plan tuple plot 
is shown in figure~\ref{fig:plan-clm}.
\end{mdframed}

%--------------
% Relaxation psuedocode
\begin{algorithm}[t]
  \begin{algorithmic}
    \FOR{each cell $n$} 
      \STATE{$x_n \gets f(x_n)$}
    	\COMMENT{logistic update} 
    \ENDFOR
        \FOR{$n = 1 \mbox{ to } N$}
    	\STATE{}\COMMENT{excess propagation}
      \IF{$x_n > x_*$}
        \STATE{$\delta \gets x_n - x_*$; $x_n \gets x_*$; $x_{n+1} \gets x_{n+1} + \delta$}
      \ENDIF
    \ENDFOR
  \end{algorithmic}
  \caption[Threshold coupled lattice timestep]{Threshold coupled lattice timestep.}
  \label{alg:threshold_lattice_relaxation}
\end{algorithm}
%--------------

%........................................................
\clearpage
\subsection{Reservoir computer dynamics}\label{sec:plan:reservoir}

Reservoir computing \cite{Dambre2012,Verstraeten2007} is often used as a computational model for unconventional substrates \cite{Dale2017}.
It comprises an underlying dynamical system (discrete time, continuous space) usually described as a recurrent neural network, plus a training method.
Here we look just at the internal dynamics, of a relatively simplified variant (various parameters set to convenient values).
The equation used here is:
\begin{equation}
\mathbf{p}_{t+1} = \tanh \left( \rho \mathbf{p}_{t}.\mathbf{W} + u_t \mathbf{v}  \right)
\end{equation}
where \textbf{p} is the system state vector (each component thought of as a node, or neuron, in the network);
$t$ is the timestep;
$\rho$ is the gain parameter;
\textbf{W} is a random weight matrix connecting the neurons, with weights initially drawn from $U[-1,1]$, then the matrix normalised so that its largest singular value is $1$;
$u$ is the input, or driving, signal;
\textbf{v} is a random input weight vector, with weights drawn from $U[-1,1]$.

\begin{figure}[tp]
\centering
\includegraphics[width=0.49\columnwidth]{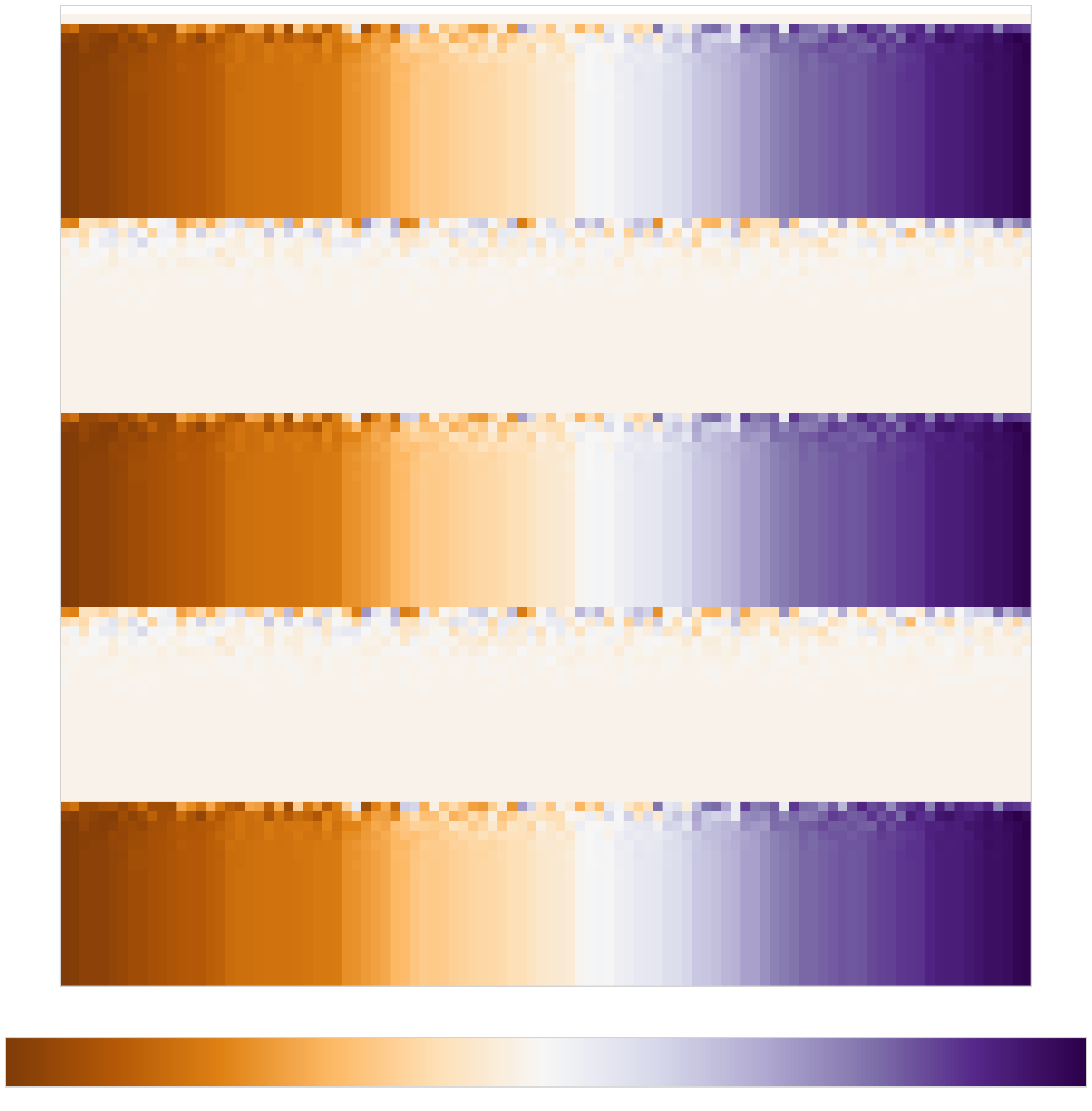}
\includegraphics[width=0.49\columnwidth]{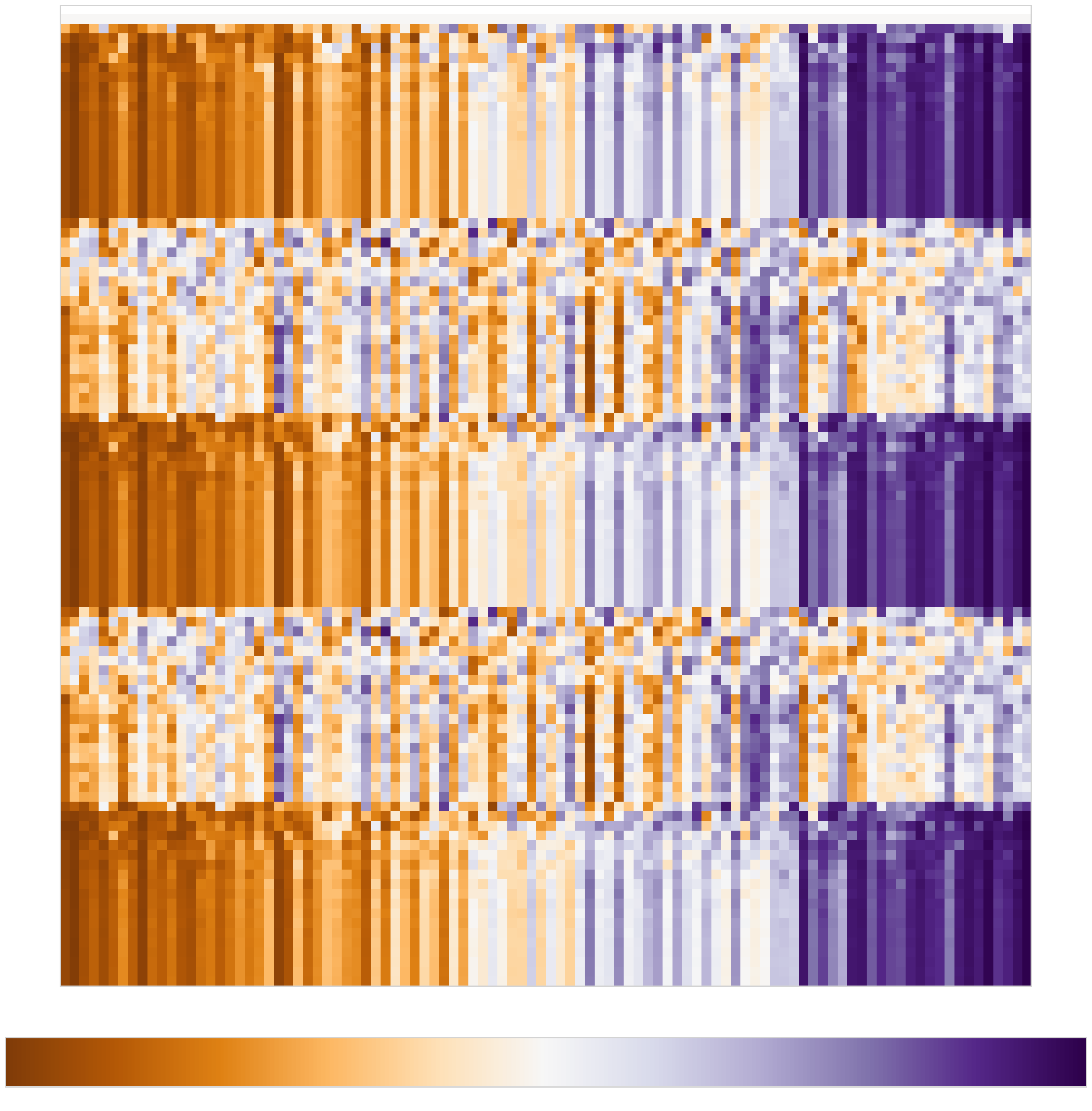}
\includegraphics[width=0.49\columnwidth]{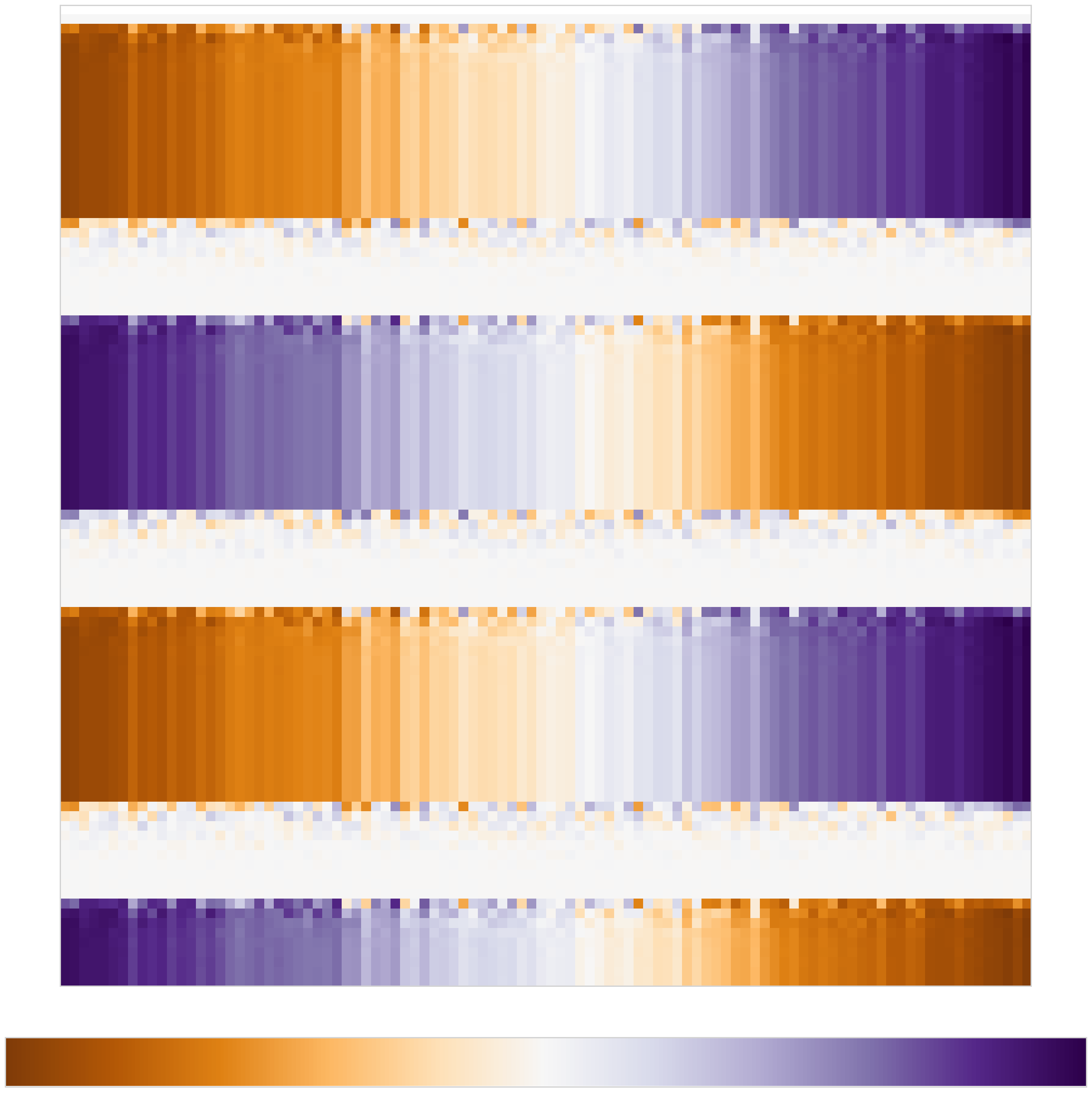}
\includegraphics[width=0.49\columnwidth]{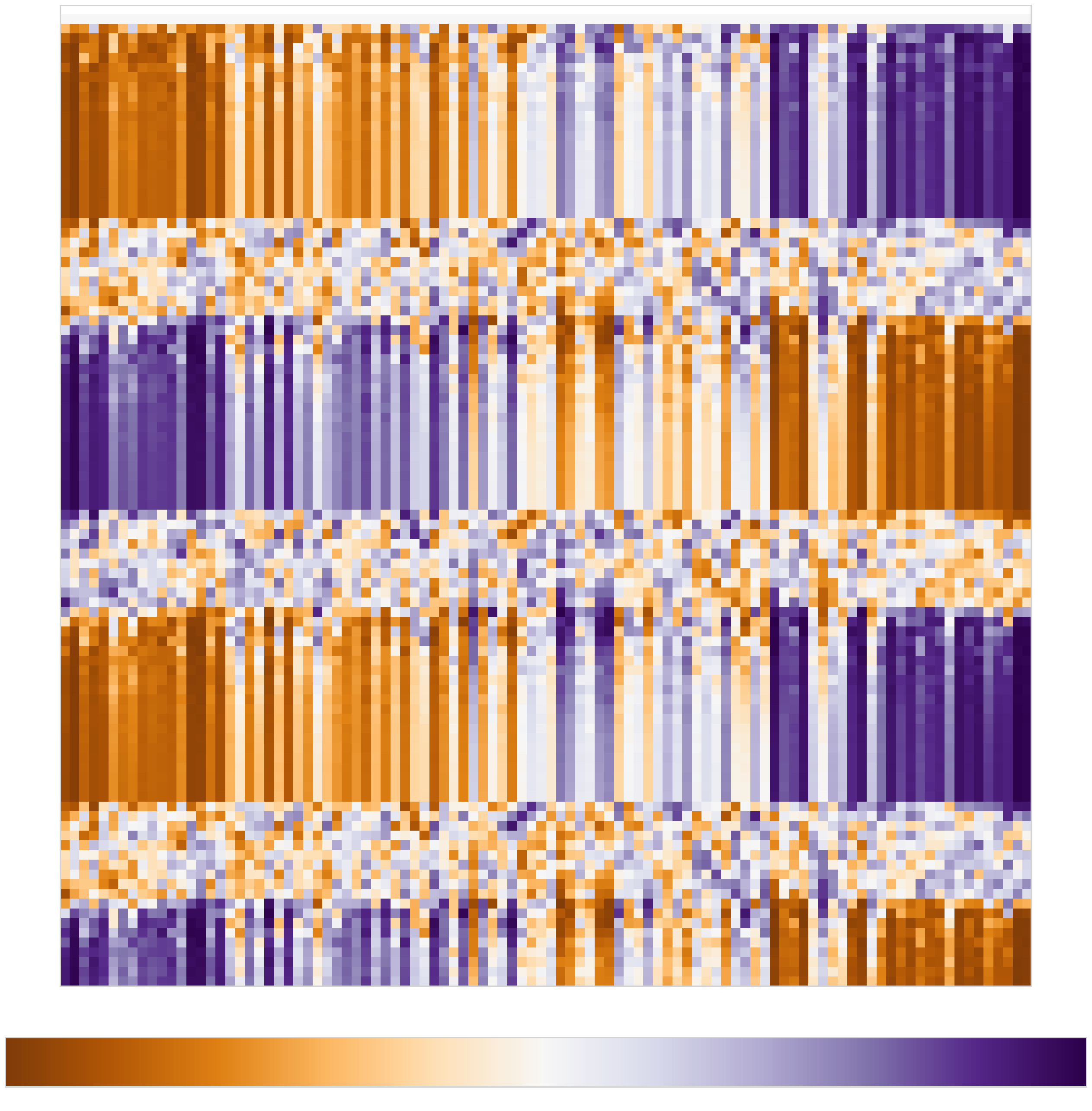}
\caption{Plan tuple plots of a reservoir dynamics,
$N = 100$ nodes,
initial state $\mathbf{x} = \mathbf{0}$; 
%$\rho = 2$;
100 timesteps,
Each node's value $x_n \in [-1,1]$ 
is indicated by its colour, as shown in the lower colour bar
(brown for $-1$, through white for $0$, to purple for $1$).
Two different input functions are shown.
In the top plot the input cycles through $u=1$ for 20 timesteps followed by $u= 0$ for 20 timesteps.
In the bottom plot it cycles through $u=1$ for 20 timesteps, $u= 0$ for 10 timesteps, $u=-1$ for 20 timesteps, $u= 0$ for 10 timesteps.
In the left plots $\rho = 1$ (fading memory region);
the transitions between driven dynamics ($|u|=1$) and decaying dynamics ($u=0$) are clear.
In the right plots $\rho = 2$ (chaotic region);
the transitions between driven dynamics ($|u|=1$) and chaotic free dynamics ($u=0$) are clear.
} 
\label{fig:plan-reservoir}
\end{figure}

\begin{mdframed}[style=exmpl,frametitle={Example: Reservoir plan tuple plot},nobreak=false]
Consider
\begin{compactenum}
\item an $N$D state space $V^N = [-1,1]^N$
\item an indexed set of $T$ state space points $\{ {\bf p}_t \}$ forming a time series of the time evolution of the reservoir
\item a plotting function that maps a component value $p\in [-1,1]$ to a colourscale with extremes $-1 \mapsto orange, 1 \mapsto purple$, passing through $0 \mapsto white$
\item a position function on the axes' indexes that plots mostly negative values nodes to the left, positive to the right
\item the reverse position function on the points' indexes $t \in 1..T$, so that time runs down the plot.
\end{compactenum}
The resulting trajectory plan tuple plots 
are shown in figure~\ref{fig:plan-reservoir}.
\end{mdframed}
%Instead of plotting the state values, we can plot the difference in state values between timesteps; see 
%figure~\ref{fig:plan-reservoir-diff}.

%\begin{figure*}[tp]
%\centering
%\includegraphics[width=0.45\textwidth]{figs/plan_reservoir10-r1-diff.pdf}~
%\includegraphics[width=0.45\textwidth]{figs/plan_reservoir10-r2-diff.pdf}
%\\[1mm]
%\includegraphics[width=0.45\textwidth]{figs/plan_reservoir10-10-r1-diff.pdf}~
%\includegraphics[width=0.45\textwidth]{figs/plan_reservoir10-10-r2-diff.pdf}
%\caption{Plan tuple plots of a reservoir dynamics,
%$N = 100$ nodes,
%initial state $\mathbf{x} = \mathbf{0}$; 
%%$\rho = 2$;
%100 timesteps,
%Each node's \textit{difference in value from its previous value} 
%is indicated by its colour.
%In the left plots $\rho = 1$ (fading memory region);
%the transitions between driven dynamics ($|u|=1$) and decaying dynamics ($u=0$) are clear.
%In the right plots $\rho = 2$ (chaotic region);
%the transitions between driven dynamics ($|u|=1$) and chaotic free dynamics ($u=0$) are clear.
%Compare with figure~\ref{fig:plan-reservoir}.
%} 
%\label{fig:plan-reservoir-diff}
%\end{figure*}

%........................................................
\clearpage
\subsection{Search algorithms}\label{sec:plan-tsp}
Here we show some examples of visualising the trajectory of a search algorithm.
The task is to search for a permutation that finds the shortest route through several points: the Travelling Salesman Problem.

\begin{figure}[tp]
\centering
%trim=l b r t
(a){\includegraphics[width=0.35\columnwidth, clip, trim=2.3cm 4cm 6.8cm 4.5cm]{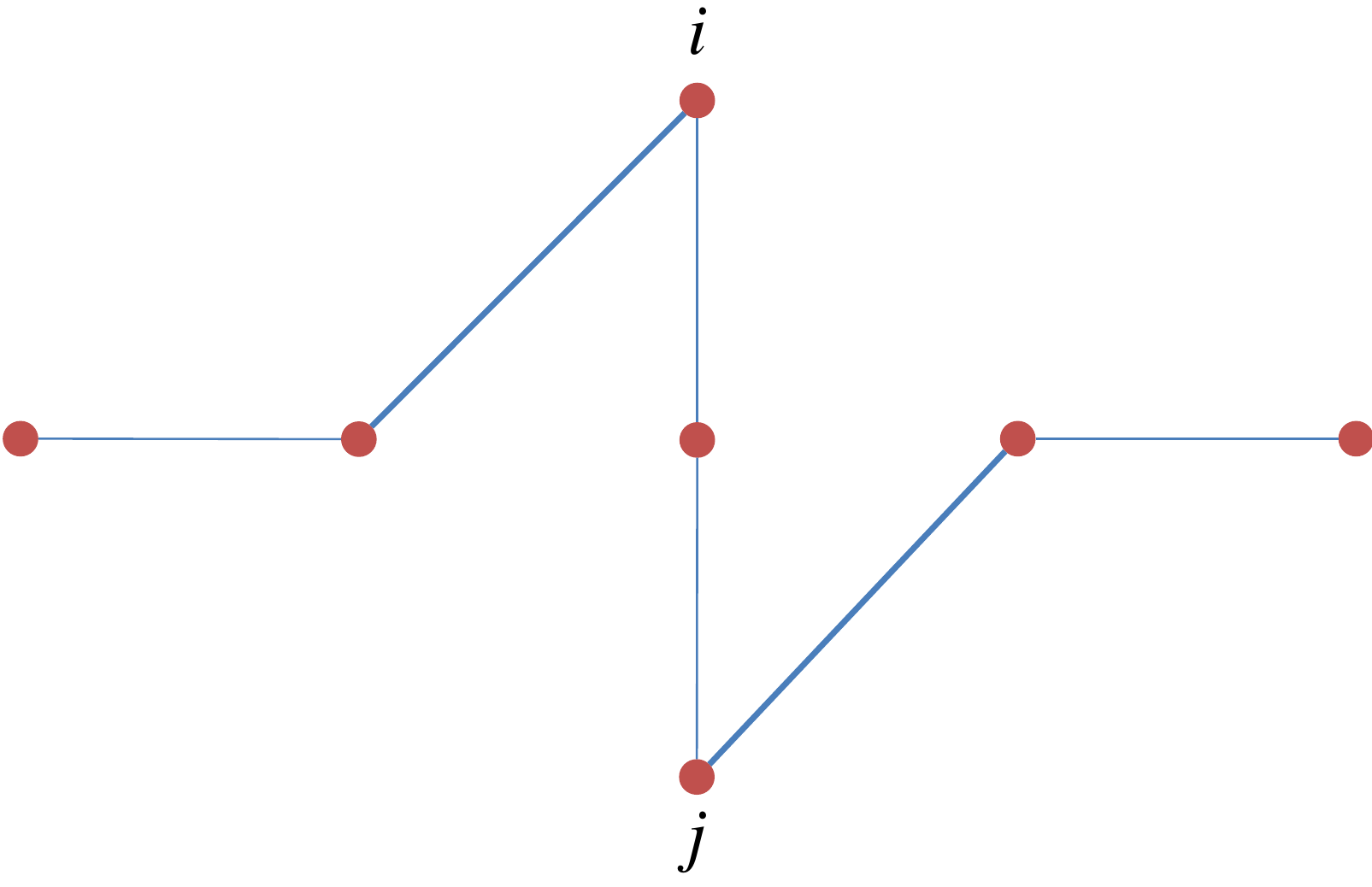}}~~~
(b){\includegraphics[width=0.35\columnwidth, clip, trim=2.3cm 4cm 6.8cm 4.5cm]{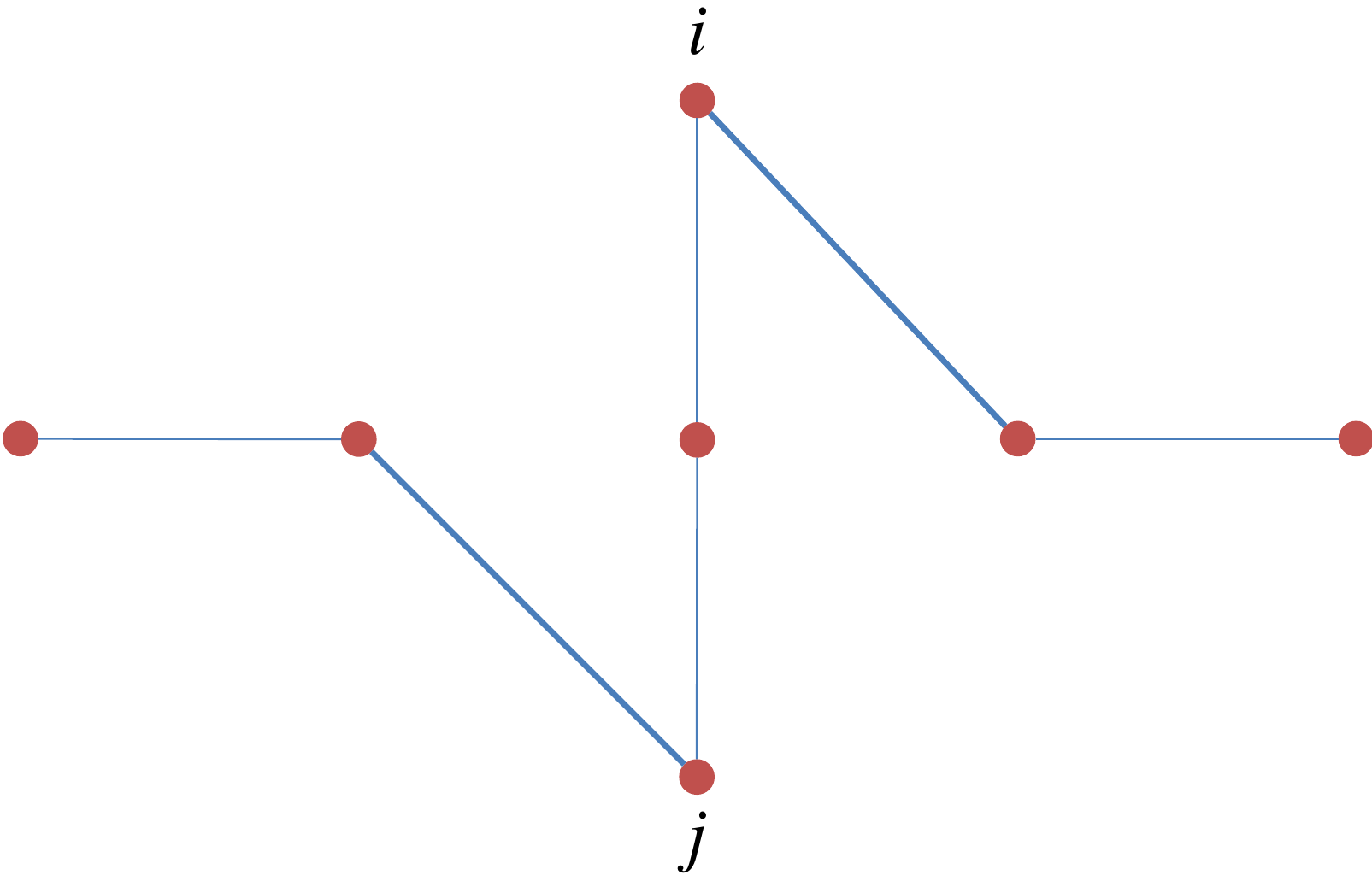}}
\caption{Cross mutation (reversing the central path segment: 2-opt).
(a) original path $\ldots, i-1, i, i+1, \ldots, j-1, j, j+1, \ldots$;
(b) final path $\ldots, i-1, j, j-1, \ldots, i+1, i, j+1, \ldots$
} 
\label{fig:plan-tsp-hill-cross}
\end{figure}

\begin{figure}[tp]
\centering
(a)~{\includegraphics[width=0.35\columnwidth, clip, trim=2.3cm 3.8cm 6.8cm 5.2cm]{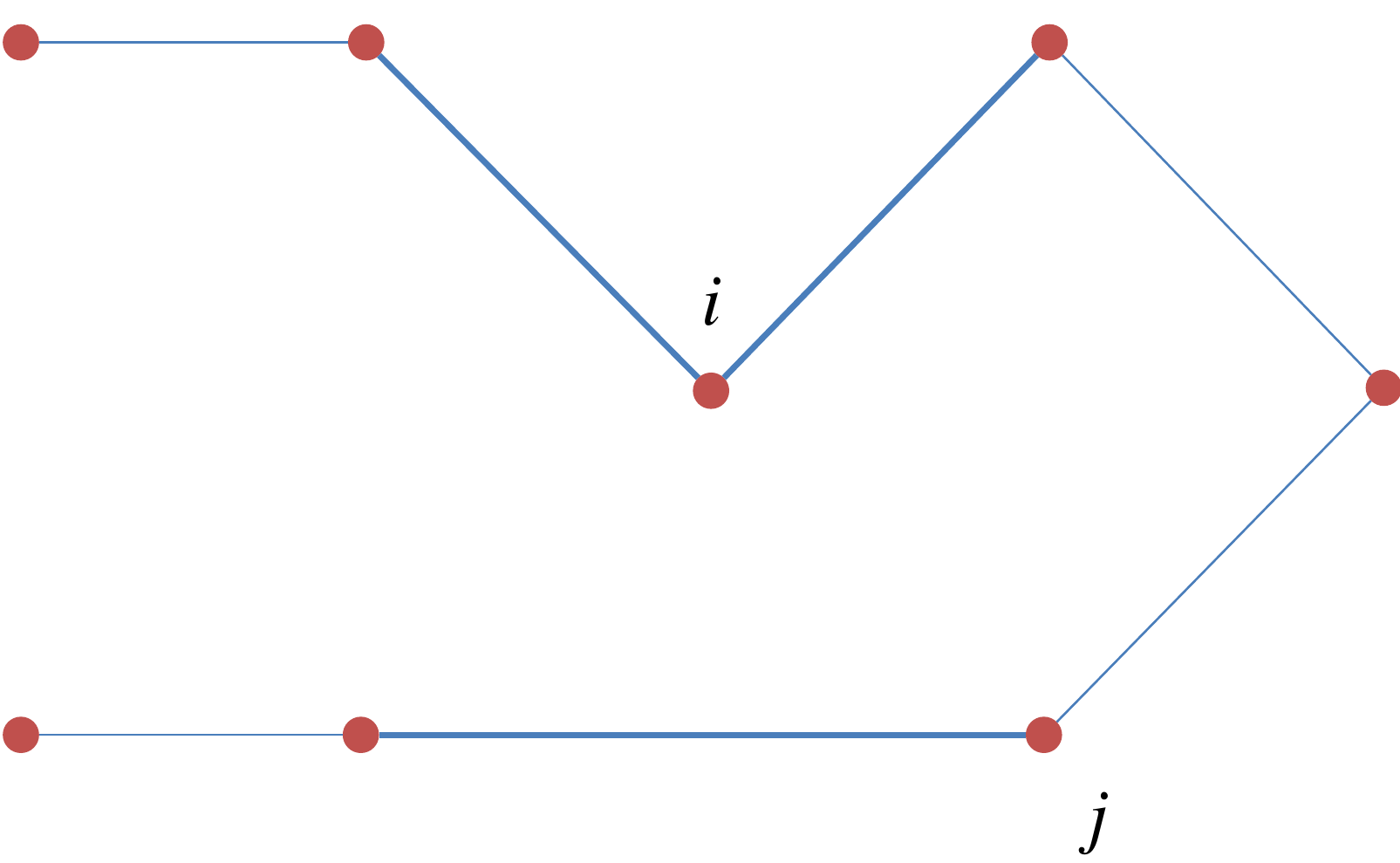}}~~~
(b)~{\includegraphics[width=0.35\columnwidth, clip, trim=2.3cm 3.8cm 6.8cm 5.2cm]{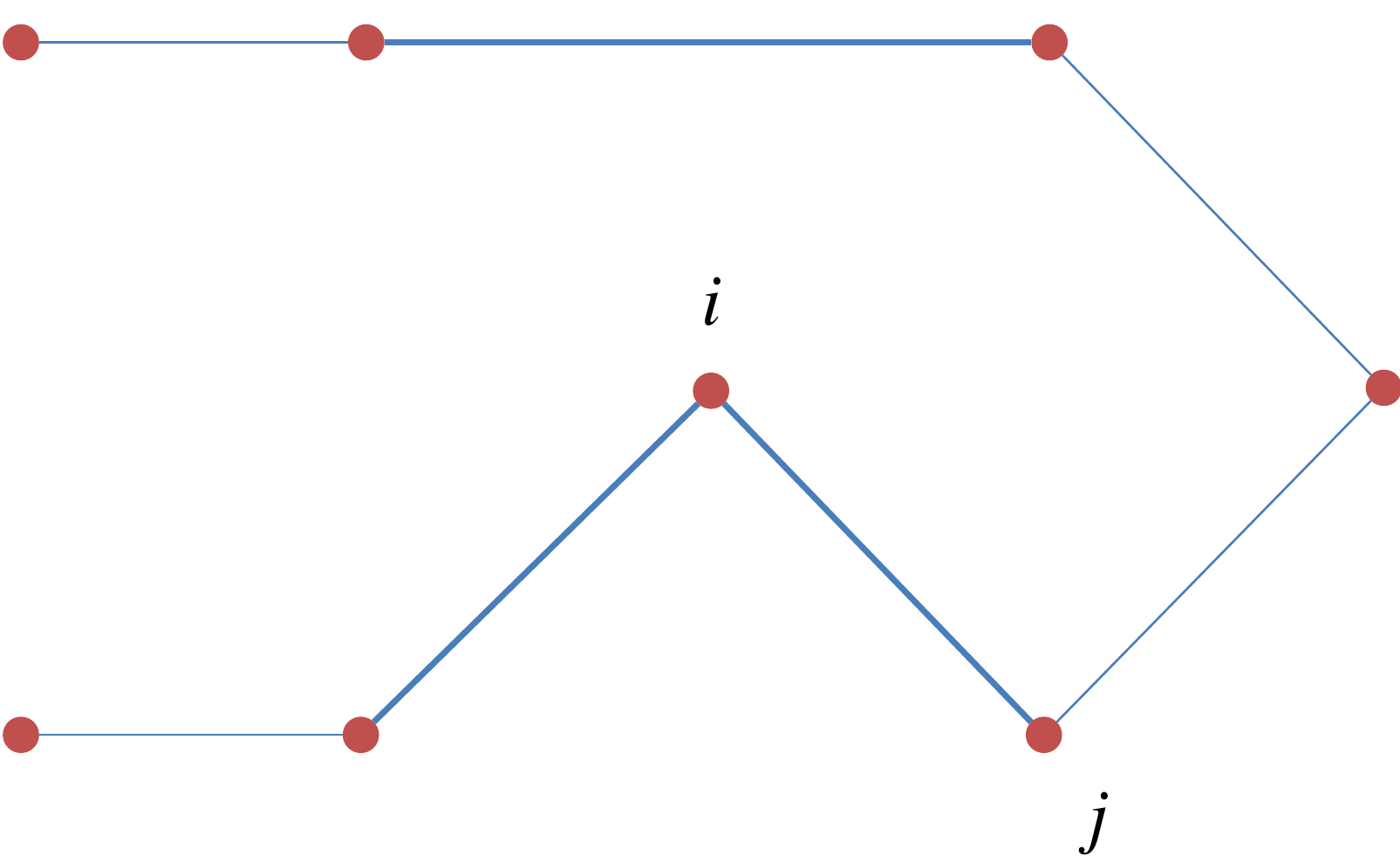}}
\caption{Reroute mutation (a restricted form of 3-opt)
(a) original path $\ldots, {i-1}, i, {i+1}, \ldots, {j-1}, j, {j+1}, \ldots$;
(b) final path $\ldots, {i-1}, {i+1}, \ldots, {j-1}, j, i, {j+1}, \ldots$
} 
\label{fig:plan-tsp-hill-reroute}
\end{figure}

%\subsubsection{TSP mutations}
The search uses two different permutation mutations, as shown in
 figures~\ref{fig:plan-tsp-hill-cross} and \ref{fig:plan-tsp-hill-reroute}.

\begin{figure*}[tp]
\centering
{\footnotesize(a)~}\includegraphics[width=0.28\textwidth]{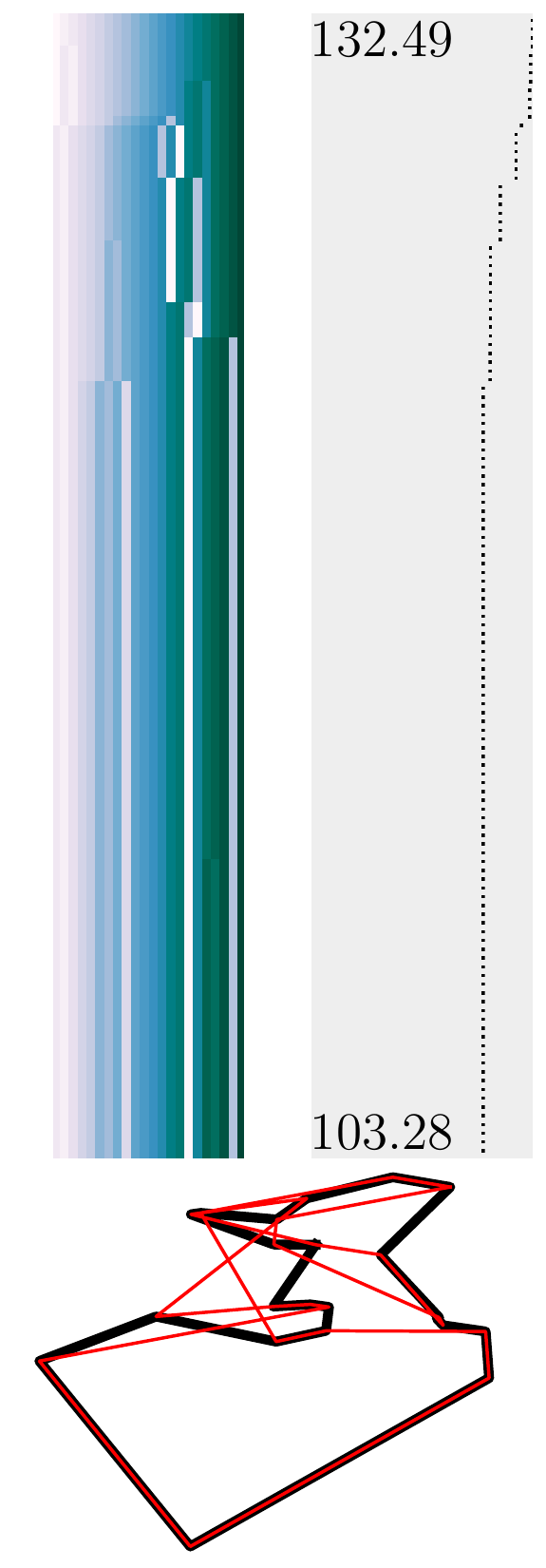}
{\footnotesize(b)~}\includegraphics[width=0.28\textwidth]{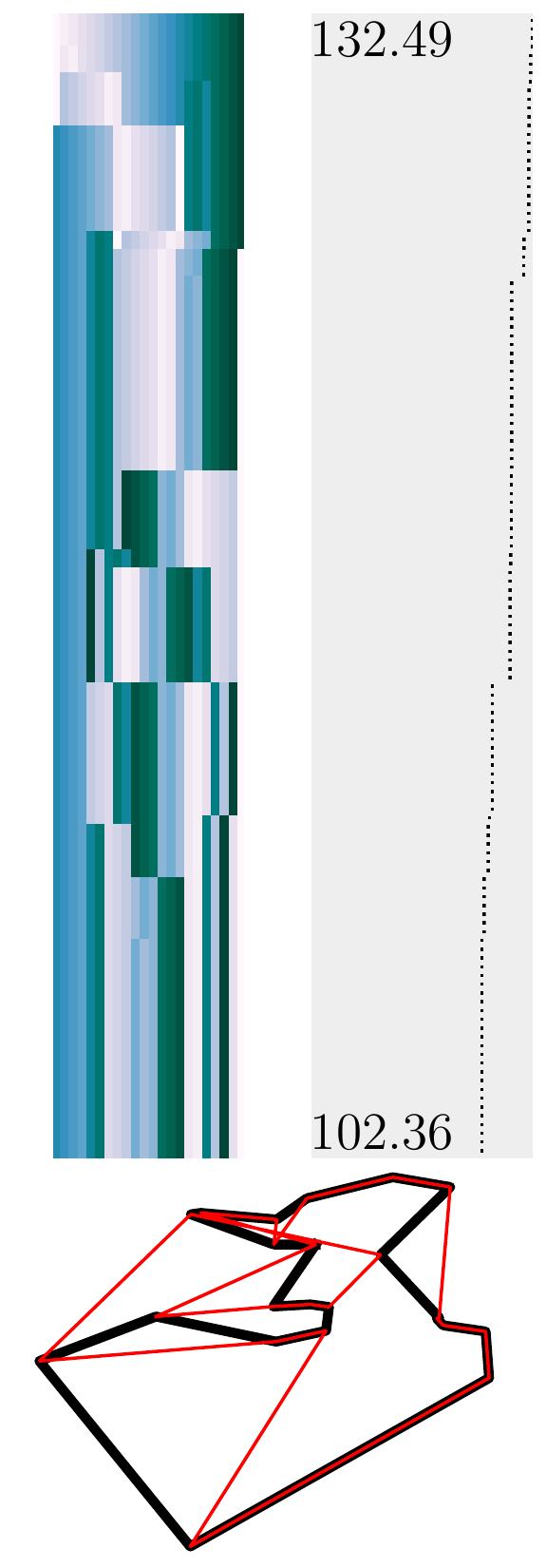}
{\footnotesize(c)~}\includegraphics[width=0.28\textwidth]{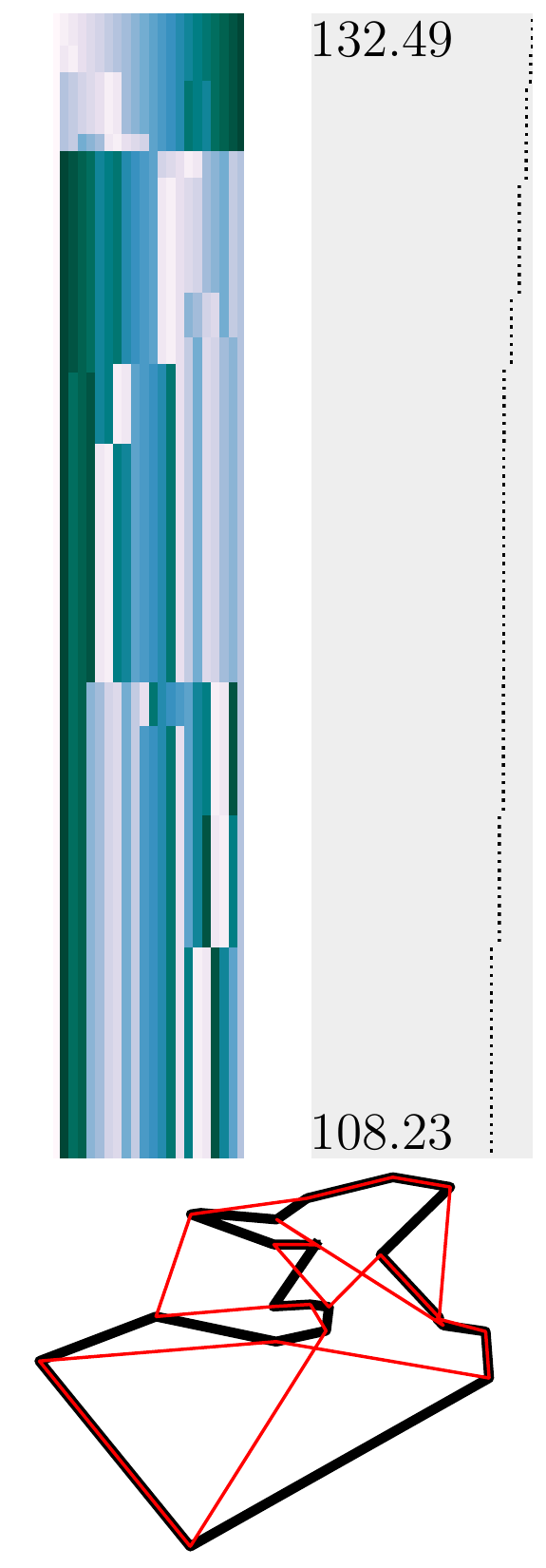}
\caption{22 city TSP searched by hill climbing, 130 steps, in plan tuple plot.
Black curve is optimal path; red curve is found path. 
(a) reroute mutations only; 
(b) 50/50 reroute and cross mutations;
(c) cross mutations only
%(See code in \S\ref{code:plan-tsp-hill}.)
} 
\label{fig:plan-tsp-hill}
\end{figure*}

\begin{figure*}[tp]
\centering
{\footnotesize(a)~}\includegraphics[height=0.84\textheight]{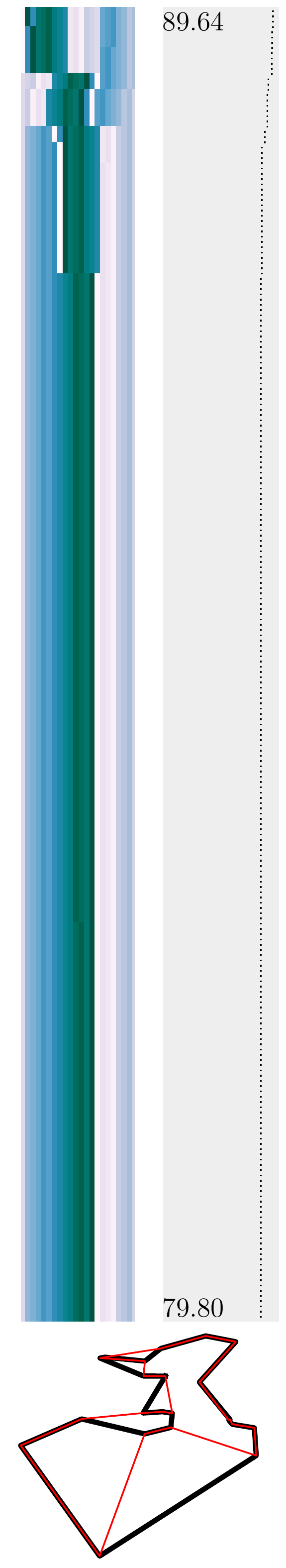}
{\footnotesize(b)~}\includegraphics[height=0.84\textheight]{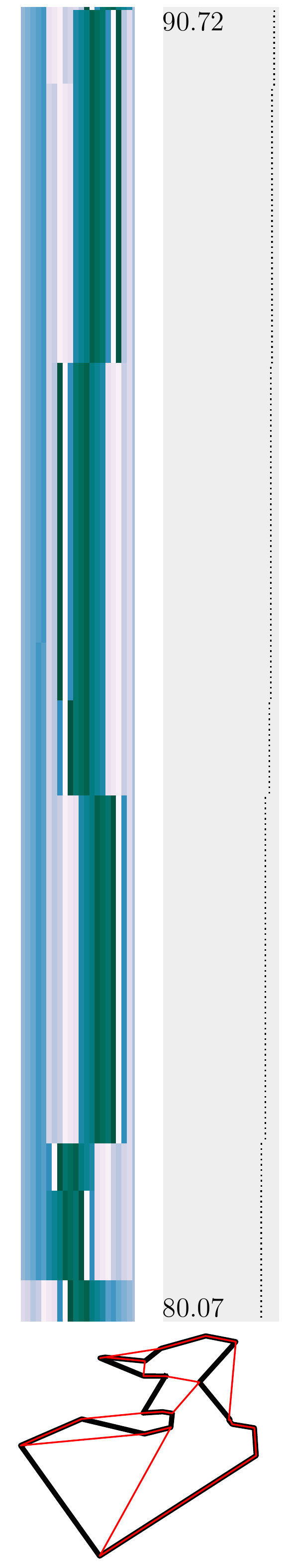}
{\footnotesize(c)~}\includegraphics[height=0.84\textheight]{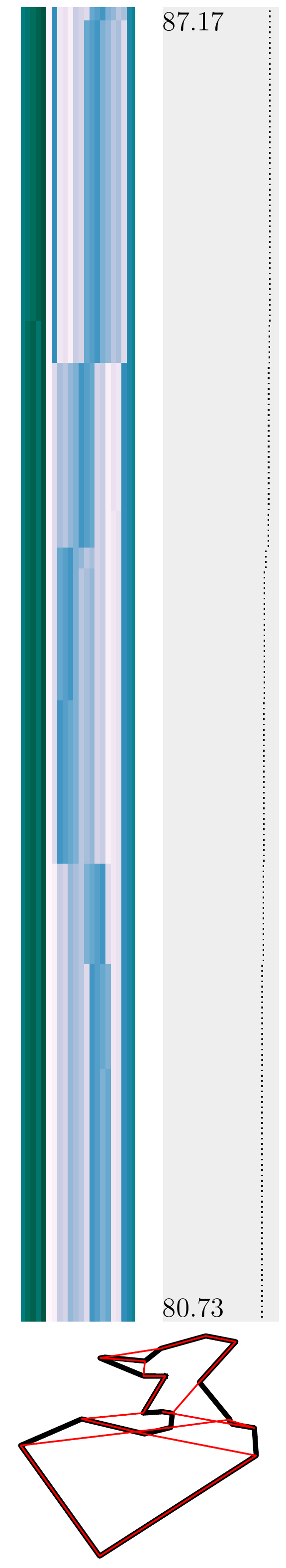}
\caption{22 city TSP searched by hill climbing, 250 steps, in plan tuple plot, with 50/50 cross and reroute mutations, greedy initial condition. Black curve is optimal path; red curve is found path. 
Colour map chosen such that optimal path has uniform change in colour.
Greedy initialisation starts at city: 
(a) 0; 
(b) 8;
(c) 16.
} 
\label{fig:plan-tsp-hill-greedy}
\end{figure*}

%\subsubsection{TSP hill climbing}
Figures~\ref{fig:plan-tsp-hill} and \ref{fig:plan-tsp-hill-greedy} illustrate the trajectories of two different forms of hill-climbing.

%---------------------------------------------------------------------
%\newpage
\section{Axis ordering}
\label{sec:axisorder}
We have seen one example of re-ordering the axes to highlight the dynamical structure,
in the RBN example, \S\ref{sec:plan:rbn}.
Additionally, the ECA example, \S\ref{sec:plan:rbn}, is naturally in the order that repects the topology.
But these are special cases.  There is a way to get a ``good'' (but not necessarily the best) order automatically,
by examining the structure of the system.

%\balance

Consider an $N$D dynamical system expressed as a set of first order ordinary differential equations
(the same argument applies to systems defined with a set of discrete time difference equations).
In general, we have:
\begin{align*}
    \dot{x}_1 &= f_1(x_1, x_2, \ldots, x_N) \\
    \dot{x}_2 &= f_2(x_1, x_2, \ldots, x_N) \\
    \ldots \\
    \dot{x}_N &= f_N(x_1, x_2, \ldots, x_N)
\end{align*}
This general case allows every $\dot{x}_i$ to depend on all $N$ of the $x_j$.
However, specific cases can have a more restricted dependence.
Instead of considering one $N$D dynamical system,
we can consider these as networks of $N$ coupled 1D dynamical systems, with the network nodes corresponding to the variables $x_i$, and edges providing the coupling information linking the nodes that appear in the corresponding $f_i$.

An example is the MSEIR model of infection\footnote{%
\url{en.wikipedia.org/wiki/Epidemic_model#The_MSEIR_model}
}, with equations:
\begin{align*}
    \dot{M} &= B - \delta M S - \mu M \\
    \dot{S} &= \delta M S - \beta S I - \mu S \\
    \dot{E} &= \beta S I - (\epsilon + \mu) E \\
    \dot{I} &= \epsilon E  - (\gamma + \mu) I \\
    \dot{R} &= \gamma I - \mu R 
\end{align*}

\begin{figure}[tp]
\centering
% l b r t
{\includegraphics[trim = 15mm 25mm 15mm 35mm, clip, width=0.7\columnwidth]{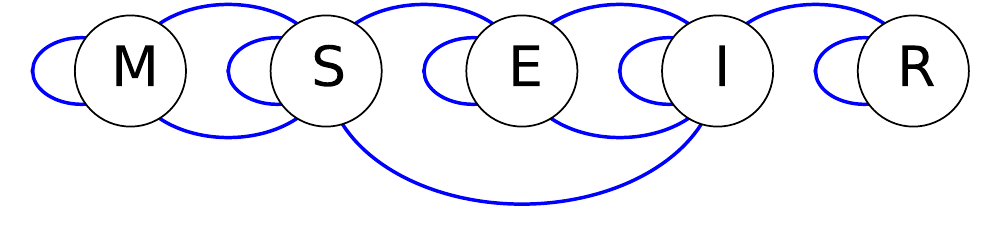}}
\caption{MSEIR model in network form.
}
\label{fig:MSEIR}
\end{figure}

We draw a network with the variables as nodes, and an edge from variable $X$ to variable $Y$ if the equation for $\dot{Y}$ contains $X$. 
We then lay out this network in a line in a way that minimises the length of the edges (since we are interested in edge length, not direction, we form an undirected graph; figure~\ref{fig:MSEIR}).
This provides a natural order for the axes in the tuple plot,
as it best captures the flow of information between the variables represented by each axis.

%\todo{CAs are a form of lattice dynamical systems;
%also threshold coupled lattice example;
%illustrate relationship between this approach and "core" approach for RBNs}

\begin{figure}[tp]
\centering
% l b r t
(a)\includegraphics[trim = 8mm 20mm 1mm 13mm, clip, width=0.8\columnwidth]{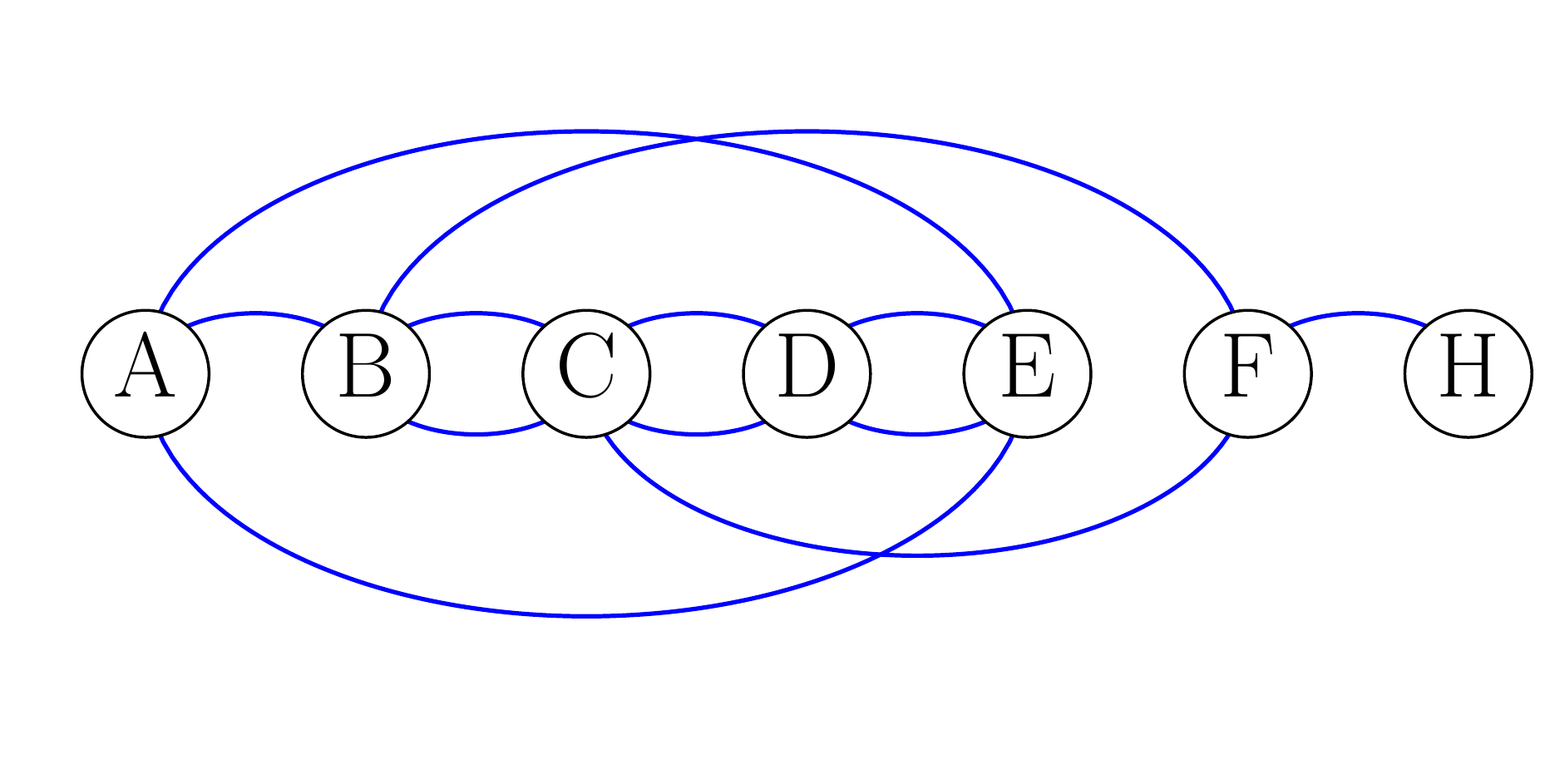}
(b)\includegraphics[trim = 8mm 30mm 1mm 13mm, clip, width=0.8\columnwidth]{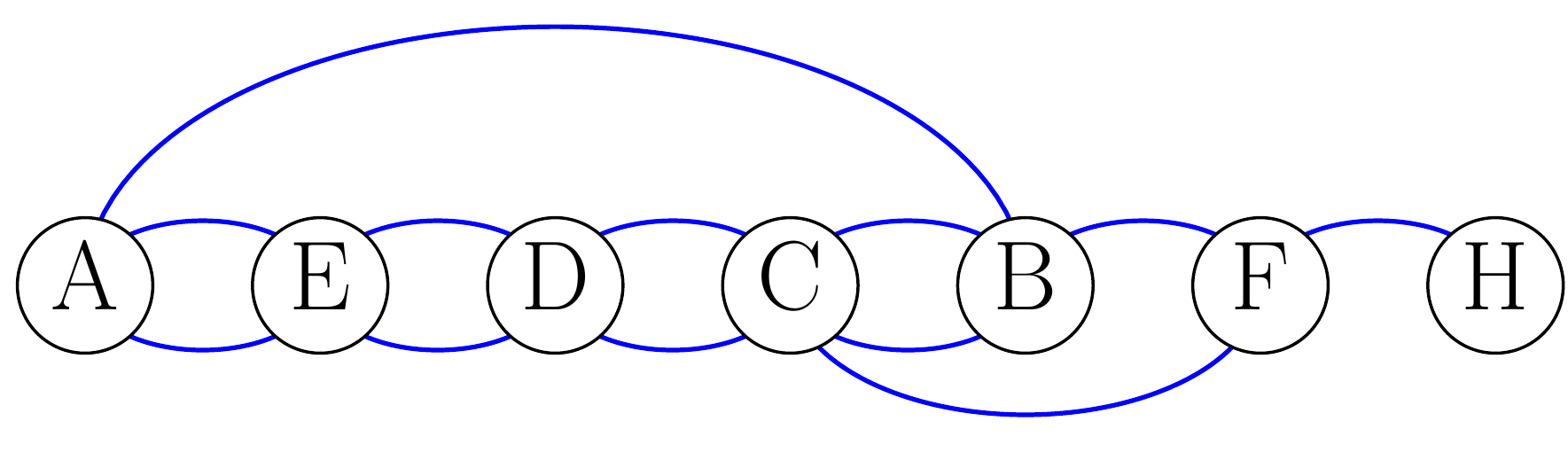}
\caption{6 state TM Busy Beaver contender: (a) alphabetical node order; (b) revised node order to minimise communication distance.
}
\label{fig:TM-BB}
\end{figure}

Consider the Turing Machine that is the current best contender for the 6-state, 2-symbol Busy Beaver\footnote{%
\url{en.wikipedia.org/wiki/Busy_beaver#Examples}
}.
Considering just the states (ignoring symbols and head movements), it has the following possible state transitions:
$
A \rightarrow B;
A \rightarrow E;
B \rightarrow C;
B \rightarrow F;
C \rightarrow D;
C \rightarrow B;
D \rightarrow E;
D \rightarrow C;
E \rightarrow A;
E \rightarrow D;
F \rightarrow H;
F \rightarrow C
$.
Drawing these nodes in alphabetical order results in the network shown in figure~\ref{fig:TM-BB}a.
Rearranging to minimise information communication distance results in figure~\ref{fig:TM-BB}b,
a candidate axis order (see \S\ref{sec:hybrid:tm}).

\begin{figure}[tp]
\centering
% l b r t
(a)~\includegraphics[trim = 0mm 0mm 152mm 0mm, clip, width=0.53\columnwidth]{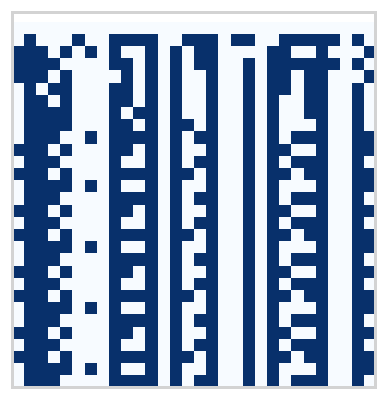}
(b)~\includegraphics[trim = 76mm 0mm 76mm 0mm, clip, width=0.53\columnwidth]{plan-rbn-3.pdf}
(c)~\includegraphics[trim = 152mm 0mm 0mm 0mm, clip, width=0.53\columnwidth]{plan-rbn-3.pdf}
\caption{RBN: 
(a) random; 
(b) revised node order to minimise communication distance;
(c) sorted to expose frozen core.
}
\label{fig:rbn_min_3}
\end{figure}

\begin{figure*}[tp]
\centering
% l b r t
(a){\includegraphics[width=0.932\textwidth]{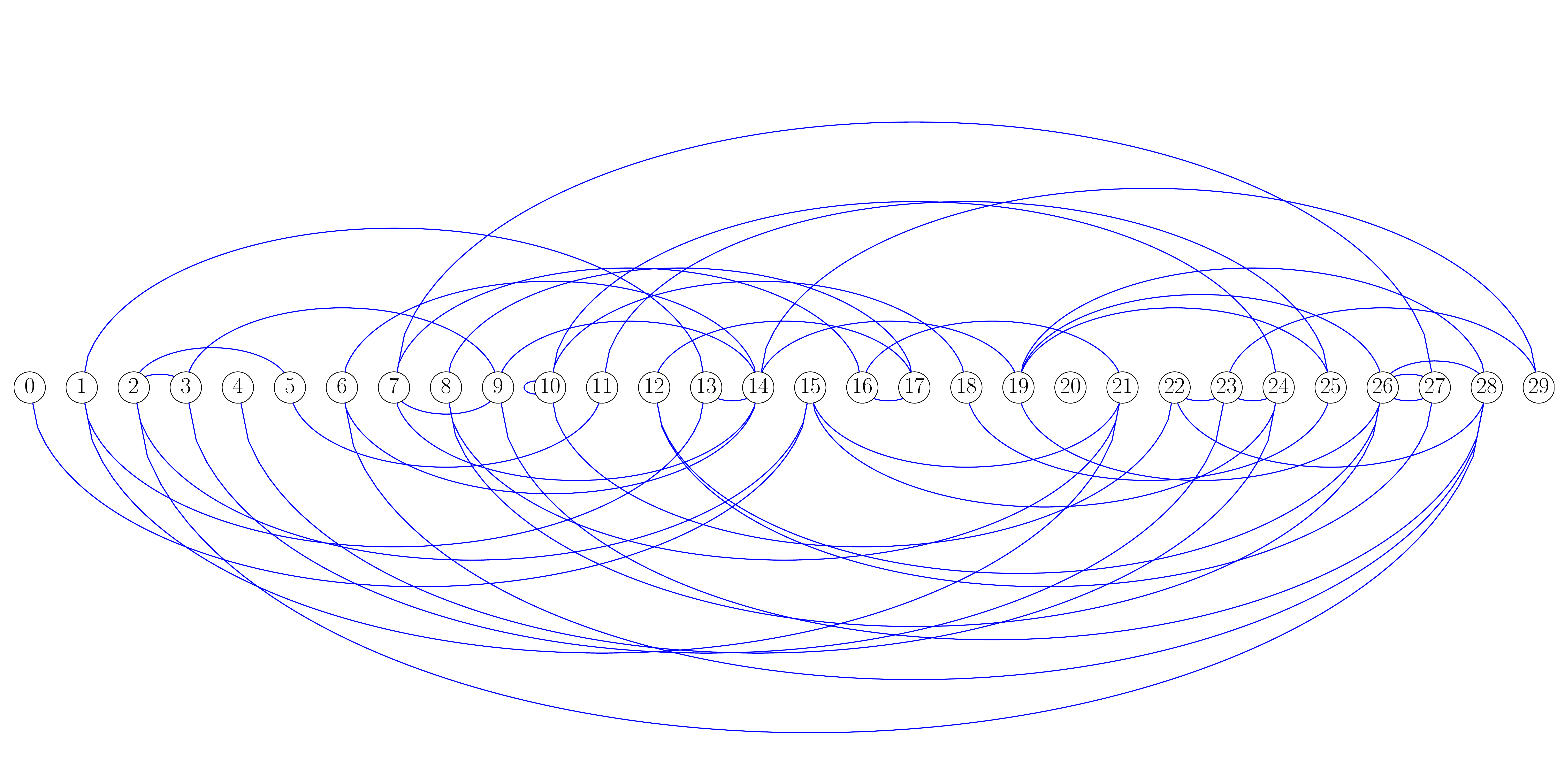}}
(b){\includegraphics[trim = 0 20mm 0 100mm, clip, width=0.932\textwidth]{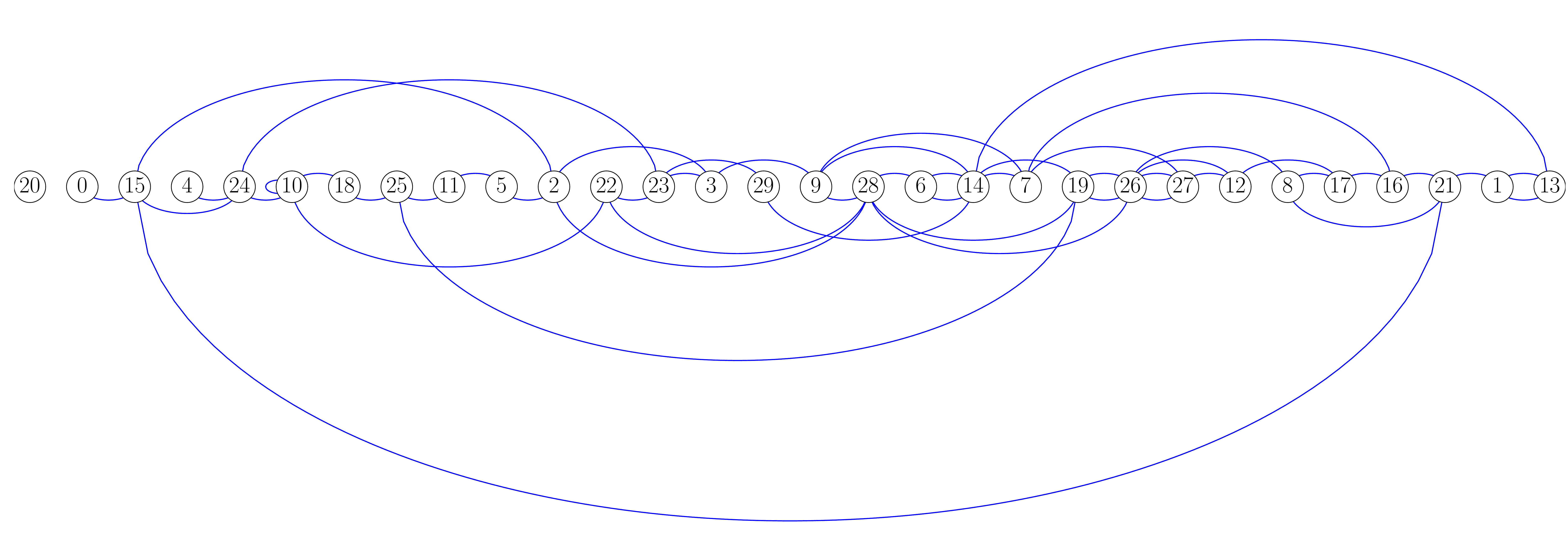}}
(c){\includegraphics[trim = 0 30mm 0 0mm, clip, width=0.932\textwidth]{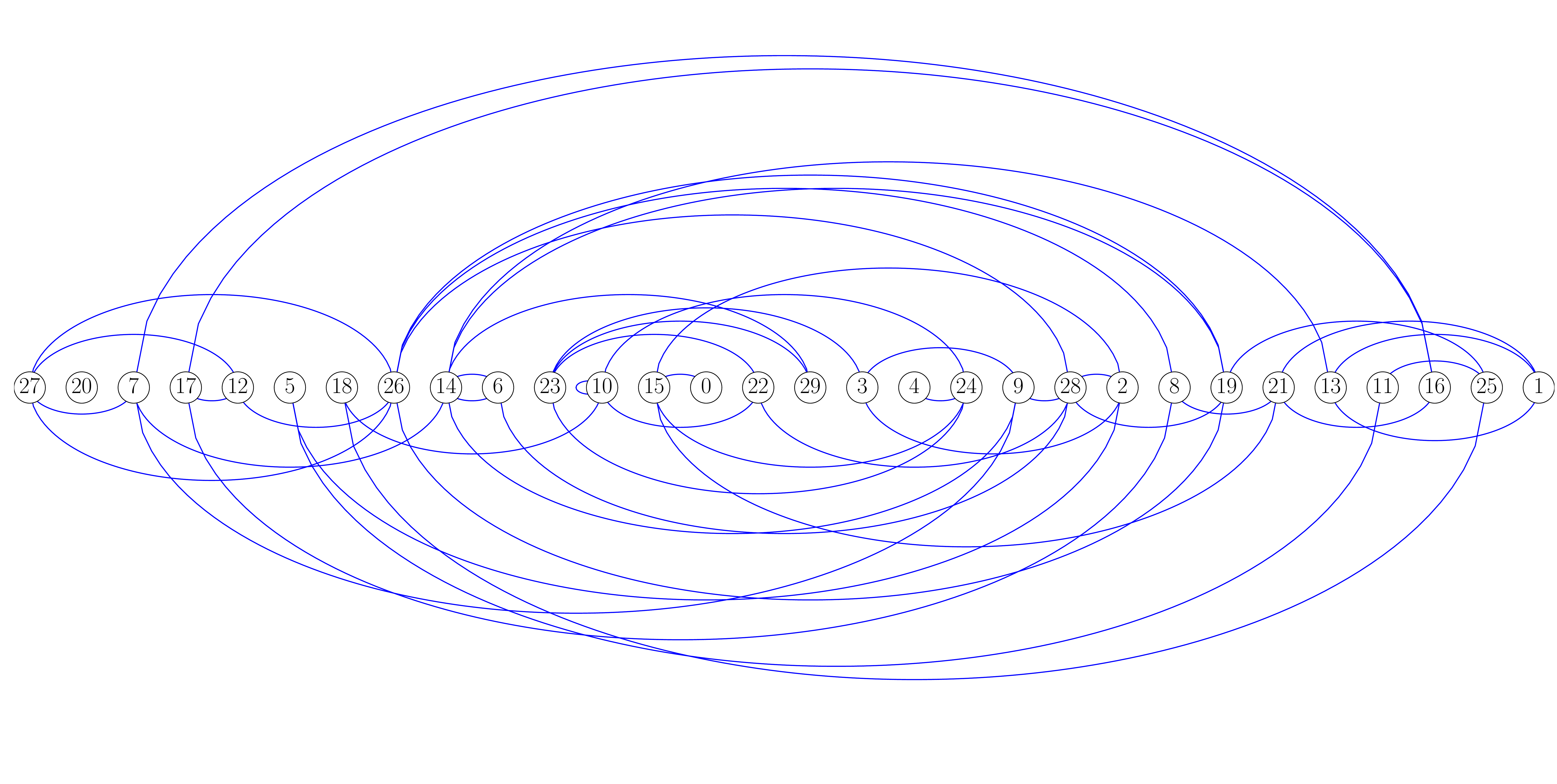}}
\caption{RBN: 
(a) random; 
(b) revised node order to minimise communication distance;
(c) sorted to expose frozen core.
}
\label{fig:rbn_min}
\end{figure*}

We can use a similar approach to order the nodes in an RBN.
For example,  
figure~\ref{fig:rbn_min_3} shows Random Boolean network plan plots
with three different node orderings shown in 
figure~\ref{fig:rbn_min}.
The ordering by minimising the communication distance results in a better plot than random,
but in the RBN case there is a further improvement possible, based on the frozen core.
The minimal communication distance ordering is a first choice, but the structure of the system may provide a better choice.

%======================================================================
\chapter{Side Tuple Plots}\label{sec:side}
\nobalance
%---------------------------------------------------------------------
\section{Looking sideways}\label{sec:side:look}
%........................................................
\subsection{Motivation}\label{sec:side:look:motiv}

\begin{figure*}[tp]
\includegraphics[width=0.95\textwidth]{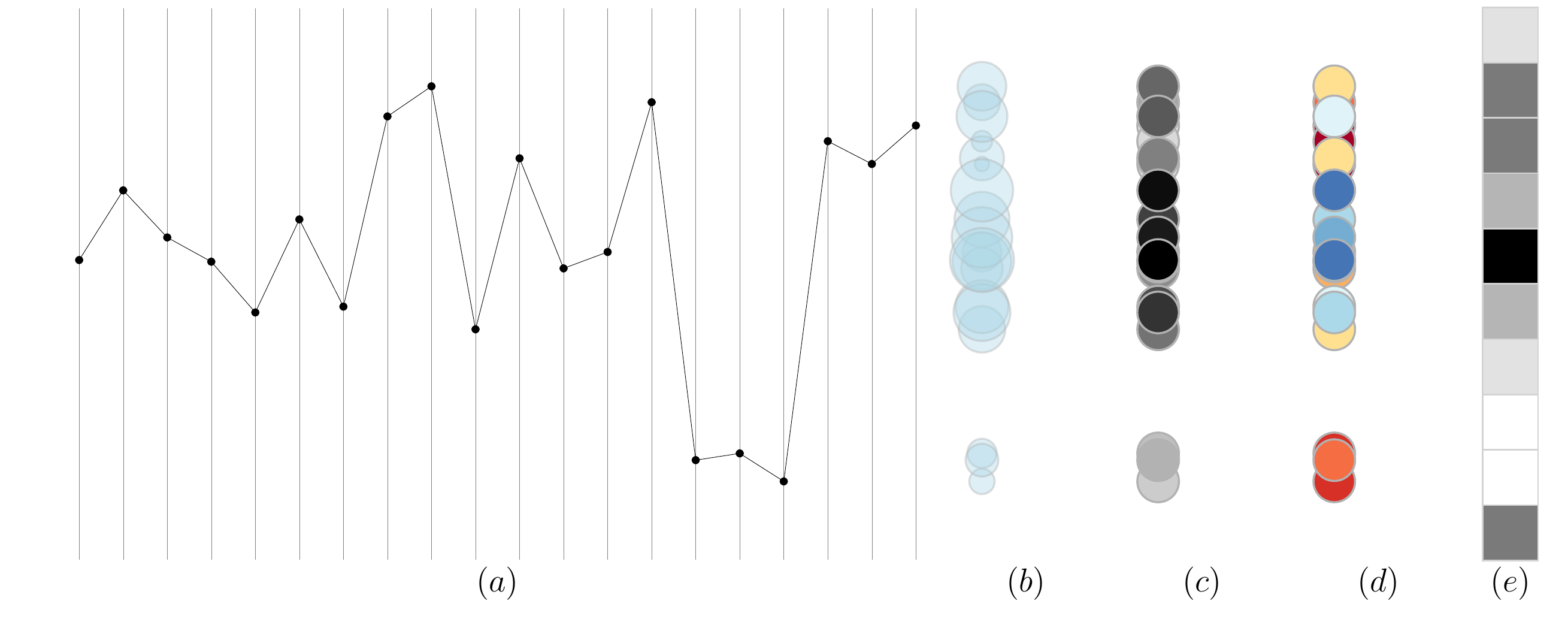}
\caption{20D real-valued random data (so $N=7, V=\Re$):
$(a)$ parallel coordinates plot;
$(b)$ the `side' view, from projecting the values onto a single vertical axis: 
side tuple plot using data points with sizes diminishing with higher dimension,
and semi-transparent markers; 
$(c)$ side tuple plot, using data points with grey levels diminishing with higher dimension;
$(d)$ side tuple plot, using data points with a colour range, useful if different dimensions have categorically different meanings;
$(e)$ side tuple plot, using a density histogram of the data points.
}
\label{fig:side-motivate}
\end{figure*}

Again consider a system with $N$ identical dimensions each with values drawn from an ordered set $V$, 
 plotted using $N$ parallel coordinates.  
Consider now looking ``across'' these coordinates, from the side elevation (figures~\ref{fig:3d3projections}c, \ref{fig:side-motivate}).  
 The various coordinate lines are overlaid, and seen as one.  

All the values are projected onto this resulting single merged coordinate line.
The higher-numbered dimensions are further away, so we might think that perspective would make the symbols representing these points look smaller (figure~\ref{fig:side-motivate}b).
 Alternatively, the further away dimensions might appear to be ``fainter'' (figure~\ref{fig:side-motivate}c).
 Or we could simply ``paint'' a suitable colours on each axis (figure~\ref{fig:side-motivate}d).
Note that values on nearer axes may occlude those on further axes.

Once there are many dimensions, the side tuple plot is in danger of becoming cluttered with multiple overlapping symbols.
An alternative is a density side tuple plot, where the symbol used represents a density histogram of the number of axes that have a value in the given range (figure~\ref{fig:side-motivate}e).

%........................................................
\subsection{Definition of side tuple plot}\label{sec:side:defn}
We use this ``looking across'' idea to define side tuple plots.
We plot one 2D point for each component of the tuple;
its position represents its value, and its symbol (size, shape, colour) represents (a) the axis index 
or (b) the density of axis indexes falling within the spatial extent of the symbol.

\begin{figure*}[tp]
\centerline{
\includegraphics[scale=0.9]{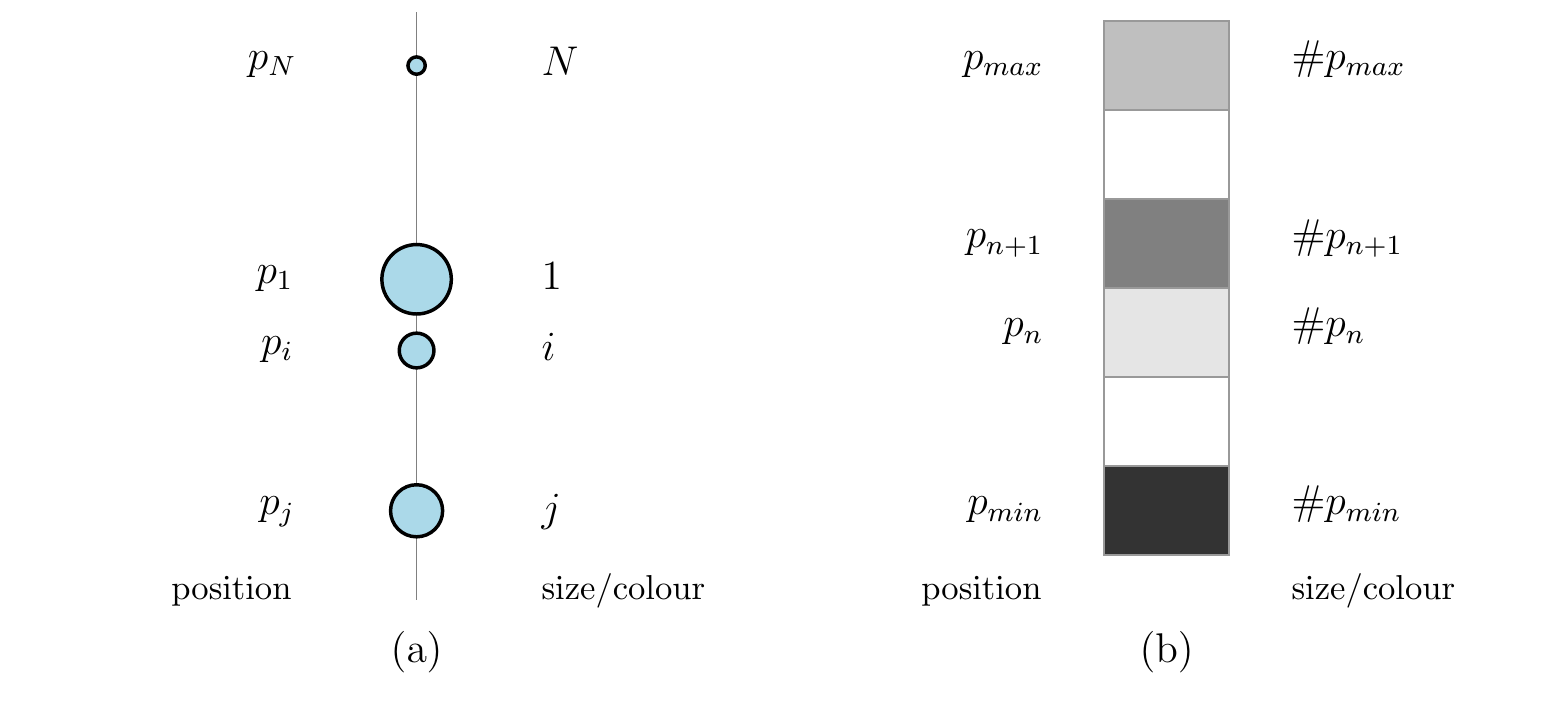}
}
\caption{Side tuple plot of the point
${\bf p} = (p_1, \ldots, p_i, \ldots, p_j, \ldots, p_{N}) \in V^N$.
(a)~Symbol side tuple plot.
There is a symbol representing each dimension $n\in 1\dots N$.
The line is an axis with positions corresponding to values $p_n \in V$.
(The vertical line itself may be elided.)
Each $p_n$ is plotted on the line at a position representing the value of $p_n$
with a symbol representing the dimension $n$.
Symbols for higher dimensions may be occluded by symbols for lower dimensions with the same value.
(b) Density side tuple plot.
The axis is divided into equal-sized bins, and the colour of each bin indicates the number of dimensions whose value falls in that bin.
}
\label{fig:side_defn}
\end{figure*}

\begin{mdframed}[style=defn,frametitle={Definition: Symbol side tuple plot},nobreak=false]
Given
\begin{compactenum}
\item an $N$D state space $V^N$
\item an $N$D point ${\bf p} = (p_n)_{n=1}^{N} \in V^N$
\item a plotting function $sym:1..N\rightarrow S$ 
\item a position function $\varpi_v : V \rightarrow \Re$ that maps values to a position in the plotting plane
\end{compactenum}
\vspace{2mm}
A {\bf symbol side tuple plot} displays ${\bf p}$ on the vertical $y$-axis line as the set of points 
\[ \{\, {\bf x}_n  \,\}_{n=1}^{N} = \{\, (1, \varpi_v(p_n) )  \,\}_{n=1}^{N} \]
where each ${\bf x}_n$ point is plotted using the symbol $sym(n)$.

See figure~\ref{fig:side_defn}a, where the symbol for each dimension has a different size;
alternatively, it might have a different colour, or shape, or combination of these.
\end{mdframed}

The position function $\varpi_v$ allows values to be scaled before being plotted.

\begin{mdframed}[style=defn,frametitle={Definition: Density side tuple plot},nobreak=false]
Given
\begin{compactenum}
\item an $N$D state space $V^N$
\item an $N$D point ${\bf p} = (p_n)_{n=1}^{N} \in V^N$
\item a number of bins $B$
\item a binning function $\beta: V \rightarrow 1..B$, that partitions $V$; if $V$ has an order $\leq$, then the partition respects the order \\
$v_1 \leq v_2 \Rightarrow \beta(v_1) \leq \beta(v_2)$
\item a density function $\rho : 1..B \rightarrow [0, 1]$ \\
$\rho(b) = \# \{~ p_n ~|~ \beta(p_n) = b ~\}_{n=1}^{N} / N$
\item a plotting function $sym: [0,1] \rightarrow S$ 
\end{compactenum}
\vspace{2mm}
A {\bf density side tuple plot} displays ${\bf p}$ on the vertical $y$-axis line as the set of bins 
\[ \{\, {\bf x}_b  \,\}_{b=1}^{B} = \{\, (1,b)  \,\}_{b=1}^{B} \]
where each bin ${\bf x}_b$ is plotted using the symbol $sym(\rho(b))$.
(See figure~\ref{fig:side_defn}b.)
\end{mdframed}

%\todo{binning only works if bins are same size??}

%.....................................................................
%\subsection{Extended side tuple plot}\label{sec:side:ext}

%\todo{ What's the $posn$ or $perm$, and use here?  Use idea that ensemble/traj plots are just isomorphic to side with a suitable $posn$ mapping?}

%........................................................
\subsection{Ensembles and trajectories}\label{sec:side:traj}
We might wish to display an \textit{ensemble} of points, representing some population in $N$D space,
or a \textit{time series} of points, representing some trajectory through $N$D space.
If we animated a side tuple plot of a sequences of states, the $N$ symbols representing the $N$D point would stay fixed in size or colour (designating their axis position), and change in position (as their value changes).
We can turn the animation into a static display if we use a 1D plan tuple plot for a single point, and use the second dimension of the plot to display the different points (see figure~\ref{fig:3d3projections}c).

\balance

\begin{mdframed}[style=defn,frametitle={Definition: ensemble symbol side tuple plot}]
Given
\begin{compactenum}
\item an $N$D state space $V^N$
\item a set of $M$ $N$D points to be plotted: $\{\, {\bf p}_m \,\}_{m = 1}^{M} = \{\, p_{mn} \,\}_{m = 1;}^{M}{}_{n = 1}^{N}$ where each $N$D point ${\bf p} = (p_n)_{n=1}^{N} \in V^N$
\item a plotting function $sym:1..N\rightarrow S$  
\item  a position function $\varpi_v: V \rightarrow \Re$ that maps values to a position in the plotting plane
\item a position function of the points' indexes $\varpi_m: 1..M \rightarrow \Re$ 
\end{compactenum}
\vspace{2mm}
An {\bf ensemble symbol side tuple plot} displays each of these ${\bf p}_m$ in the plotting plane as a side tuple plot, with 
\begin{align} 
&\{ {x}_{mn} \}_{m = 1;}^{M}{}_{n = 1}^{N} =  \nonumber \\
&\quad \{ ( \varpi_m(m), \varpi_v(p_{mn}) ) \}_{m = 1;}^{M}{}_{n = 1}^{N} \nonumber
\end{align}
where each ${p}_{mn}$ point is plotted using the symbol $sym(p_{mn})$.
\end{mdframed}

There is an analogous definition for an ensemble density side tuple plot.

\begin{figure*}[tp]
\centerline{
\includegraphics[width=0.95\textwidth]{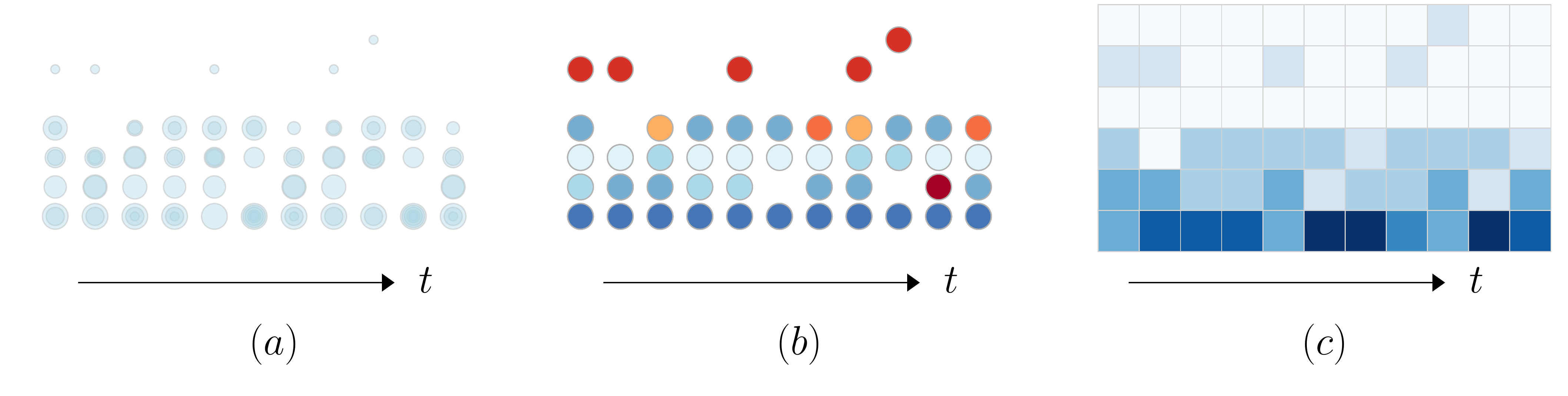}
}
\caption{A trajectory, or time series, as a side tuple plot:
(a) symbol plot, using symbol size;
(b) symbl plot, using symbol colour;
(c) density plot.
}
\label{fig:side-traj}
\end{figure*}

A trajectory through this state space can be visualised by showing consecutive states in consecutive columns as a time series (figure~\ref{fig:side-traj}). This trajectory visualisation does not necessarily make cycles in the trajectory
immediately visible, but they can be inferred as repeated patterns of states.

%\todo{ Interpretation}

%\todo{ Density plots instead?}

%\todo{Are density plots the only ones that make sense?}

\begin{figure*}[tp]
\centering
\includegraphics[width=0.99\textwidth]{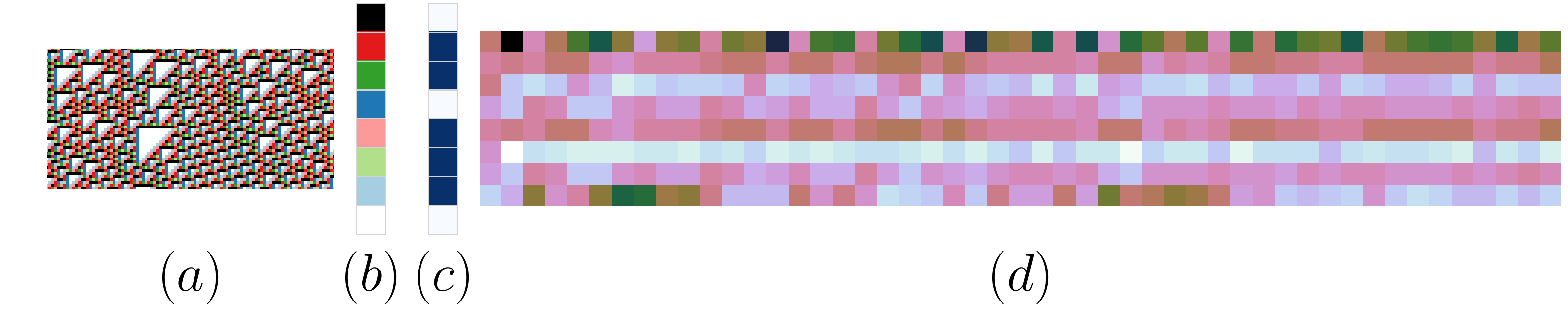}
\caption{Side tuple plot of the lookup table access, of
Elementary CA rule 110 with $N=400$ cells, $T=200$ timesteps, 
and a random (50\% 0s, 50\% 1s) initial condition (right column).
The plan tuple plot is repeated from figure~\ref{fig:plan-ca110-table}, for reference.
}
\vspace{4cm}
\label{fig:side-ca}
\end{figure*}
%---------------------------------------------------------------------
%\clearpage
\section{Information loss}
%---------------------------------------------------------------------
Unlike the plan tuple plot, the side tuple plot can lose some information.
For a given point, if different dimensions have the same value,
the symbols will be overlayed, and the `higher' numbered dimension data occluded;
this can be mitigated to some degree by the use of transparent symbols.
The density plot allows the values of all dimensions to be captured, but loses the information of which dimension is which.

%---------------------------------------------------------------------
\section{Examples}\label{sec:side-egs}
%........................................................
\subsection{Hypersphere surface}\label{sec:side-hyper}

\begin{figure*}[tp]
\centering
\includegraphics[width=0.84\textwidth]{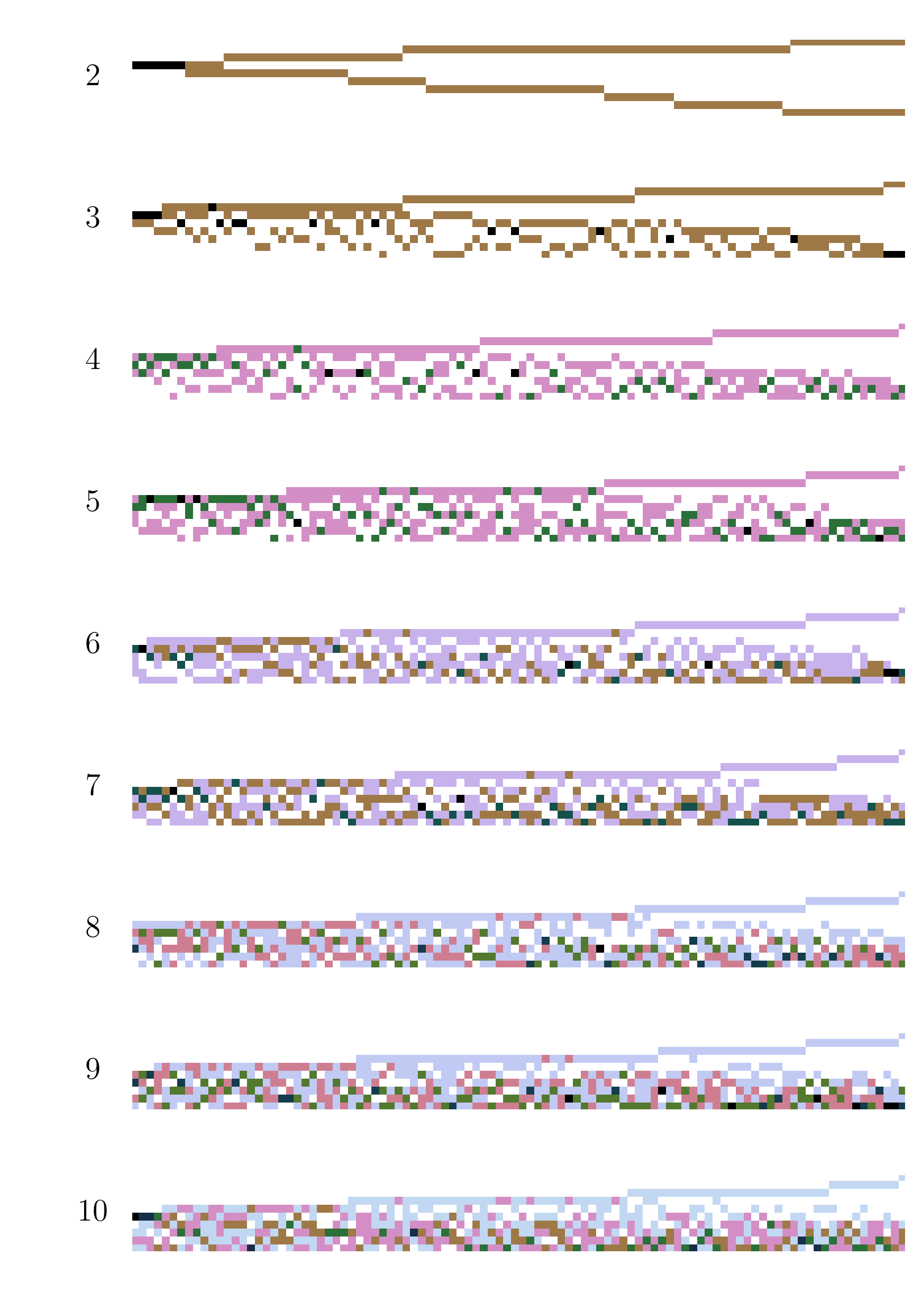}
\caption{Ensemble density side tuple plot of 100 points on the surface of the positive hyper-octant of an $N$-dimensional hypersphere, $N=2$--$10$.
The ensemble of points are sorted in order of the
maximum coordinate value.
}
\label{fig:side-hyper}
\end{figure*}

Figure~\ref{fig:side-hyper} plots 100 points from the surface of a hypersphere in an ensemble side tuple plot.  Contrast with the plan tuple plot of figure~\ref{fig:plan-hyper}.

%........................................................
%\clearpage
%\subsection{Iris data in a side tuple plot}\label{sec:side-iris}
%\todo{all data, and in three colours}

%........................................................
%\clearpage
\subsection{Elementary cellular automata}\label{sec:side-ca}

\begin{figure*}[tp]
\centering
\includegraphics[width=0.93\textwidth]{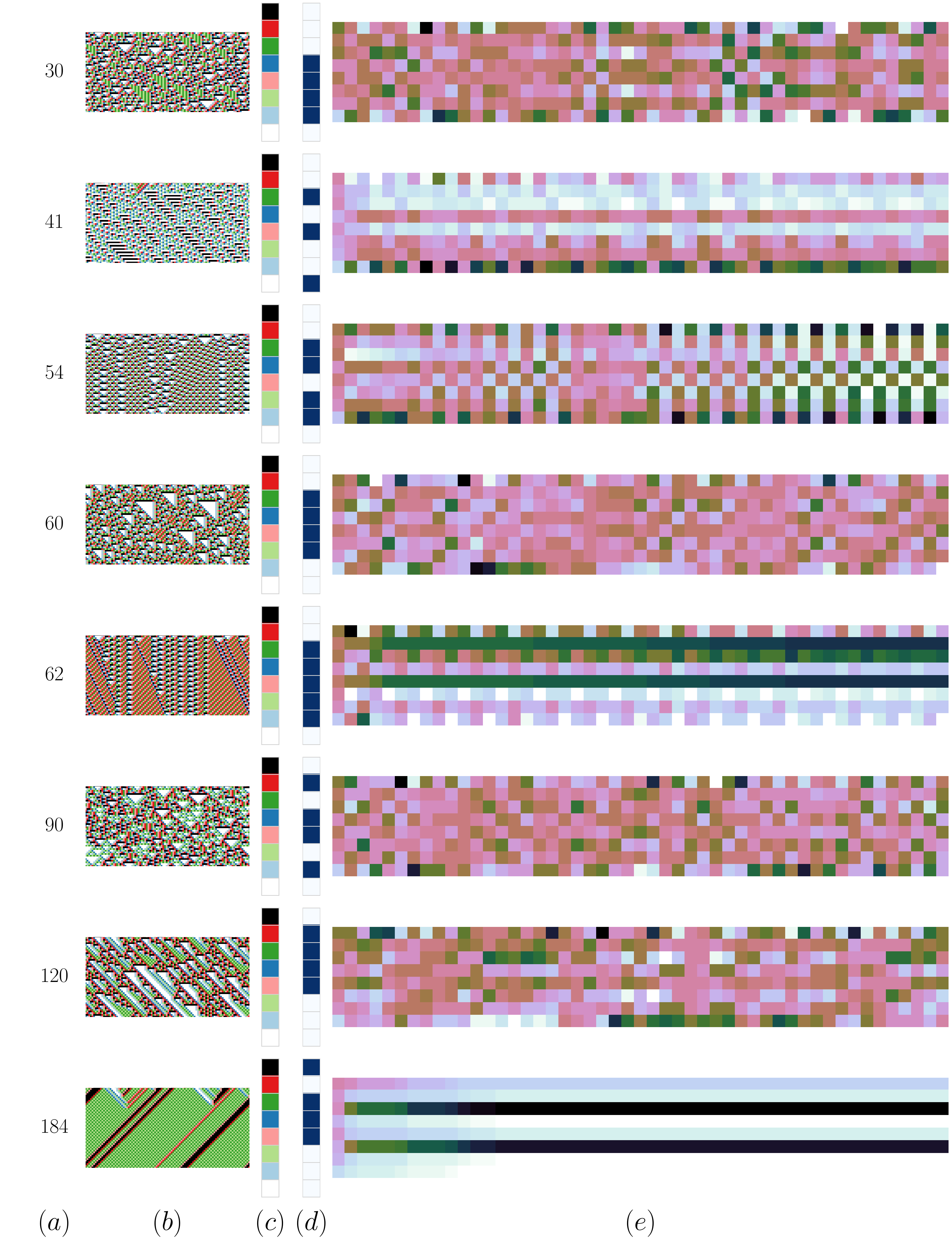}
\caption{Side tuple plot of the lookup table access, of
Elementary CA rules with $N=100$ cells, $T=40$ timesteps, 
and a random (50\% 0s, 50\% 1s) initial condition (right column).
The plan tuple plot, and colour bar, and lookup table bar, are repeated from figure~\ref{fig:plan-ca-multi}, for reference.
}
\label{fig:side-ca-multi}
\end{figure*}

See figure~\ref{fig:side-ca}, \ref{fig:side-ca-multi}.
The more chaotic class 3 ECAs have a more uniform distribution in the side tuple plot,
whereas the class 2 ECAs have a more varied distribution, indicating that rules are being accessed differently because of their increased structure.

%---------------------------------------------------------------------
\subsection{Coupled logistic maps}\label{sec:side-clm}

\begin{figure*}[tp]
\centering
\includegraphics[width=0.9\textwidth]{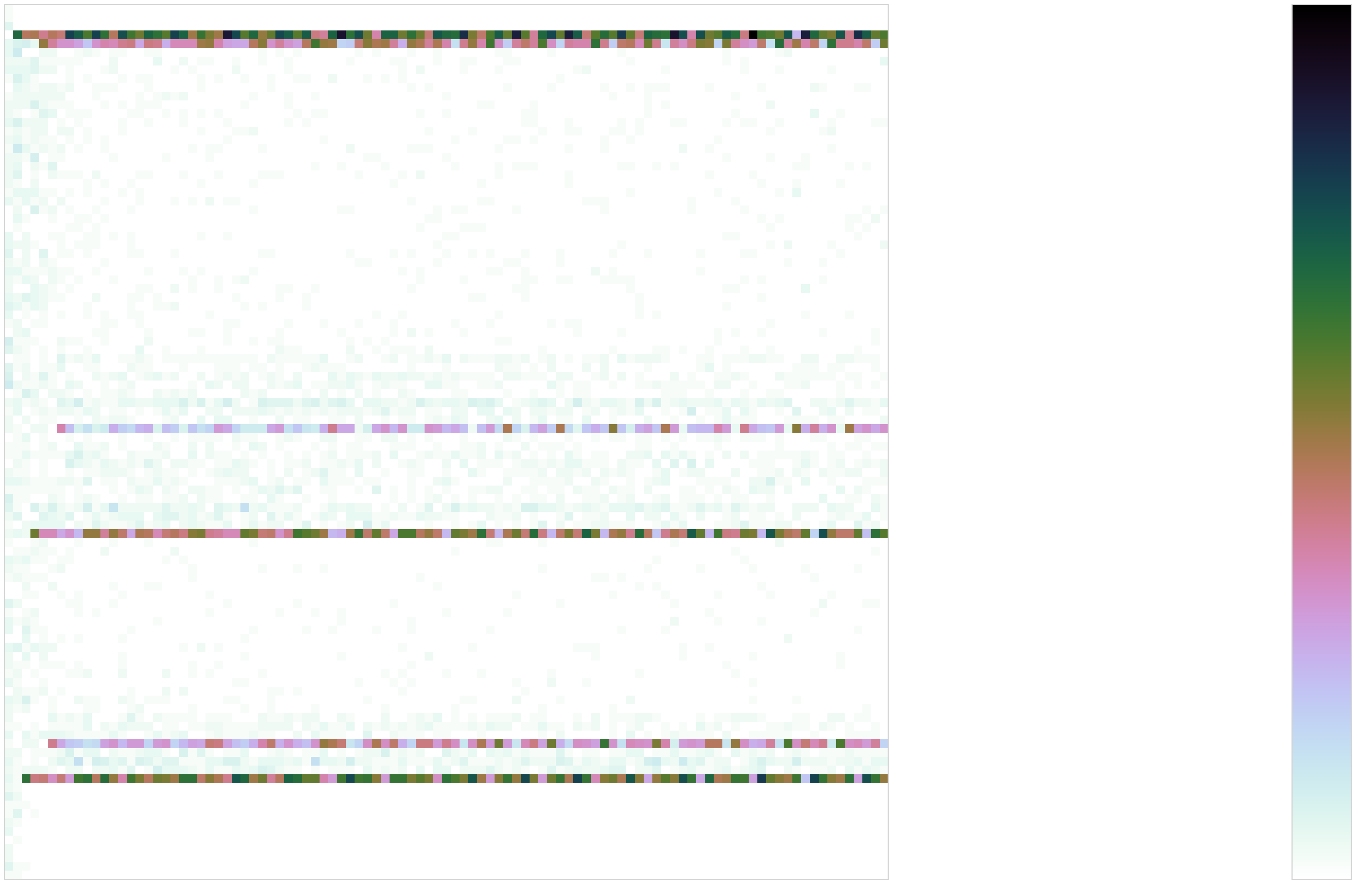}
\caption{Side tuple plot of the threshold coupled lattice,
threshold $x_* = 0.971$, number of cells $N = 200$, 50 timesteps:
the $x$ axis shows time running across the page;
the $y$ axis shows the histogrammed value of each of the 200 cells. 
} 
\label{fig:side-clm}
\end{figure*}

A particular instance of a threshold coupled lattice is shown in figure~\ref{fig:side-clm},
 in a side tuple plot.  Contrast with the plan tuple plot of figure~\ref{fig:plan-clm}.
The different views can be used to highlight different aspects of the system's trajectory through its state space.

%======================================================================
\chapter{Hybrid Tuple Plots}
\label{sec:hybrid}
\nobalance
%........................................................
\section{Motivation}\label{sec:hybrid:motiv}

The plan and side tuple plots defined above assume homogeneous dimensions.
Complicated and complex systems can have heterogeneous dimensions,
for example, a mix of continuous and discrete dimensions.
We can use a hybrid of different appropriate techniques for these different dimensions
to get a visualisation of the whole state space.
%........................................................
\section{Definition of hybrid tuple plots}\label{sec:hybrid:defn}

Assume the full heterogeneous $N$D state space  $V^N$ can be partitioned into $d$ subspaces:
$V_1^{N_1} \times V_2^{N_2} \times \ldots \times V_d^{N_d}$, where each of the subspaces $V_i^{N_i}$ is homogeneous.
A hybrid tuple plot then is a concatenation of individual plots appropriate for each of the subspaces.

%\todo{ The ``points'' are now not atomic, but are plots themselves -- link to Grammar of Graphics}
%-------------------------------------
\section{Examples}\label{sec:hybrid:egs}
%............................................
\subsection{Turing machines}\label{sec:hybrid:tm}

Consider a Turing machine (TM) (see, eg, \cite[ch.3]{Sipser1997}).  
It has a finite set of (machine) states $Q$,
a finite tape alphabet $\Gamma$,
and a transition function $\delta : Q \times  \Gamma \rightarrow Q \times \Gamma \times \{L,R\}$
which defines how the machine state changes and how the head reads, writes and moves on the tape.
The full state space (configuration space) of a TM is
$Q \times \mathbb Z \times (\mathbb Z \rightarrow \Gamma)$,
which captures the machine state, the head position, and the tape state.

\begin{figure*}[tp]
\centerline{
\scalebox{0.55}{\includegraphics{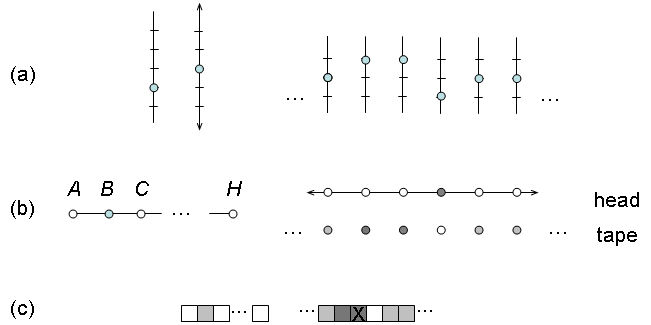}}
}
\caption{Hybrid tuple plots of a 3 tape-symbol TM:
(a) the parallel coordinates view (leaving the polyline implicit): 
the state space $q \in Q$, the head position $h \in \mathbb Z$,
the tape $t \in \mathbb Z \rightarrow \Gamma$;
(b)~the various coordinates rearranged, and the tape as a plan tuple plot; 
(c) combining these into a hybrid visualisation of the state space.}
\label{fig:tmpar}
\end{figure*}

\begin{figure}[tp]
\centerline{
\includegraphics[height=0.12\textheight,valign=t]{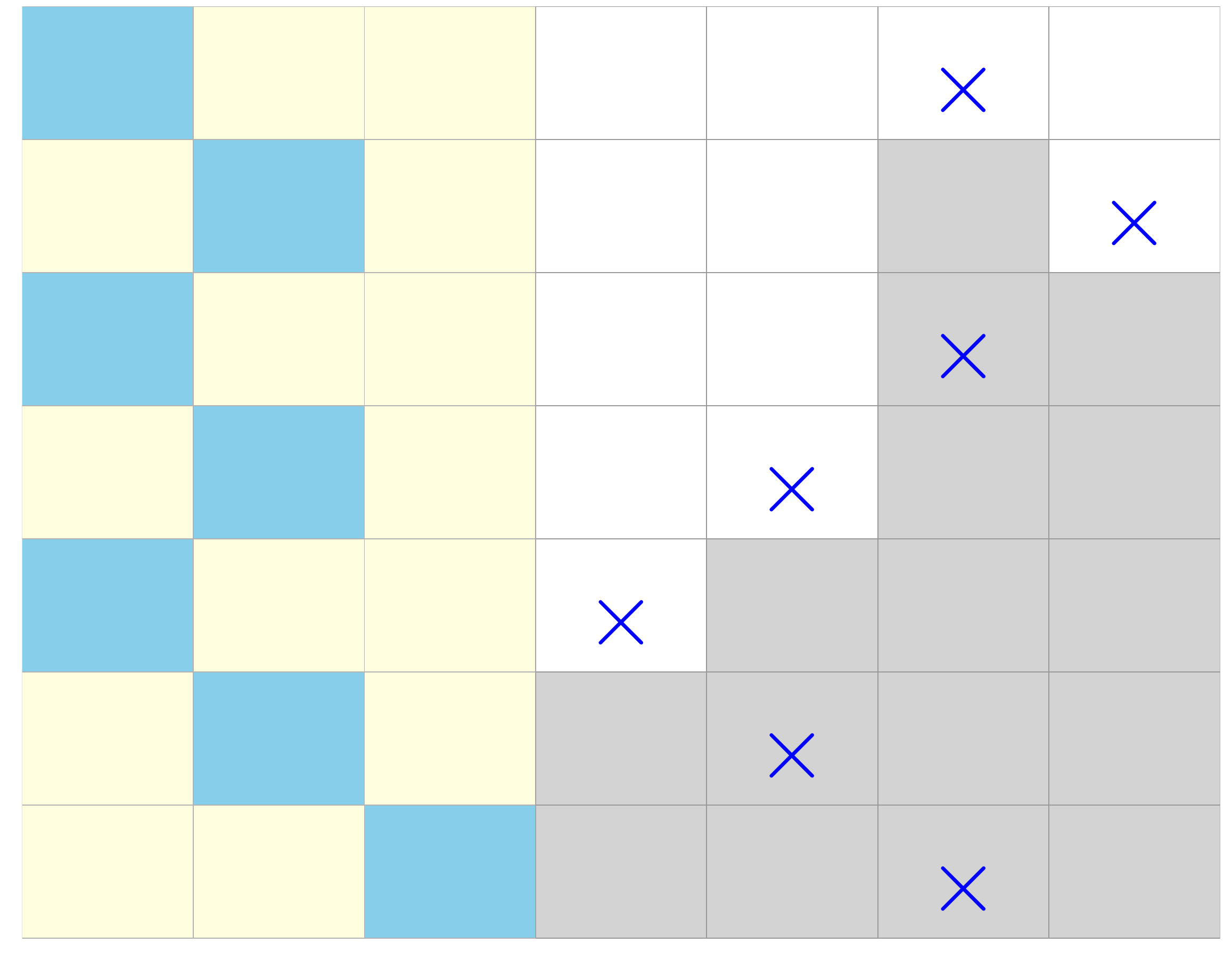}
\includegraphics[height=0.24\textheight,valign=t]{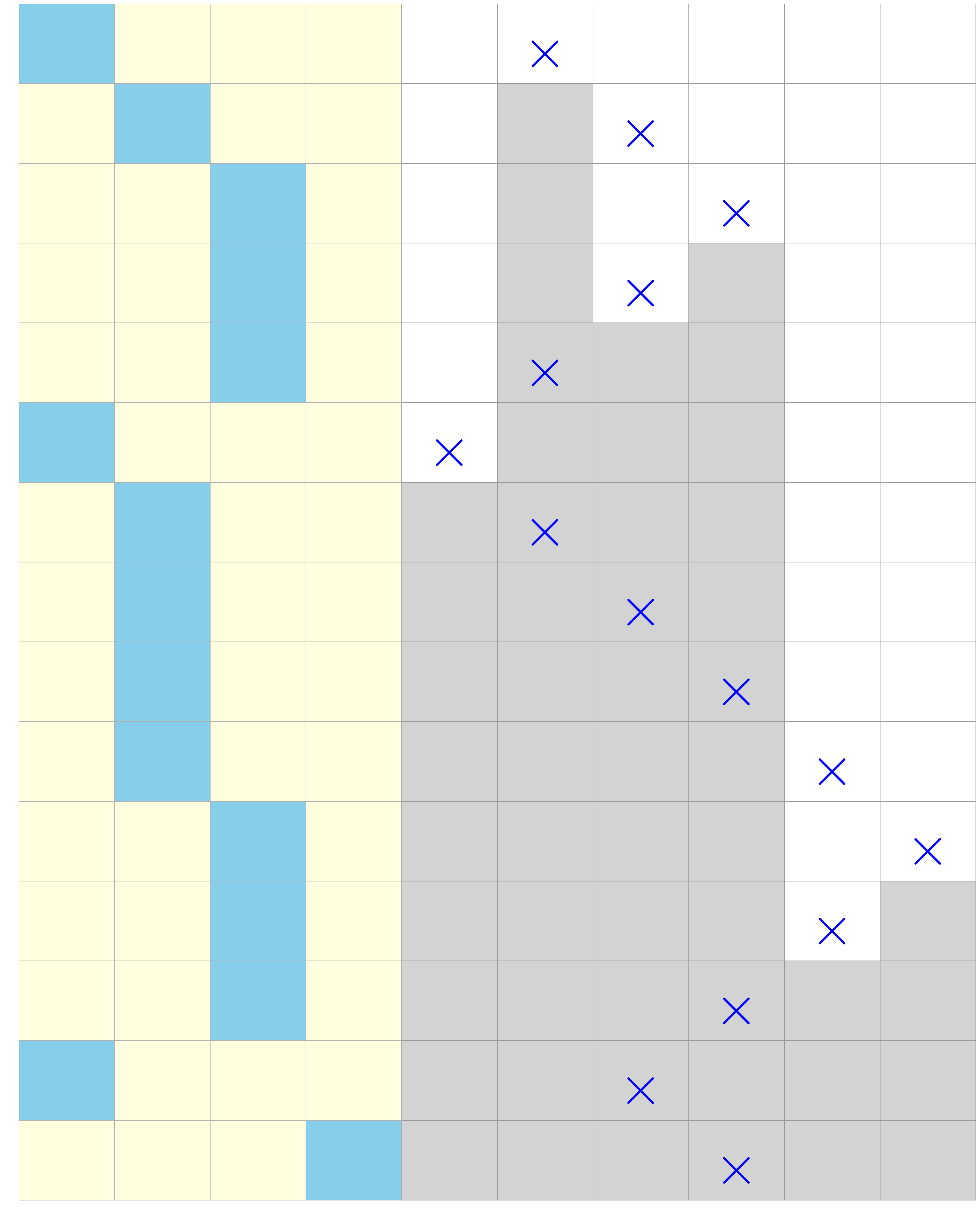}
}
\caption{Hybrid tuple plot time series view of a 2 (machine) state ``Busy Beaver'' TM, and a 3 state BB.
}
\label{fig:tm23}
\end{figure}

\begin{figure}[tp]
\centerline{
\includegraphics[height=0.85\textheight,valign=t]{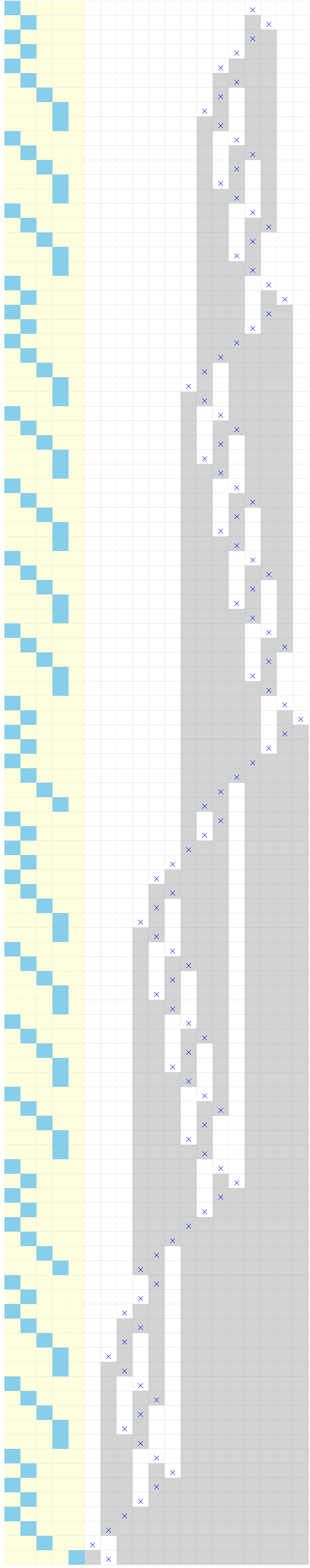}
}
\caption{Hybrid tuple plot time series view of a 4 (machine) state ``Busy Beaver'' TM.
}
\label{fig:tm4}
\end{figure}

\begin{figure}[tp]
\centerline{
\includegraphics[height=0.92\textheight,valign=t]{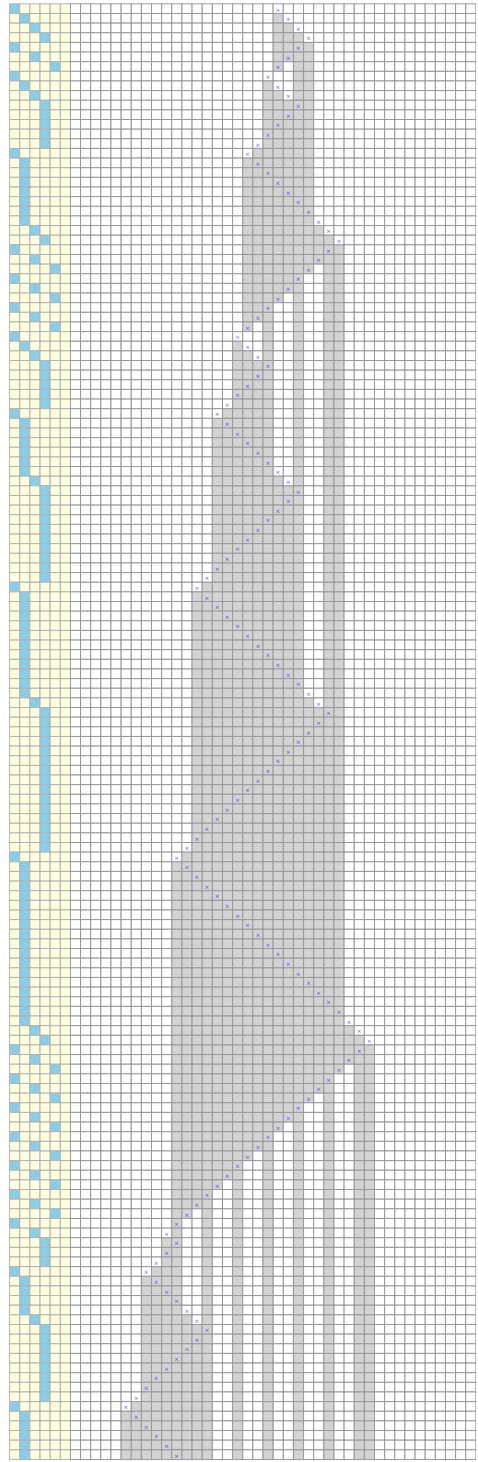}
}
\caption{Hybrid tuple plot time series view of the first 150 steps of a 5 (machine) state ``Busy Beaver'' TM.
}
\label{fig:tm5}
\end{figure}

Thus we have a heterogeneous set of coordinates: one for the machine state, $q \in Q$,
one for the head position, $h \in \mathbb Z$, and an infinite (unbounded) set for the state of the tape at each position, $t \in \mathbb Z \rightarrow \Gamma$.
If we draw the tape state in a plan tuple plot, and suitably arrange the other two coordinates,
we can produce a 1D representation of a TM state (figure~\ref{fig:tmpar}),
and a visualisation of a TM execution trajectory by a
time series of these states (for example, figures \ref{fig:tm23}--\ref{fig:tm5}).
This allows visualisation of the relationship between machine state changes and head position, for example.

Note how this involves a design decision on the state component.
The $\#Q$ states could be displayed using one axis with a $\#Q$-valued state (as in the leftmost axis of figure~\ref{fig:tmpar}a), or as $\#Q$-axes of binary values with the constraint that exactly one of the values may be `on' (the leftmost part of figure~\ref{fig:tmpar}b).
We use the latter for the hybrid tuple plots, as it is more visually accessible.

%............................................
\subsection{Flocking}\label{sec:hybrid:flock}

\begin{figure*}[t]
\centering
(a)\includegraphics[width=0.8\textwidth]{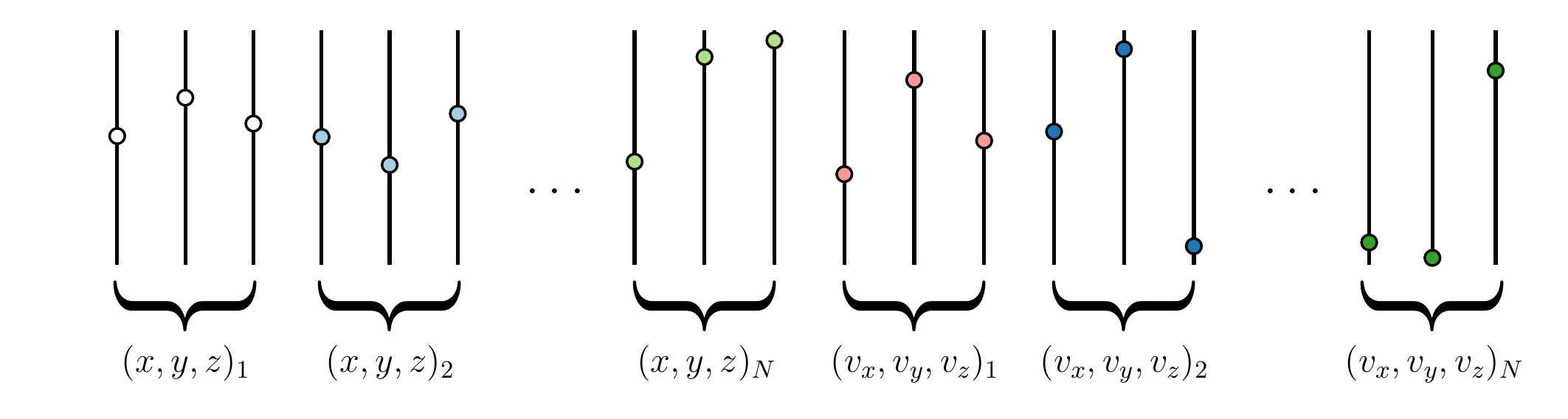}

(b)\includegraphics[width=0.8\textwidth]{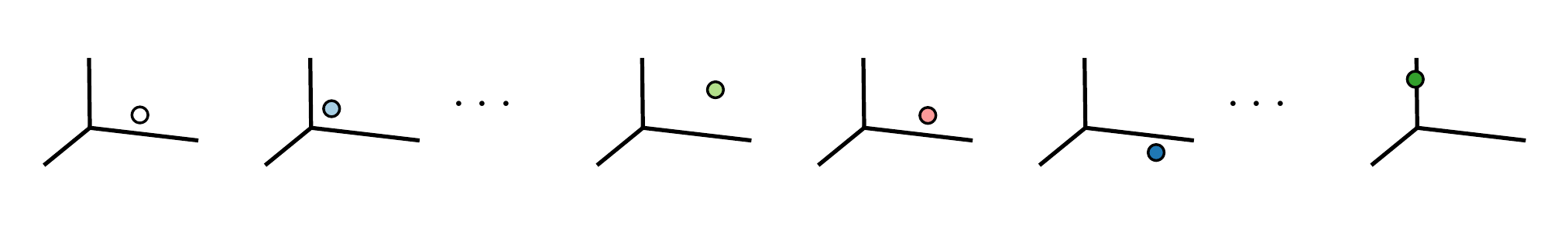}

(c)\includegraphics[width=0.8\textwidth]{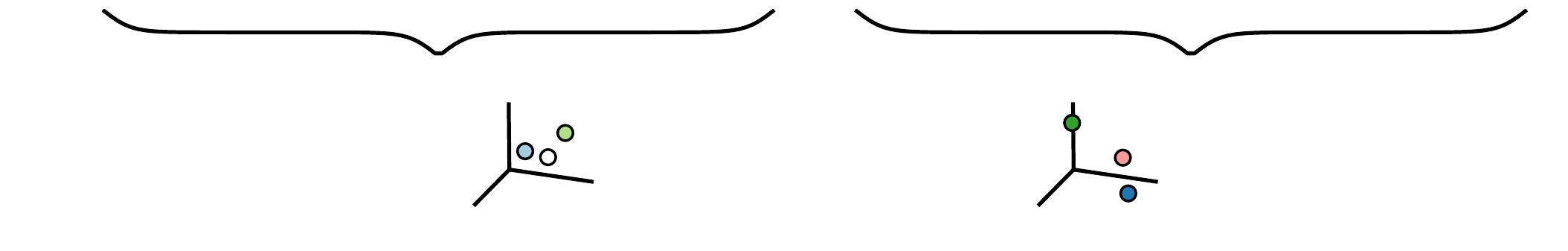}
\caption{Views of a $6N$ dimensional phase space comprising 3D positions and velocities of $N$ particles:
(a) parallel coordinates;
(b) hybrid tuple plot: each individual particle 3D position and velocity set is combined into an orthogonal coordinate set, and the different particles' spaces are shown in parallel;
(c) side tuple plot}
\vspace{4cm}
\label{fig:hyproj}
\end{figure*}

Consider an $N$ particle flocking simulation, with a $6N$D state (phase) space, 
comprising 3D positions and 3D velocities of the $N$ particles.
The parallel coordinates are $6N$ lines, with values from $\mathbb R$ (figure~\ref{fig:hyproj}a).
However, these have structure.  $3N$ of the lines represent position, $3N$ represent velocity.
Moreover, triples of lines represent 3D position and 3D velocity of a single particle.
First, consider drawing each of these triples in conventional orthogonal coordinates
instead of parallel coordinates,
reserving the parallel form for the $N+N$ copes of these orthogonal triples (figure~\ref{fig:hyproj}b).
Then use the side projected form to overlay the $N$ position sets, and the $N$ velocity sets separately (figure~\ref{fig:hyproj}c).
The resulting hybrid orthogonal/side tuple plot is the same as the ordinary 3D view of $N$-particle positions and $N$-particle velocities in two separate 3D spatial plots. (Note that in typical flocking visualisations, only the spatial aspect is shown; here it is clear that the whole state-space plot includes the velocity part too.)

As with homogeneous side tuple plots, this plot loses information about which point represents which set of axes (which particle).  This is particularly important in relating between the position and velocity plots. Colour may be used to highlight particular particles across the plots.

Particle space superficially looks homogeneous in the dimensions,
whereas our hybrid form is not homogeneous.
In particle state space, rotating the axes in the individual 3D spaces make sense: the choice of 3D spatial axes is arbitrary.  However, arbitrary rotations of the axes in the full $6N$-D phase space do not make the same kind of sense, as this would merge ``individual'' particles.  The choice of the $N$ sets of 6D axes is {\sl not} arbitrary.  Once the $N$ subsets of pairs of triples have been identified, however, the $N$ sets can be rearranged (essentially relabelling the particles),
which demonstrates that a side plot is suitable.   The ``lost'' information is just the arbitrary particle label. The topology on the $N$ sets is of a totally connected graph: all particles communicate with all other particles.

%================================================================
\chapter{Further work}\label{sec:further}
\balance
This report defines how to plot a state space of fixed size at fixed resolution.
If this approach helps to make the structure of complex state spaces clear, then there are techniques enhancements that could be developed.

%\section{Zooming or coarse graining}

It should be possible to `zoom out', to get a coarse-grained overview of a state structure.
In plan tuple plots, zooming out could make (some) axes move closer together and merge, so they are not individually resolved, but display the combined value, or some kind of average value, of the combined state components.
Some axes might, for some reason, be `closer together' than others, and so merge first;
for example, in a boids plot, the three axes of a single boid's position or velocity might merge to give a scalar position or speed.
This violates Inselberg's original `equidistant axes' rule, but that rule is used to allow geometrical reasoning over the space, so need be followed only in such cases.

In the trajectory plan or side plots, the time dimension could be coarse grained in a similar manner.
If the coarse graining timescale is an attractor cycle length, this would covert a multi-step dynamic behaviour into a single static value.

Correspondingly, it should be possible to `zoom in' to restore the original.
In such a case, it might not be immediately obvious at first glance what the underlying dimensionality of the space is: there might be a further zoom in available that would reveal more structure.

Dimensions might be lost on zooming out because they become somehow too small to see:
a toroidal tube looks like a 3D tube when close, a 2D torus when zoomed out somewhat, a 1D ring when zoomed out further, a 0D point from a distance, and invisible from a great distance.
A flock from a distance might be describable as one macroscopic emergent point entity;
closer in the individual boids become apparent, and their microstate dimensions appear.

The number of dimensions might change because the system state becomes larger:
a new tape square is added to a Turing Machine; a boid is born; a new object is created in a program.

How to change the number of dimensions the case of discrete time is relatively straightforward: a new dimension axis is added at the relevant timestep.  For a case like a Turing Machine or a growing CA, this can be seen as a `lazy' extension of a finite but growing state space.
The state space might be seen as hierarchical: a macro-dimension for an object with micro-dimensions for its internals.

For continuous time systems, the mechanism(s) of state space growth are not so clear:
a new dimension might initially be very small (and so invisible unless zoomed in sufficiently), then steadily grow or uncurl; there might be some fractal structure; it might ``fade in''.

%Uncurling -- dim getting longer (genuine modulo, v lazy extension)
%
%New dims (eg zoom in on Cantor Dust, as a plan parallel repn?)

%hierarchical?  eg each macro-coord is the state space of an object - look inside, micro-coords of the object

%================================================================
%\section{Distributions and landscapes}
%
%Prob distribution, fitness landscape : height on a landscape -- scalar field -- just another dimension?  (But of multiple points -- contour approach) -- attractor basins?
%
%%================================================================
%\section{Quantum spaces}
%
%Parallel Bloch spheres -- does this work?
%
%Hilbert curve repn of hypercube -- talk about earlier, in Plan chapter?
%
%%================================================================
%\section{Time}
%
%1D CA approach
%
%Animation
%
%%================================================================
%\section{Further issues}
%
%Polar coords??  (ie, non-linear dimensions) -- a different form of hybrid -- or of original parallel coords?
%
%Relative v absolute config space
%
%Visualising neighbourhoods/adjacent possible (look at von Neumann neighbourhood of a hypercube in parallel, in plan -- Hilbert?)
%
%Visualising reconstructing the attractor

%================================================================
\chapter{Summary and Conclusions}

As new unconventional computational devices are developed, new tools are needed to help explore their behaviour.  
Visualisation is a powerful tool, and parallel coordinates are one way of visualising high dimensional spaces.

We have described the use of parallel coordinates for visualising state space, and trajectories through that space.
Two generalised forms introduced here, plan tuple plots and side tuple plots, can be used to simplify, condense, and clarify the plots.

When the various dimensions of the state space can be interpreted as individuals, such as cells in a CA, or nodes in an RBN or neural network,
then the tuple plots coincide with some conventional plots in the literature.

Plan tuple ensemble plots in these cases plot the individuals along a (typically horizontal) line, and indicate their state values through symbols or colour scales.
These plots coincide with conventional ways of plotting the time evolution of ECAs (\S\ref{sec:plan-ca}) and RBNs (\S\ref{sec:plan:rbn}).
Here we generalise, both by highlighting the role of axis ordering to clarify structure (\S\ref{sec:axisorder}), and by applying to non-boolean systems (\S\ref{sec:plan:clm}, \S\ref{sec:plan:reservoir}).

Side tuple plots in these cases plot the individuals along the same (typically vertical) value axis, and distinguish between individuals through symbols or colour scales. 
These plots coincide with conventional ways of plotting multiple individuals in a single individual's state space, such as flocking particles (\S\ref{sec:hybrid:flock}).
Here we expose that structure, and generalise to ensembles, typically ordered time series, of such plots. 

We have unified these conventional cases into a generalised common framework.
With careful design of hybrid coordinate systems that incorporate multiple forms of plot,
much information about the dynamics of a system can be compactly encoded and visualised (\S\ref{sec:hybrid:tm}).

Using these plots in a standardised way should help communicate complex systems dynamics more clearly, with less cognitive load needed for interpreting the various plots.

%\subsection*{Acknowledgements}
%
%Simon P -- radial plot idea
\appendix
\footnotesize
\bibliographystyle{plain}
\bibliography{report}

\end{document}